\documentclass[apj,twocolumn,numberedappendix]{openjournal}

\usepackage{amsmath}
\usepackage{graphicx}
\usepackage{booktabs}
\usepackage{xcolor}
\definecolor{famNEF}{HTML}{E08214}\definecolor{famPopIII}{HTML}{C51B8A}\definecolor{famXRB}{HTML}{238B45}\definecolor{famAGN}{HTML}{8C510A}
\usepackage{xspace}
\usepackage{hyperref}

\hypersetup{
  colorlinks=true,
  citecolor=blue,
  linkcolor=blue,
  urlcolor=blue
}

\newcommand{\orcidlink}[1]{\href{https://orcid.org/#1}{%
  \raisebox{-0.2ex}{\includegraphics[height=1.45ex]{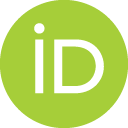}}}}
\newcommand{\authorname}[2][]{\textsc{#2}\if\relax\detokenize{#1}\relax\else\,\orcidlink{#1}\fi}
\makeatletter
\def\frontmatter@authorformat{\small\scshape\centering}
\def\frontmatter@title@below{\vspace*{-2.63\baselineskip}\vspace*{0.25in}}
\makeatother
\newcommand{\JWST}{\textit{JWST}\xspace}
\newcommand{\NIRSpec}{\textit{NIRSpec}\xspace}
\newcommand{\NIRCam}{\textit{NIRCam}\xspace}
\newcommand{\MUSE}{\textit{MUSE}\xspace}
\newcommand{\Cloudy}{\textsc{Cloudy}\xspace}
\newcommand{\lya}{Ly$\alpha$\xspace}
\newcommand{\hb}{H$\beta$\xspace}
\newcommand{\ha}{H$\alpha$\xspace}
\newcommand{\heii}{He\,\textsc{ii}\xspace}
\newcommand{\oii}{[O\,\textsc{ii}]\xspace}
\newcommand{\oiii}{[O\,\textsc{iii}]\xspace}
\newcommand{\neiii}{[Ne\,\textsc{iii}]\xspace}
\newcommand{\sii}{[S\,\textsc{ii}]\xspace}
\newcommand{\ciii}{C\,\textsc{iii}]\xspace}
\newcommand{\civ}{C\,\textsc{iv}\xspace}

\newcommand{\muv}{M_{\mathrm{UV}}}
\newcommand{\Zgas}{Z_{\mathrm{gas}}}
\newcommand{\Zstar}{Z_{\star}}

\makeatletter
\renewcommand\title[2][]{%
  \def\@title{#2}%
  \def\@shorttitle{#1}%
  \let\@AF@join\@title@join}
\makeatother

\begin{document}

\title[Active Pop~III star formation in AMORE6]{%
  JWST/NIRSpec observations of AMORE6 suggest active Pop~III star formation in a self-polluted galaxy}
\shorttitle{Active Pop~III star formation in AMORE6}
\shortauthors{Curti et al.}

\author{
\authorname[0000-0002-2678-2560]{Mirko Curti}\altaffilmark{1},
\authorname[0009-0006-4326-6097]{Elka Rusta}\altaffilmark{2,3},
\authorname[0000-0001-7298-2478]{Stefania Salvadori}\altaffilmark{3,4},
\authorname[0000-0002-6719-380X]{Stefano Carniani}\altaffilmark{5},
\authorname[0000-0003-2388-8172]{Francesco D'Eugenio}\altaffilmark{6,7},
\authorname[0009-0004-7069-1326]{Bianca Moreschini}\altaffilmark{3,4,8},
\authorname[0000-0002-4985-3819]{Roberto Maiolino}\altaffilmark{6,7,9},
\authorname[0000-0002-9889-4238]{Alessandro Marconi}\altaffilmark{3,4},
\authorname[0000-0001-8349-3055]{Giacomo Venturi}\altaffilmark{4},\\
\authorname[0000-0002-2545-5752]{Francesco Belfiore}\altaffilmark{8},
\authorname[0009-0005-5187-4001]{Elisa Cataldi}\altaffilmark{3,4},
\authorname[0000-0001-5487-0392]{Viola Gelli}\altaffilmark{8,10,11},
\authorname[0000-0001-7730-8634]{Yuki Isobe}\altaffilmark{6,7,12},
\authorname[0000-0002-1660-9502]{Xihan Ji}\altaffilmark{6,7},
\authorname[0000-0001-6251-649X]{Mingyu Li}\altaffilmark{6,7,13},
\authorname[0000-0001-9820-5773]{Robert G. Pascalau}\altaffilmark{6,7},
\authorname[0000-0002-9081-2111]{James A. A. Trussler}\altaffilmark{14},
\authorname[0000-0003-4891-0794]{Hannah \"Ubler}\altaffilmark{15}
}

\affiliation{%
\altaffilmark{1} INAF -- Osservatorio di Astrofisica e Scienza dello Spazio di Bologna, Via Piero Gobetti 93/3, I-40129 Bologna, Italy\\
\altaffilmark{2} Universit\`a di Firenze, Dipartimento di Fisica e Astronomia, Via G. Sansone 1, I-50019 Sesto Fiorentino, Florence, Italy\\
\altaffilmark{3} DTU Space, Technical University of Denmark, Elektrovej 327, DK-2800 Kgs. Lyngby, Denmark\\
\altaffilmark{4} INAF -- Osservatorio Astrofisico di Arcetri, Largo E. Fermi 5, I-50125 Florence, Italy\\
\altaffilmark{5} Scuola Normale Superiore, Piazza dei Cavalieri 7, I-56126 Pisa, Italy\\
\altaffilmark{6} Kavli Institute for Cosmology, University of Cambridge, Madingley Road, Cambridge CB3 0HA, UK\\
\altaffilmark{7} Cavendish Laboratory, University of Cambridge, 19 JJ Thomson Avenue, Cambridge CB3 0HE, UK\\
\altaffilmark{8} European Southern Observatory, Karl-Schwarzschild-Strasse 2, D-85748 Garching bei M\"unchen, Germany\\
\altaffilmark{9} Department of Physics and Astronomy, University College London, Gower Street, London WC1E 6BT, UK\\
\altaffilmark{10} Cosmic Dawn Center (DAWN), Copenhagen, Denmark\\
\altaffilmark{11} Niels Bohr Institute, University of Copenhagen, Jagtvej 128, DK-2200 Copenhagen N, Denmark\\
\altaffilmark{12} Waseda Research Institute for Science and Engineering, Faculty of Science and Engineering, Waseda University, 3-4-1 Okubo, Shinjuku, Tokyo 169-8555, Japan\\
\altaffilmark{13} Department of Astronomy, Tsinghua University, Beijing 100084, People's Republic of China\\
\altaffilmark{14} Center for Astrophysics $|$ Harvard \& Smithsonian, 60 Garden Street, Cambridge, MA 02138, USA\\
\altaffilmark{15} Max-Planck-Institut f\"ur extraterrestrische Physik (MPE), Gie{\ss}enbachstra{\ss}e 1, D-85748 Garching, Germany
}

\email{mirko.curti@inaf.it}

\begin{abstract}
AMORE6 is a strongly lensed ($\mu\simeq78$), intrinsically faint galaxy at $z=5.725$, previously identified as a candidate for metal-free star formation from the absence of \oiii$\lambda5007$ in \NIRCam slitless spectroscopy. Here, we present deep \JWST/\NIRSpec spectroscopy of its brighter image, AMORE6-B. We detect \heii$\lambda1640$ and \heii$\lambda4686$, with \heii$\lambda4686$/\hb$\,=0.080\pm0.015$ and EW(\heii$\lambda4686)=35\pm8$\,\AA, weak \oiii ($R_3=$ \oiii$\lambda5007$/\hb $=0.25$) and \civ$\lambda\lambda1548,51$, and no other metal lines.
The hydrogen and helium lines give a negligible attenuation, $T_{\rm e}\simeq2\times10^{4}$\,K, and a helium mass fraction of $Y=0.226^{+0.022}_{-0.024}$. The oxygen abundance, $12+\log(\mathrm{O/H})\simeq6.08^{+0.24}_{-0.80}$ (i.e. in the range $0.04\%-0.4\%~Z_\odot$, depending on the assumed calibration), is the lowest among any known galaxies of similar stellar mass, while C/O is close to or above the solar value. 
Based on photoionization modeling of different families of ionizing sources, the emission lines favour metal-free stars with a top-heavy initial mass function (IMF), as in the self-polluted phase of simulated Population~III (Pop~III) galaxies, or a metal-poor population with a luminous hard X-ray source; ordinary metal-poor populations, very massive stars, and accretion-based sources are disfavoured. 
A full SED fitting under the fiducial Pop~III self-polluted scenario delivers a total mass of $1.3\times10^{5}\,M_\odot$ of Pop~III stars formed within the last 4\,Myr, which ionize mildly enriched gas while letting 86\% of their ionizing photons escape. 
The \heii-to-Balmer ratio constrains the slope $x$ and the characteristic mass $M_{\rm ch}$ of the Pop~III IMF jointly, to $90x-268\lesssim M_{\rm ch}/M_\odot\lesssim172x-330$ ($3\sigma$; $M_{\rm ch}\lesssim70\,M_\odot$ for a Salpeter slope), and excludes the most top-heavy IMFs allowed by the near-field constraints.
X-ray binaries are a possible alternative, but require either a metal-poor population too young (1\,Myr) to host compact objects and an X-ray luminosity orders of magnitude above that of metal-poor galaxies, or $2\times10^{6}\,M_\odot$ of older Pop~III.2 stars whose metals were mostly expelled. 
\end{abstract}


\maketitle

\section{Introduction}
\label{sec:introduction}

The first generation of stars, the so-called Population~III (Pop~III), formed out of gas composed exclusively of hydrogen, helium, and traces of lithium produced in the Big Bang. In the absence of metals, cooling is inefficient and fragmentation is suppressed, so that Pop~III stars are expected to be, on average, considerably more massive than present-day stars \citep{KlessenGlover2023}. Their extremely hot photospheres would produce copious hydrogen- and helium-ionizing photons, leaving a distinctive imprint on the nebular spectrum of their host systems, such as very large Balmer-line equivalent widths, strong \heii recombination emission, and the complete absence of collisionally excited metal lines \citep{Schaerer2002,Schaerer2003}. Although Pop~III star formation is generally expected to peak at very early epochs, theoretical models predict that inefficient metal mixing can preserve pockets of pristine gas in which metal-free stars continue to form well into the epoch of reionization, and possibly down to $z\sim6$ \citep{Trenti2009,Venditti2023,Venditti2025}. Finding such systems would provide a direct window on the conditions of the first star formation, on the primordial initial mass function, and on the first chemical enrichment of the Universe, whose imprint (e.g.\ super-solar C/O ratios in metal-poor gas) is now starting to be detected at $z>7$ \citep{Pollock2026,DEugenio2024gsz12,Nakajima2025}.

Identifying a genuine Pop~III system is, however, observationally challenging, because no single spectral feature constitutes an unambiguous signature \citep{NakajimaMaiolino2022,Trussler2023}. Strong \heii emission requires a hard ionizing spectrum, but this can also be supplied by very massive stars at non-zero metallicity \citep{Schaerer2025}, by stripped or binary stars \citep{Eldridge2017}, by X-ray binaries, or by accretion onto black holes, including direct-collapse black holes \citep{NakajimaMaiolino2022}; indeed, hard ionizing spectra producing strong high-ionization UV lines are not uncommon among metal-enriched galaxies in the reionization era \citep[e.g.][]{Topping2025}. Conversely, weak or undetected \oiii emission does not by itself imply an extremely low oxygen abundance: the strength of the forbidden lines also depends on the ionization parameter \citep{KewleyDopita2002,Kewley2019}, on the shape of the ionizing spectrum \citep{Gutkin2016,Steidel2016}, on the gas density (through collisional de-excitation, which can mimic a low O/H in the dense nebulae common at high redshift; \citealt{Harikane2025,Martinez2026,Moreschini2026,Hsiao2026}), and on the geometry of the ionized region, e.g.\ in density-bounded nebulae leaking ionizing photons \citep{JaskotOey2013,Zackrisson2013,Plat2019}. In addition, the presence of metal lines does not by itself rule out Pop~III stars, since galaxies hosting only metal-free stars can display metal emission once their gas has been polluted by the first supernovae \citep{Rusta2025}. Moreover, at the oxygen abundances of interest ($12+\log(\mathrm{O/H})\lesssim7$), metallicity diagnostics are largely uncalibrated and must rely on photoionization modelling or extrapolations \citep{Morishita2024,Nakajima2022empress, Isobe2026mzr}. A convincing case for Pop~III-like star formation therefore requires the simultaneous combination of several independent constraints: a hard ionizing spectrum, the joint absence of lines from multiple heavy elements (O, C, N, Ne, S), and the quantitative exclusion of alternative explanations.

The \JWST has dramatically advanced the search for such objects. Deep spectroscopy of strongly lensed sources has revealed star-forming complexes with oxygen abundances of only a few per cent of the solar value or less, such as LAP1 at $z=6.6$, with $12+\log(\mathrm{O/H})\approx6.5$--6.8 \citep{Vanzella2023,Nakajima2025,Scholtz2026}, while other extremely metal-poor systems have been reported both in lensed and blank fields \citep[e.g.][]{Willott2025,Cullen2025,Maiolino2026BHnearpristine,Li2026arXiv260924589L}. Slitless \NIRCam spectroscopy has enabled systematic searches for galaxies with weak or absent \oiii relative to the Balmer lines \citep{Hsiao2025}, and photometric selections targeting the colours expected for metal-free stellar populations have identified ultra-faint Pop~III candidates, providing the first constraints on the Pop~III UV luminosity function \citep{Fujimoto2025}. At the same time, spectroscopic follow-up has demonstrated how fragile these identifications can be: deeper \NIRSpec observations of the GLIMPSE-D candidate detected \oiii and ruled out a metal-free interpretation \citep{Fujimoto2026}. At higher redshift, a \heii emitter with no detected metal lines has been confirmed in the vicinity of GN-z11 at $z=10.6$, with the pristine nature of the gas supported by deep limits on the metal lines and a Pop~III interpretation favouring a top-heavy IMF, although non-stellar origins of its ionizing spectrum cannot yet be fully excluded \citep{Maiolino2026,Ubler2026,Rusta2026}. These results well illustrate both the relevance and the challenges of searching for Pop~III galaxies: candidates are now being found, but their confirmation requires deep, multi-line spectroscopy and a careful treatment of the degeneracies between metallicity and other properties of the ionized gas.

In this context, AMORE6 (Abell 2744 Metal-poor Object REdshift 6) stands out as one of the most compelling candidates identified to date \citep{Morishita2026}. AMORE6 is a doubly imaged source behind the galaxy cluster Abell 2744 at $z=5.725$, first identified through its \lya emission with \MUSE \citep{Mahler2018,Richard2021}, and magnified by factors of $\mu\simeq39$ and $\simeq78$ in its two images, AMORE6-A and AMORE6-B \citep{Bergamini2023a,Bergamini2023b}. Its intrinsic properties are extreme: an intrinsic UV magnitude of $\muv\simeq-14.5$, a stellar mass of only $\sim4\times10^{5}\,M_\odot$, a very blue UV slope, a light-weighted age of $\sim1.5$\,Myr, and a rest-frame \hb equivalent width of $\sim1000$\,\AA. Using \NIRCam wide-field slitless spectroscopy (WFSS) from the ``All the Little Things'' programme \citep{Naidu2024}, \citet{Morishita2026} detected \hb in both images but no \oiii$\lambda\lambda4959,5007$ emission, and translated the resulting line-ratio limit into an oxygen abundance of $12+\log(\mathrm{O/H})<5.97$ ($3\sigma$), i.e.\ below $\sim0.2\%$ of the solar value, making AMORE6 the most metal-poor galaxy candidate known.
This abundance limit, however, is inferred from a single line ratio assuming a specific set of photoionization-models, namely a fixed ionization parameter ($\log U=-0.5$), a gas density of $10^{3}$\,cm$^{-3}$, a 1\,Myr binary stellar population, and equal stellar and gas-phase metallicities \citep{Morishita2026}. A lower ionization parameter, a higher density approaching the critical density of the \oiii transitions, a different ionizing spectrum, or a density-bounded geometry could all suppress the \oiii emission at metallicities appreciably higher than the quoted limit. Moreover, the WFSS data cover only the observed 3--4\,$\mu$m range, and therefore provide no access to \heii, to the rest-frame UV metal lines, or to the density- and temperature-sensitive diagnostics required to break these degeneracies. The central question addressed in this paper is thus whether deep, multi-line spectroscopy confirms that AMORE6 hosts genuinely near-pristine gas, or whether unusual density, ionization, geometry, or a non-standard ionizing source can account for its spectrum.

To answer this question, in this work we analyze deep \JWST/\NIRSpec multi-object spectroscopy of AMORE6 obtained as part of the GO~9645 programme \citep{TreuGO9645,Morishita2026b}, covering the full rest-frame UV-to-optical range from $\sim0.14$ to $\sim0.78\,\mu$m with the G140M, G235M, and G395M gratings at $R\sim1000$. These data give access, within a single slit spectrum, to \heii$\lambda1640$ and $\lambda4686$, the UV lines of carbon, nitrogen, and oxygen, the \oii and \neiii lines, the Balmer series, the auroral \oiii$\lambda4363$ line, and the strong optical lines of oxygen, nitrogen, and sulphur, allowing us to test the Pop~III hypothesis through the joint behaviour of several independent diagnostics. 

The paper is organised as follows. In Section~\ref{sec:target} we present the imaging of the Abell 2744 field, the properties of AMORE6 known from previous work, and the strong-lensing model, and in Section~\ref{sec:photometry} the morphology and photometry of AMORE6-B. Section~\ref{sec:observations} presents the \NIRSpec observations and their reduction, and Section~\ref{sec:measurements} the emission-line and continuum measurements. In Section~\ref{sec:conditions} we derive the properties of the gas that follow directly from the spectrum: the attenuation, the electron temperature and density, the helium abundance, the ionization parameter, the ionizing-photon production efficiency, and the oxygen and carbon abundances. In Section~\ref{sec:models} we compare the spectrum with photoionization models of a set of competing ionizing sources, first in diagnostic diagrams and then by comparing all the lines with the models, and in Section~\ref{sec:sed} we model the spectral energy distribution with the families favoured by the lines to derive the properties of the stellar population. We discuss the implications of these results for the nature of AMORE6 and its chemical enrichment, for the initial mass function of Pop~III stars, and for alternative ionizing sources and the escape of ionizing photons in Section~\ref{sec:discussion}, and summarise our conclusions in Section~\ref{sec:conclusions}.

\section{Available data}
\label{sec:target}

\subsection{Imaging of the Abell 2744 field}
\label{sec:imaging}

Abell 2744 ($z=0.308$) is a massive galaxy cluster and one of the best-studied strong-lensing fields. It was one of the six Hubble Frontier Fields \citep{Lotz2017}, and since the start of \JWST operations it has been imaged with \NIRCam by several programmes: GLASS-JWST \citep[ERS~1324;][]{Treu2022}, UNCOVER \citep[GO~2561;][]{Bezanson2024}, DD~2756, GO~2883, All the Little Things \citep[ALT, GO~3516;][]{Naidu2024}, GO~3538, BEACON \citep[GO~3990;][]{Morishita2025}, and Medium Bands, Mega Science \citep[GO~4111;][]{Suess2024}. Together with the \textit{HST}/ACS F435W, F606W, and F814W images, these data provide 23 filters between 0.4 and 4.8\,$\mu$m at the position of AMORE6: all eight \NIRCam broad-band filters from F070W to F444W and 12 medium-band filters from F140M to F480M. The $5\sigma$ point-source depths are 28.5--29.1\,AB mag in the \NIRCam broad bands and 26.8--28.6\,AB mag in the medium bands \citep{Morishita2026}.

We use the mosaics of the third data release of UNCOVER and Medium Bands, Mega Science (DR3; \citealt{Bezanson2024,Weaver2024,Suess2024}), reduced with \textsc{grizli} v7.2 \citep{Brammer2023grizli} from all public \NIRCam imaging of the field, drizzled to $0.02''$ (short-wavelength channel) and $0.04''$ (long-wavelength channel) pixels on a common grid, and registered to the \textit{Gaia}~DR3 astrometric frame. Our fiducial photometry of AMORE6-B is measured on the DR3 versions of these mosaics in which models of the bright cluster galaxies and of the local intracluster light have been subtracted, at the native resolution of each band, using the empirical PSFs released with them; the original mosaics are used for the RGB image of Fig.~\ref{fig:overview}, PSF-matched to F444W with the released kernels. Around AMORE6 we verified the astrometry of the mosaics independently. The UNCOVER and GLASS-JWST catalogue positions agree to within $0.001''$ (median, 37 sources within $20''$; scatter $0.009''$ per axis), the F444W centroids of 72 isolated stars agree with the UNCOVER positions to $\simeq5$\,mas, and the three secure \textit{Gaia}~DR3 matches, propagated to the epoch of the observations, agree to $\simeq0.01''$. Any astrometric offset between the imaging and the frame of the MSA configuration, which was designed on the same mosaics, is therefore $\lesssim0.01''$. These images are used for the morphology and photometry of AMORE6-B (Section~\ref{sec:photometry}), and for modelling the scene within the MSA slitlet and the sources in overlapping shutters (Section~\ref{sec:observations}).

The field is also covered by extensive spectroscopy. This includes \MUSE integral-field spectroscopy of the cluster core \citep{Mahler2018,Richard2021}, \NIRCam F356W wide-field slitless spectroscopy (WFSS) from ALT \citep{Naidu2024}, and \NIRSpec prism spectroscopy from UNCOVER \citep{Bezanson2024}. These surveys supply the spectroscopic redshifts of the multiple images that constrain the lens model (Section~\ref{sec:lensmodel}) and of the sources whose spectra overlap that of AMORE6 on the \NIRSpec detectors (Section~\ref{sec:observations}).

\subsection{AMORE6}
\label{sec:amore6}

\begin{figure*}
  \centering
  \includegraphics[width=\textwidth]{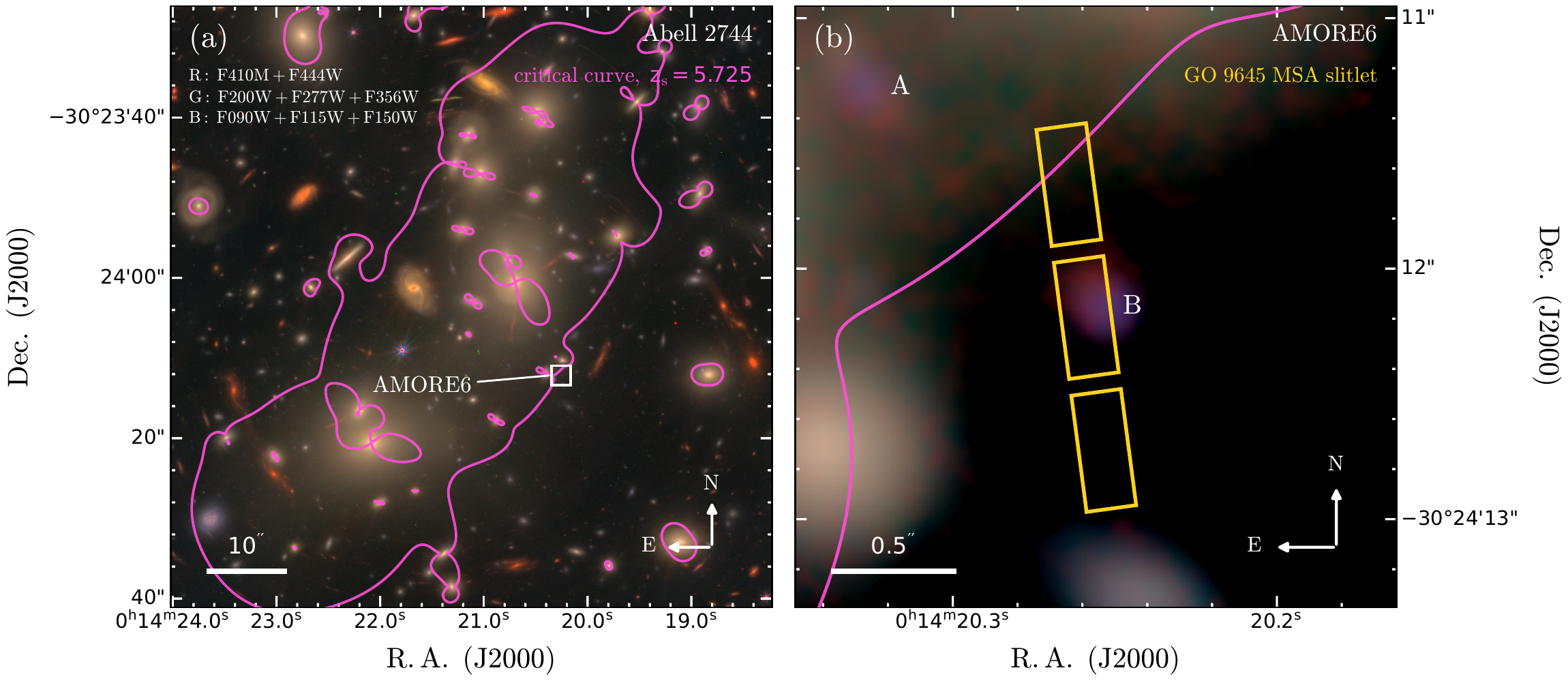}
  \caption{\textit{(a)} \NIRCam RGB image of the core of Abell 2744 from the UNCOVER mosaics \citep{Bezanson2024,Weaver2024}, with F090W+F115W+F150W in blue, F200W+F277W+F356W in green, and F410M+F444W in red. The magenta line marks the critical curve for a source at $z=5.725$ from the best-fit \textsc{Lenstool} strong-lensing model of \citet{Bergamini2023b}, and the white square marks the region shown in panel (b). \textit{(b)} Zoom on the two images of AMORE6, with all bands PSF-matched to F444W. AMORE6-A and AMORE6-B lie on opposite sides of the critical curve. The yellow rectangles show the GO~9645 three-shutter MSA slitlet containing AMORE6-B at the first nod position, from the flight MSA metadata (Section~\ref{sec:observations}).}
  \label{fig:overview}
\end{figure*}

AMORE6 was first identified through its \lya emission in the \MUSE data \citep{Mahler2018,Richard2021}, which detected the line at the positions of both images and established a redshift $z=5.725$. Its two images, AMORE6-A and AMORE6-B, lie $\simeq1.3''$ apart, west of the main cluster core, on opposite sides of the critical curve for a source at that redshift (Fig.~\ref{fig:overview}). The known properties of the system, from \citet{Morishita2026} unless otherwise stated, are listed in Table~\ref{tab:amore6_literature}, together with the values measured in this work.

\begin{deluxetable*}{llcccc}
  \tablecaption{Properties of AMORE6.\label{tab:amore6_literature}}
  \tablehead{
    \colhead{Quantity} & \colhead{Unit} & \colhead{AMORE6-A} & \colhead{AMORE6-B} & \colhead{A+B\tablenotemark{a}} & \colhead{AMORE6-B, this work}
  }
  \startdata
  \multicolumn{6}{l}{\textit{Positions}} \\
  RA, Dec (J2000)                & deg & 3.584723, $-30.40316$ & 3.584413, $-30.40340$ & \nodata & \nodata \\
  GO~9645 MSA target             & deg & \nodata & \nodata & \nodata & 3.5843994, $-30.4033756$ \\
  \addlinespace
  \multicolumn{6}{l}{\textit{Lensing \citep{Bergamini2023b}}} \\
  $\mu$ \citep{Morishita2026}    & \nodata & $39.3^{+3.7}_{-3.5}$ & $77.7^{+8.4}_{-5.9}$ & \nodata & \nodata \\
  $\mu$ (this work)\tablenotemark{b} & \nodata & $39.2^{+4.4}_{-3.7}$ & $77.8^{+8.2}_{-6.9}$ & \nodata & \nodata \\
  \addlinespace
  \multicolumn{6}{l}{\textit{Spectroscopy (\NIRCam WFSS and \MUSE; this work: \NIRSpec)}} \\
  $z$(\hb)\tablenotemark{f}      & \nodata & $5.7253^{+0.0011}_{-0.0010}$ & $5.7253\pm0.0005$ & $5.7253\pm0.0005$ & $5.72354\pm0.00002$ \\
  $F$(\hb)                       & $10^{-19}$\,erg\,s$^{-1}$\,cm$^{-2}$ & $15.7\pm3.8$ & $22.6\pm3.7$ & \nodata & $20.3\pm1.8$ \\
  $F$(\oiii$\lambda5007$)          & $10^{-19}$\,erg\,s$^{-1}$\,cm$^{-2}$ & $<9.5$ & $<9.8$ & \nodata & $5.1\pm0.7$ \\
  EW$_0$(\hb)                    & \AA & $1289^{+456}_{-386}$ & $939^{+248}_{-210}$ & $1122^{+260}_{-220}$ & $478^{+70}_{-62}$ \\
  $R_3=$\,\oiii$\lambda5007$/\hb     & \nodata & $<0.60$ & $<0.43$ & $<0.33$ & $0.249\pm0.025$ \\
  $12+\log(\mathrm{O/H})$\tablenotemark{c} & \nodata & $<6.24$ & $<6.09$ & $<5.97$ & $6.08^{+0.24}_{-0.80}$ \\
  $F$(\lya)/$F$(\hb)               & \nodata & \nodata & \nodata & $14.7\pm2.0$ & \nodata \\
  $\Delta v$(\lya$-$\hb)           & km\,s$^{-1}$ & \nodata & \nodata & $14\pm150$ & \nodata \\
  FWHM$_\mathrm{intr}$(\lya)     & km\,s$^{-1}$ & \nodata & \nodata & $69.3^{+0.8}_{-0.7}$ & \nodata \\
  \addlinespace
  \multicolumn{6}{l}{\textit{Photometry and SED of AMORE6-B, image plane}\tablenotemark{d}} \\
  $\muv-2.5\log\mu$              & AB mag & \nodata & $-19.25^{+0.07}_{-0.08}$ & \nodata & $-19.95\pm0.08$ \\
  $\beta_\mathrm{UV}$            & \nodata & \nodata & $-2.77^{+0.07}_{-0.09}$ & \nodata & $-3.01\pm0.23$ \\
  $\log(\mu M_\star/M_\odot)$    & \nodata & \nodata & $7.53^{+0.18}_{-0.08}$ & \nodata & $6.99\pm0.04$ \\
  Light-weighted age             & Myr & \nodata & $1.5^{+1.2}_{-1.0}$ & \nodata & \nodata \\
  Age of the oldest stars        & Myr & \nodata & \nodata & \nodata & 4 \\
  Mass-weighted age              & Myr & \nodata & \nodata & \nodata & 1.3 \\
  $\log(\mu\,\mathrm{SFR})$ from Balmer lines\tablenotemark{e} & $M_\odot$\,yr$^{-1}$ & \nodata & $0.71^{+0.06}_{-0.08}$ & \nodata & $0.64\pm0.03$ \\
  $\log(\mu\,\mathrm{SFR}_\mathrm{10\,Myr})$ & $M_\odot$\,yr$^{-1}$ & \nodata & $0.42^{+0.10}_{-0.12}$ & \nodata & \nodata \\
  $\log\xi_\mathrm{ion}$         & Hz\,erg$^{-1}$ & \nodata & \nodata & \nodata & $25.52^{+0.06}_{-0.07}$ \\
  $r_\mathrm{e}$ (image plane, F115W) & mas & $39.5\pm3.1$ & $47.6\pm1.9$ & \nodata & \nodata \\
  $R_\mathrm{e}$ (source plane)  & pc & \nodata & \nodata & $36.0^{+9.4}_{-7.5}$ & \nodata \\
  $\log\Sigma_\star$             & $M_\odot$\,pc$^{-2}$ & \nodata & \nodata & $2.26^{+0.18}_{-0.08}$ & \nodata \\
  $\log\Sigma_\mathrm{SFR}$      & $M_\odot$\,yr$^{-1}$\,kpc$^{-2}$ & \nodata & \nodata & $2.16^{+0.06}_{-0.08}$ & \nodata
  \enddata
  \tablecomments{The columns labelled ``AMORE6-A'', ``AMORE6-B'', and ``A+B'' are from \citet{Morishita2026}, except the magnifications labelled ``this work''; the last column is from the \NIRSpec spectrum and the photometry of Section~\ref{sec:photometry} (lines and photometry on the total-flux scale) and from the SED model of Section~\ref{sec:sed}. Uncertainties are $1\sigma$ and upper limits $3\sigma$; those of the absolute \NIRSpec line fluxes, and of the quantities derived from them, include the uncertainty of the path-loss correction (Table~\ref{tab:line_fluxes}). Line fluxes and photometry are not corrected for magnification. The \lya quantities come from the \MUSE spectra integrated over both images.}
  \tablenotetext{a}{Stack of the WFSS spectra of A and B, each divided by its magnification; the source-plane size comes from the joint reconstruction of both images.}
  \tablenotetext{b}{At the model-predicted image positions (Section~\ref{sec:lensmodel}; systematic uncertainty $\pm0.10$\,dex).}
  \tablenotetext{c}{\citet{Morishita2026}: from $R_3$ with \Cloudy models for a 1\,Myr BPASS population, $\log U=-0.5$, and $n_\mathrm{H}=10^3$\,cm$^{-3}$. This work: median and range of the $R_3$ calibrations (Section~\ref{sec:empirical_oh}).}
  \tablenotetext{d}{Quantities that scale with the magnification are given in the image plane, i.e.\ multiplied by $\mu$; the values of \citet{Morishita2026}, corrected for $\mu_\mathrm{B}=78$, are multiplied by 78. Intrinsic values follow from division by $\mu_\mathrm{B}$; for $\mu_\mathrm{B}=77.8$ (this work) they are given in Section~\ref{sec:sed}. The source-plane size scales as $\mu^{-1/2}$, and the surface densities do not depend on $\mu$.}
  \tablenotetext{f}{This work: nebular redshift from all the lines (statistical uncertainty).}
  \tablenotetext{e}{\citet{Morishita2026}: from \hb. This work: from \ha, corrected for the attenuation of Section~\ref{sec:attenuation}, with the conversion $\log\mathrm{SFR/(M_\odot~yr^{-1})}=\log L(\mathrm{H}\alpha)/\mathrm{(erg~s^{-1})}-41.67$ implied by theirs.}
\end{deluxetable*}

\paragraph{Previous measurements.}
The redshift and the emission-line constraints of Table~\ref{tab:amore6_literature} come from the F356W/GRISM~R WFSS observations of ALT \citep{Naidu2024}, which cover 3--4\,$\mu$m at the position of AMORE6. These were taken at two position angles ($56^\circ$ and $60^\circ$), each with 24.7\,ks on source. \citet{Morishita2026} subtracted the continuum with a median filter and extracted the spectra of both images separately at each position angle, to identify spectral contamination from neighbouring sources. The PA\,$=56^\circ$ spectrum of B is contaminated by a bright cluster galaxy $0.4''$ to the south-east, while that at PA\,$=60^\circ$ is mostly clean. \hb is detected in both images and at both position angles, at the wavelength expected from the \MUSE \lya redshift ($\Delta v=14\pm150$\,km\,s$^{-1}$), whereas \oiii$\lambda\lambda4959,5007$ is undetected in all spectra. Stacking the two images, after dividing each by its magnification, gives $R_3<0.33$ ($3\sigma$). With \Cloudy models powered by a 1\,Myr BPASS population of the same metallicity as the gas, at $\log U=-0.5$ and $n_\mathrm{H}=10^3$\,cm$^{-3}$, this translates into $12+\log(\mathrm{O/H})<5.97$. Adopting $\log U=-2.5$ or $n_\mathrm{H}=10^6$\,cm$^{-3}$ would raise this limit by $\simeq0.4$ and $\simeq0.6$\,dex, respectively. The \lya/\hb ratio of $14.7\pm2.0$, compared with 23.5 for Case~B recombination, implies a \lya escape fraction of $63^{+8}_{-9}\%$.

The broad-band properties are derived from AMORE6-B alone, because A lies close to two bright foreground galaxies at $z\sim0.5$ and to a compact source of uncertain association ($z_\mathrm{phot}=5.5^{+0.3}_{-0.4}$) to its north-east. \citet{Morishita2026} reduced the 20 \NIRCam bands consistently, on a common $0.0315''$ grid registered to \textit{Gaia}~DR3. They measured fluxes in $r=0.16''$ apertures on images matched to the F444W PSF, with sources detected on an F150W+F200W stack. They fitted the photometry with \textsc{gsf} using BPASS v2.2.1 binary templates at $Z_\star=10^{-5}Z_\odot$, a non-parametric star-formation history, nebular templates of the same metallicity, and SMC dust. The resulting SED is that of an extremely young ($\simeq1.5$\,Myr), low-mass ($\simeq4\times10^5\,M_\odot$), and very blue ($\beta_\mathrm{UV}\simeq-2.8$) stellar population. The size comes from a \textsc{lenstruction} reconstruction of both images in F115W with an elliptical S\'ersic source, which gives $R_\mathrm{e}=36$\,pc, consistent with image-plane S\'ersic fits to A and B separately. \citet{Morishita2026} also noted that the observed flux ratio between the two images, $1.44\pm0.42$ in \hb and $1.38\pm0.05$ in F115W, is lower than the magnification ratio of the model, $1.98\pm0.29$. If the magnification of A is the more robust of the two, this would imply $\mu_\mathrm{B}\simeq55$; we return to this point in Section~\ref{sec:lensmodel}.

\paragraph{Position of AMORE6-B.}
In the bCG-subtracted UNCOVER images, the centroid of AMORE6-B is offset from its published coordinate by $0.11$--$0.13''$ towards the north-west, consistently in F115W, F356W, and F444W, and lies within $0.02$--$0.04''$ of the GO~9645 MSA target position. Given the astrometric checks of Section~\ref{sec:imaging}, this offset is local to the published coordinate and not a property of the mosaics. It matters for the photometry: an $r=0.16''$ aperture at the published coordinate reproduces the published fluxes (e.g.\ $37.5\pm1.9$ against $38.8\pm4.6$\,nJy in F115W), whereas the same aperture centred on the centroid collects 1.37--1.40 times more flux in all three bands. The colours, and hence $\beta_\mathrm{UV}$ and the age, are thus unaffected, but the absolute fluxes, and with them $\muv$ and $M_\star$, may be underestimated by $\simeq0.35$\,mag. The cause of the offset is still to be confirmed.

\paragraph{Quantities adopted and remeasured.}
The \NIRSpec spectra provide a new and independent measurement of the systemic redshift, of the \hb and \oiii fluxes and their equivalent widths, and of all the other diagnostics discussed in Section~\ref{sec:measurements}. We use the WFSS measurements only for comparison, and the \MUSE \lya profile to measure the velocity offset of \lya from the nebular systemic redshift (Section~\ref{sec:leakage}). We re-derive the morphology and photometry of AMORE6-B from PSF-convolved models of the bCG-subtracted images (Section~\ref{sec:photometry}), and the magnification from the lens model (Section~\ref{sec:lensmodel}). The size is adopted from \citet{Morishita2026} and rescaled to our magnification; the stellar mass and star-formation history are derived in Section~\ref{sec:sed}.

\subsection{Strong-lensing model}
\label{sec:lensmodel}

As our fiducial lens model of Abell 2744 we adopt the \textsc{Lenstool} \citep{Jullo2007} model of \citet{Bergamini2023b}, the same model used by \citet{Morishita2026}. It is constrained by the positions of 149 spectroscopically confirmed multiple images (including individual knots) of sources at $1.0\lesssim z\lesssim9.8$, including both images of AMORE6 (system 33.1), and reproduces them with an rms of $0.43''$ (and $0.17''$ for AMORE6). The total mass is described by dual pseudo-isothermal elliptical (dPIE) profiles \citep{Eliasdottir2007}: four cluster- and group-scale haloes, five individually optimised galaxies (the two BCGs and three bright members of the infalling groups), and 171 cluster members whose velocity dispersions and truncation radii follow luminosity scaling relations. All lensing quantities are computed in the flat $\Lambda$CDM cosmology of the model ($H_0=70\,\mathrm{km\,s^{-1}\,Mpc^{-1}}$, $\Omega_\mathrm{m}=0.3$), for which $D_{ds}/D_s=0.850$ at $z=5.725$.

We evaluate the model directly from the published best-fit parameters and MCMC chain, using an independent implementation of the dPIE deflection field \citep{KassiolaKovner1993} that reproduces the \textsc{Lenstool} magnification maps to better than $10^{-5}$. AMORE6 is a fold-type image pair: A and B lie on opposite sides of the $z=5.725$ critical curve, with B the closer of the two and of positive parity (Fig.~\ref{fig:overview}). Following \citet{Morishita2026}, the reference magnifications are computed at the model-predicted image positions for 5000 random realisations of the posterior, giving $\mu_\mathrm{A}=39.2^{+4.4}_{-3.7}$ and $\mu_\mathrm{B}=77.8^{+8.2}_{-6.9}$, in agreement with the published $39.3^{+3.7}_{-3.5}$ and $77.7^{+8.7}_{-5.9}$.

Because AMORE6-B lies close to the critical curve, its magnification is sensitive to sub-arcsecond positional differences. The model-predicted images are offset from the observed ones by $0.15''$ (A) and $0.23''$ (B), and evaluated at the observed positions the same posterior gives $\mu_\mathrm{A}=33.5^{+4.5}_{-3.8}$ and $\mu_\mathrm{B}=99.5^{+19.7}_{-14.3}$; this difference reflects the residual of the lens model, not an astrometric error, since the published, catalogue, and MSA-target positions of B agree to within $0.1''$ (Appendix~\ref{app:lensing}). For all absolute quantities we adopt the whole-image magnification at the model-predicted position, $\mu_\mathrm{B}=77.8^{+8.2}_{-6.9}$, with a systematic uncertainty of $\pm0.10$\,dex that spans the observed-position estimate (Appendix~\ref{app:lensing}). This applies to the photometric quantities, which are measured over the whole image, and to the \NIRSpec line fluxes, which we first correct to total with the slit-loss model (Section~\ref{sec:observations}) and then divide by $\mu_\mathrm{B}$. The region of B sampled by the shutter is magnified by only $\simeq3\%$ more than the whole image (Appendix~\ref{app:lensing}). The flux ratio observed between the two images (Section~\ref{sec:amore6}) favours, if anything, a lower $\mu_\mathrm{B}$ than predicted, a point also discussed in Appendix~\ref{app:lensing}. The magnification affects only absolute quantities (stellar mass, star-formation rate, $\muv$, and size); line ratios and abundances are independent of the magnification, apart from the small differential effects discussed in Appendix~\ref{app:lensing}. Image A is contaminated by nearby foreground light and is not targeted by GO~9645, so we analyse AMORE6-B alone and use A only as a consistency check.

\section{Morphology and photometry of AMORE6-B}
\label{sec:photometry}

Before turning to the spectroscopy, we model the light distribution of AMORE6-B in the \NIRCam images and remeasure its photometry. The same morphological models define the slit-loss corrections of the \NIRSpec spectra (Section~\ref{sec:slitloss}) and the total fluxes fitted in Section~\ref{sec:sed}, so that spectroscopy and photometry refer to the same total-flux scale.

\subsection{Intrinsic morphology}
\label{sec:morphology}

We model the intrinsic morphology of AMORE6-B with elliptical Gaussians, convolved with the empirical PSF of each band and fitted to the UNCOVER images, rather than with S\'ersic profiles, because AMORE6-B is barely resolved ($r_\mathrm{e}\simeq48$\,mas), so that the S\'ersic index is unconstrained and its extended wings would make the total flux and the slit-loss correction depend on an extrapolation into the intracluster light. In the short-wavelength bands F115W, F150W, and F200W the continuum requires a compact component with $\sigma\simeq0.02''$ plus an extended component with $\sigma=0.07$--$0.09''$, each carrying approximately half of the flux. In the long-wavelength bands a single Gaussian suffices, with $\sigma\simeq0.06''$ in F277W and $0.08$--$0.09''$ in F356W and F444W, and axis ratios of 0.5--0.65.

The nebular emission is more extended. We isolate it with continuum-subtracted maps built from the \NIRCam medium bands: F335M, which contains \hb and \oiii$\lambda\lambda4959,5007$, minus the adjacent line-free F360M; and F444W, dominated by \ha, minus the mean of F410M, F430M, and F460M, assuming a flat $f_\nu$ continuum across each set of bands.\footnote{We note that the continuum bands are not entirely free of line emission: F410M contains He\,\textsc{i}$\lambda5876$ and [O\,\textsc{i}]$\lambda6300$, F430M [O\,\textsc{i}]$\lambda\lambda6300,6363$, and F460M \sii$\lambda\lambda6716,6731$. However, these lines are extremely weak in a metal-poor system such as AMORE6, and only slightly bias the continuum estimate.} Since \oiii is weak in AMORE6, both maps trace mainly Balmer emission. They give $\sigma\simeq0.10$--$0.11''$ for the \hb+\oiii and \ha emission. All these models reproduce the flux measured within the MSA shutter in each band to $\simeq1$\,nJy.

\subsection{Photometry}
\label{sec:phot}

We measure the photometry of AMORE6-B in the 20 \NIRCam bands on the bCG-subtracted DR3 mosaics at their native resolution (Section~\ref{sec:imaging}), by fitting a model of the local scene convolved with the empirical PSF of each band. Within $1.36''$ of AMORE6-B, the model includes AMORE6-B, the three neighbouring sources that intersect this region, and a planar background. The neighbours are described by elliptical Gaussians fitted jointly to F115W, F150W, and F200W. AMORE6-B takes the morphological model of Section~\ref{sec:morphology} for the same band, or for the nearest band with a fit of its own (Table~\ref{tab:photometry}), and only its amplitude is free. The total fluxes are therefore defined in the same way as those to which the slit-loss correction refers, and agree with the totals of the morphological fits to within 5\%. The fit proceeds in two steps: the full scene fixes the fluxes of the neighbours, whose non-Gaussian wings would otherwise bias the planar background at the position of B; AMORE6-B and the plane are then fitted within $0.7''$ after subtracting the neighbours.

The random uncertainty is the scatter of the fluxes recovered when the best-fitting scene is injected at nearby positions and refitted with the same procedure, which propagates the correlated noise of the mosaics; the median bias of the recovered fluxes is included as well. We add in quadrature the changes obtained with a constant instead of a planar background, with fit radii of $0.6''$ and $0.8''$, and with point-like instead of extended neighbours, the scatter over draws of the morphology of AMORE6-B from the covariance of its fit, and a 3\% calibration uncertainty. The resulting photometry is listed in Table~\ref{tab:photometry}. AMORE6-B is detected with ${\rm S/N}>4$ in 16 bands; F140M, F430M, F460M, and F480M have ${\rm S/N}<3$.

The adopted morphologies fit the central $0.5''$ of AMORE6-B better than a single compact Gaussian ($\sigma=0.025''$) fitted jointly to F115W, F150W, and F200W, in all bands with ${\rm S/N}>7$ except F162M, where the two are indistinguishable. With that compact model the fluxes of the detected bands would be 17--48\% lower, the difference being larger in the long-wavelength bands. Our fluxes exceed the $r=0.16''$ aperture fluxes of \citet{Morishita2026} by a median factor of 1.86 in the detected bands, from 1.5--1.8 below 2.1\,$\mu$m to 1.9--2.2 above 2.5\,$\mu$m, and 2.8 in F335M, which contains \hb and \oiii. A factor of $\simeq1.4$ is due to the centring of the aperture (Section~\ref{sec:amore6}); the remainder is the flux of the extended components outside the aperture, which is larger in the long-wavelength bands, where AMORE6-B is more extended. The short- to long-wavelength colours are therefore $\simeq0.2$\,mag redder than those of \citet{Morishita2026}, whereas the colours within the short-wavelength bands, which set $\beta_\mathrm{UV}$, are nearly unchanged.

\begin{deluxetable}{lcccc}
  \tablecaption{\NIRCam photometry of AMORE6-B.\label{tab:photometry}}
  \tablehead{
    \colhead{Filter} & \colhead{$\lambda_{\rm pivot}$} & \colhead{Morphology\tablenotemark{a}} &
    \colhead{$F_\nu$\tablenotemark{b}} & \colhead{$F_\nu$ (M26)\tablenotemark{c}} \\
    \colhead{} & \colhead{($\mu$m)} & \colhead{} & \colhead{(nJy)} & \colhead{(nJy)}
  }
  \startdata
  F070W\tablenotemark{d} & 0.704 & (F115W) & $8.7\pm1.7$ & $<24$ \\
  F090W\tablenotemark{d} & 0.902 & (F115W) & $74.1\pm3.7$ & $48.4\pm1.5$ \\
  F115W & 1.154 & F115W & $65.2\pm2.8$ & $38.8\pm4.6$ \\
  F140M & 1.404 & (F150W) & $34.4\pm17.4$ ($<52$) & $31.0\pm5.6$ \\
  F150W & 1.501 & F150W & $54.8\pm3.1$ & $29.7\pm4.0$ \\
  F162M & 1.626 & (F150W) & $46.6\pm4.2$ & $27.1\pm5.4$ \\
  F182M & 1.845 & (F200W) & $38.1\pm4.3$ & $23.1\pm4.9$ \\
  F200W & 1.990 & F200W & $42.7\pm3.4$ & $25.0\pm5.3$ \\
  F210M & 2.093 & (F200W) & $30.8\pm3.9$ & $21.1\pm5.9$ \\
  F250M & 2.503 & (F277W) & $28.3\pm5.9$ & $15.1\pm4.9$ \\
  F277W & 2.786 & F277W & $32.4\pm2.4$ & $15.9\pm5.5$ \\
  F300M & 2.996 & (F277W) & $28.6\pm3.0$ & $15.4\pm5.3$ \\
  F335M & 3.365 & F335M & $52.5\pm6.7$ & $18.6\pm4.5$ \\
  F356W & 3.563 & F356W & $36.5\pm3.8$ & $16.8\pm3.9$ \\
  F360M & 3.621 & (F356W) & $22.2\pm5.0$ & $10.6\pm5.2$ \\
  F410M & 4.092 & (F444W) & $21.9\pm4.8$ & $10.0\pm4.8$ \\
  F430M & 4.280 & (F444W) & $26.4\pm12.7$ ($<38$) & $<22$ \\
  F444W & 4.421 & F444W & $59.3\pm3.7$ & $26.6\pm3.1$ \\
  F460M & 4.630 & (F444W) & $17.6\pm14.7$ ($<44$) & $<44$ \\
  F480M & 4.834 & (F444W) & $18.9\pm12.9$ ($<39$) & $<35$
  \enddata
  \tablecomments{Total fluxes in the image plane, not corrected for magnification or attenuation (Section~\ref{sec:phot}).}
  \tablenotetext{a}{Band whose morphological model of AMORE6-B (Section~\ref{sec:morphology}) is adopted; in parentheses when the filter has no fit of its own. F115W, F150W, and F200W: two Gaussian components; F277W to F444W: one.}
  \tablenotetext{b}{Uncertainties include the injection--recovery scatter and bias, the background, fit-radius, neighbour, and morphology terms, and a 3\% calibration term. For ${\rm S/N}<3$ the $3\sigma$ upper limit is given in parentheses; the measured value is used in the SED fit.}
  \tablenotetext{c}{\citet{Morishita2026}: $r=0.16''$ apertures at the published coordinate on images matched to the F444W PSF.}
  \tablenotetext{d}{Blueward of, or containing, \lya; not used in the SED fit (Section~\ref{sec:sed}). The F070W flux (rest frame 920--1160\,\AA) is the continuum transmitted through the \lya forest, with an effective transmission of $\simeq0.1$ relative to the extrapolated UV continuum.}
\end{deluxetable}

\section{JWST/NIRSpec observations}
\label{sec:observations}

\subsection{GO 9645 observing configuration}

AMORE6 was observed with the multi-object spectroscopy mode of \NIRSpec \citep{Jakobsen2022,Ferruit2022} as part of the Cycle~5 programme GO~9645 \citep[PI T.~Treu;][]{TreuGO9645}, which targets a large sample of galaxies at $z>6$ with the three medium-resolution ($R\sim1000$) grating/filter combinations G140M/F100LP, G235M/F170LP, and G395M/F290LP. A single MSA configuration was used for all three dispersers, at an aperture position angle of $187.66^\circ$ and after MSATA target acquisition. The final APT configuration, \texttt{daje-jwst-fiducial-replan (Obs 1)}, contains 82 targets in 81 targeted slitlets, one additional commanded group without an assigned target, and 23 failed-open shutters. AMORE6-B was included as a primary target and assigned a three-shutter slitlet (quadrant~1, column~281, shutters 79--81), which was nodded along the slit with the standard three-point pattern (Fig.~\ref{fig:overview}). AMORE6-A does not fall in any open shutter at any nod position. In the flight MSA metadata, the target position of AMORE6-B lies $0.042$--$0.045''$ from the centre of its shutter along the dispersion direction (i.e.\ at $\simeq0.22$ of the shutter width) and $0.031$--$0.044''$ along the slit, with differences of $\lesssim6$\,mas between the three nods; the flight placement differs from the planned one by only $\simeq6$\,mas. The metafiles of the G140M and G395M visits encode identical positions.

All exposures used the NRSIRS2 readout mode. The observations were executed between 2026-11-16 and 2026-11-24. In this article we analyse the data obtained up to 2026-11-22: the complete G140M observations (visits 1--3), and visits 6 and 8 in G235M and G395M, respectively. Their effective exposure times, summed over the three nods, are 182.9, 46.2, and 50.1\,ks (50.8, 12.8, and 13.9\,h) in G140M, G235M, and G395M, respectively (Table~\ref{tab:observations}), and no exposure was rejected (Section~\ref{sec:reduction}).
Together, the three gratings cover the rest-frame range $\simeq1440$--7840\,\AA\ at $z=5.725$. 
The main emission lines covered by the observations are listed in Table~\ref{tab:line_fluxes}. 
G140M covers \civ$\lambda\lambda1548,1551$, \heii$\lambda1640$, O\,\textsc{iii}]$\lambda\lambda1661,1666$, and N\,\textsc{iii}]$\lambda1750$ on NRS1, and [Ne\,\textsc{iv}]$\lambda\lambda2422,2424$ and [O\,\textsc{ii}]$\lambda2470$ on NRS2, while \ciii$\lambda\lambda1907,1909$ falls in the gap between the two detectors.
G235M covers \oii$\lambda\lambda3726,3729$, \neiii$\lambda3869$, H$\delta$, H$\gamma$, \oiii$\lambda4363$, and \heii$\lambda4686$ on NRS2, the last three of which are also covered by G395M. In G395M, \hb lies only $\simeq2$ pixels from the red edge of NRS1, among the edge pixels masked in the reduction, and \oiii$\lambda\lambda4959,5007$ fall in the detector gap at every nod. The trace of AMORE6 on NRS2, however, continues in G235M beyond the nominal range of the grating, up to and beyond the \oiii doublet: \hb and \oiii$\lambda\lambda4959,5007$, and thus the ratio $R_3=$\,\oiii$\lambda5007$/\hb that primarily motivated the interpretation of AMORE6 as a nearly pristine system \citep{Morishita2026}, are therefore recovered in G235M, leveraging the extended spectral calibration outlined in Section~\ref{sec:reduction}. 
Finally, G395M covers \ha, [N\,\textsc{ii}]$\lambda\lambda6548,6584$, \sii$\lambda\lambda6716,6731$, He\,\textsc{i}$\lambda5876$, and [O\,\textsc{i}]$\lambda6300$ on NRS2.

The same MSA configuration was used with long-pass filters, so that the medium-resolution spectra are longer than the separation between slitlets along the dispersion direction, and the configuration was planned allowing spectral overlaps. Projecting the final Obs~1 mask onto the detector with the optical model and the flight MSA metadata, we find that eight physical shutters belonging to three other science slitlets and two failed-open shutters fall on the detector rows of the AMORE6 slitlet in every grating. The most important is the slitlet at column~311 of the same quadrant, which contains the primary target UNCOVER 23271 at $z=6.183$ \citep{Naidu2024} and the secondary source UNCOVER 23399. Its spectrum is dispersed only 0.05--0.14\,$\mu$m redwards of AMORE6 at the same detector column. The other science slitlets, containing UNCOVER 24261 and 23918, have small or vanishing net weights after the nodded-background combination. Including the background-frame exclusions described below, the overlapping shutters raise the total effective sky background to 2.35--3.50 times that of the isolated AMORE6 slitlet and increase the line-flux uncertainties by 20--44\%, depending on the line and on the second-order efficiency. The treatment of these overlaps is described in Section~\ref{sec:reduction}.

\begin{deluxetable*}{lcccccccc}
  \tablecaption{\JWST/\NIRSpec MSA observations of AMORE6-B in GO~9645.\label{tab:observations}}
  \tablehead{
    \colhead{Disperser/Filter} & \colhead{Nominal $\lambda$} & \colhead{Detector gap\tablenotemark{a}} &
    \colhead{Rest-frame gap} & \colhead{Exposure blocks\tablenotemark{b}} &
    \colhead{$N_{\rm exp}$} & \colhead{$N_{\rm int}$} & \colhead{$t_{\rm exp}$\tablenotemark{c}} &
    \colhead{Dates} \\
    \colhead{} & \colhead{($\mu$m)} & \colhead{($\mu$m)} & \colhead{(\AA)} &
    \colhead{$N_{\rm CP}\times(N_{\rm grp},\,N_{\rm int})$} & \colhead{} & \colhead{} &
    \colhead{(ks)} & \colhead{(UT)}
  }
  \startdata
  G140M/F100LP & 0.97--1.89 & 1.189--1.289 & 1768--1916 & $6\times(19,7)+1\times(19,2)$ & 21 & 132 & 182.9 & 2026-11-17--22 \\
  G235M/F170LP & 1.66--3.17\tablenotemark{d} & 1.982--2.149 & 2947--3196 & $1\times(19,7)+1\times(13,6)$ & 6 & 39 & 46.2 & 2026-11-18--19 \\
  G395M/F290LP & 2.87--5.27 & 3.263--3.547 & 4853--5275 & $1\times(19,7)+1\times(16,6)$ & 6 & 39 & 50.1 & 2026-11-16--17
  \enddata
  \tablecomments{All exposures belong to observation~1 of GO~9645 (PI T.~Treu; MSA plan
    \texttt{daje-jwst-fiducial-replan (Obs 1)}), at aperture
    PA $=187.66^\circ$, with MSATA target acquisition, NRSIRS2 readout, and the three-shutter
    slitlet nod. AMORE6-B (APT source 100000, RA $=3.584399^\circ$, Dec $=-30.403376^\circ$) is
    in quadrant~1, column~281, shutters 79--81, and sits in S80, S79, and S81 in nods 1, 2,
    and 3. The table lists the data analysed in this paper, from visits 1--3 (G140M), 6 (G235M), and 8 (G395M); exposure parameters and the realised pointing are from the
    metadata of the executed visits.}
  \tablenotetext{a}{Wavelength range without data between NRS1 and NRS2 in the combined spectra of AMORE6-B,
    including the detector-edge pixels masked in the reduction; rest-frame values at $z_{\rm sys}=5.7235$. Lines lost in the gaps: \ciii$\lambda\lambda$1907,1909
    in G140M, and \hb, which lies 2~pixels from the physical edge of NRS1, and \oiii$\lambda\lambda$4959,5007 in G395M.}
  \tablenotetext{b}{Each configuration pointing (CP) is one exposure in each of the three nods;
    $N_{\rm grp}$ and $N_{\rm int}$ are groups per integration and integrations per exposure.}
  \tablenotetext{c}{Effective exposure time (\texttt{EFFEXPTM}), summed over the three nods. AMORE6-B is in the slitlet in every nod, so this is
    also its on-source time in the nodded combination.}
  \tablenotetext{d}{On NRS2 the AMORE6 trace continues beyond the nominal range and covers
    \hb\ and \oiii$\lambda\lambda$4959,5007, outside the standard calibration.}
\end{deluxetable*}

\subsection{Data reduction and spectral extraction}
\label{sec:reduction}

We reduced the data with the \NIRSpec GTO pipeline, developed by the ESA \NIRSpec Science Operations Team and the \NIRSpec GTO team. We use the version adopted for the fourth data release of JADES, to which we refer for a full description \citep{CurtisLake2026,Scholtz2026DR4}. Here we summarise the main steps and describe the choices specific to AMORE6. The pipeline (version 5.2, CRDS context \texttt{jwst\_1584.pmap}) starts from the count-rate maps produced by the first stage of the STScI pipeline (\texttt{calwebb\_detector1}, version 1.20), which includes the reference-pixel correction, jump detection with the flagging of large cosmic-ray events (``snowballs''), and ramp fitting. Each integration is treated as an independent exposure, giving 132, 39, and 39 exposures in G140M, G235M, and G395M, respectively. The GTO pipeline then performs, in order: the pixel-to-pixel subtraction of the background from the nodded exposures; the extraction of the two-dimensional trace of the slitlet and the assignment of its wavelength and spatial coordinates; the correction of the wavelength zero point for the position of the source within the shutter; flat-fielding; the path-loss correction; the rectification and resampling of the trace onto a regular two-dimensional grid; and the boxcar extraction of one-dimensional spectra. The individual one-dimensional spectra are finally combined with inverse-variance weights after $3\sigma$ clipping \citep{Scholtz2026DR4}. We additionally subtracted residual $1/f$ noise at the level of count rate maps and masked hot pixels. %

\paragraph{Background subtraction.}
The MSA configuration contains no slitlets dedicated to a master background, and AMORE6 lies in the core of the cluster, where intracluster light and the wings of the bright cluster galaxies contribute $\simeq27$\,nJy per shutter in F115W ($\simeq15$\,nJy in F444W); this is comparable to the continuum of AMORE6-B itself. We therefore adopt the nodded background subtraction for all measurements, both lines and continuum. Integrating the UNCOVER images over the open area of each shutter at each nod shows that the diffuse light cancels in the nodded difference to $\lesssim1$\,nJy, because it is smooth on the scale of the slitlet. In two of the six background frames, however, the shutters contain other structures: at the third nod, the lower shutter contains a foreground galaxy (UNCOVER 26201, $z_\mathrm{phot}\simeq0.1$), and at the second nod the upper shutter falls on the halo of a bright galaxy north of the slitlet. Both are excluded from the background, as is done in the pipeline for any shutter contaminated by a neighbouring source \citep{Scholtz2026DR4}. The background of the first nod is thus the mean of the other two, while that of the second and third nods is the first nod alone. 
With these exclusions the nodded continuum of AMORE6-B predicted from the images is positive at all nods ($+25$, $+32$, and $+18$\,nJy in F115W), and the residual nod-to-nod scatter of $\simeq\pm7$\,nJy, due to structure in the background shutters, is included in the systematic uncertainty of the continuum and of the equivalent widths. 
The exclusions increase the noise of the combined spectrum by $\simeq11\%$ relative to the standard three-nod subtraction. As a further consequence of nodding along the slit, the wings of AMORE6-B that extend into the adjacent shutters are partially self-subtracted, removing $\lesssim6\%$ of the flux of the source; we correct for this loss in the forward model described below. 

\paragraph{Extended wavelength range of G235M.}
\hb and \oiii$\lambda\lambda4959,5007$ fall at 3.27--3.37\,$\mu$m, on the extension of the G235M trace on NRS2, beyond the upper limit of the standard calibration (3.17\,$\mu$m for the CRDS flat-field reference files). Following \citet{Scholtz2026DR4}, we extract the trace beyond the nominal range using the wavelength solution of the instrument optical model, and calibrate it with flat-field curves extended beyond their nominal range. The F-flat of G235M/F170LP is replaced by that of G395M/F290LP, and the S-flat is linearly extrapolated from its last 0.1\,$\mu$m. \citet{Scholtz2026DR4} find that the fluxes of the medium-resolution gratings agree within 10\% over the extended ranges. For AMORE6, the H$\gamma$ fluxes measured in G235M and G395M, in their overlap at 2.87--3.17\,$\mu$m, agree within 2\%, and those of \heii$\lambda4686$ within $1.8\sigma$ (\oiii$\lambda4363$ is not detected in either grating). After the slit-loss correction, the \hb flux measured in the extended G235M range also agrees with the WFSS flux of \citet{Morishita2026} within $0.7\sigma$.

In the extended range, the detector also receives second-order light at half the first-order wavelength: \hb and \oiii$\lambda\lambda4959,5007$ coincide with second-order light at 1.64--1.68\,$\mu$m, which the F170LP filter transmits at 50--70\%. 
In the case of AMORE6, the second-order spectrum consists of the rest-frame UV continuum at 2430--2500\,\AA\ and of the [Ne\,\textsc{iv}]$\lambda\lambda2422,2424$ and [O\,\textsc{ii}]$\lambda2470$ lines, whose second-order images fall $\simeq10$\,nm bluewards of \hb and $\simeq13$\,nm bluewards of \oiii$\lambda4959$, respectively, and are thus separated from the first-order lines. The same wavelength range is observed directly in first order in G140M, which therefore provides an empirical template of the second-order spectrum. [Ne\,\textsc{iv}]$\lambda\lambda2422,2424$ is not detected in G140M ($<2.0\times10^{-19}$\,erg\,s$^{-1}$\,cm$^{-2}$ at $3\sigma$, i.e.\ $<10\%$ of \hb before the reduction by the second-order efficiency and by the filter), so that the second-order light reduces to a faint continuum, which is absorbed by the local continua of the line fit (Section~\ref{sec:linefit}).

\begin{figure*}
  \centering
  \includegraphics[width=\textwidth]{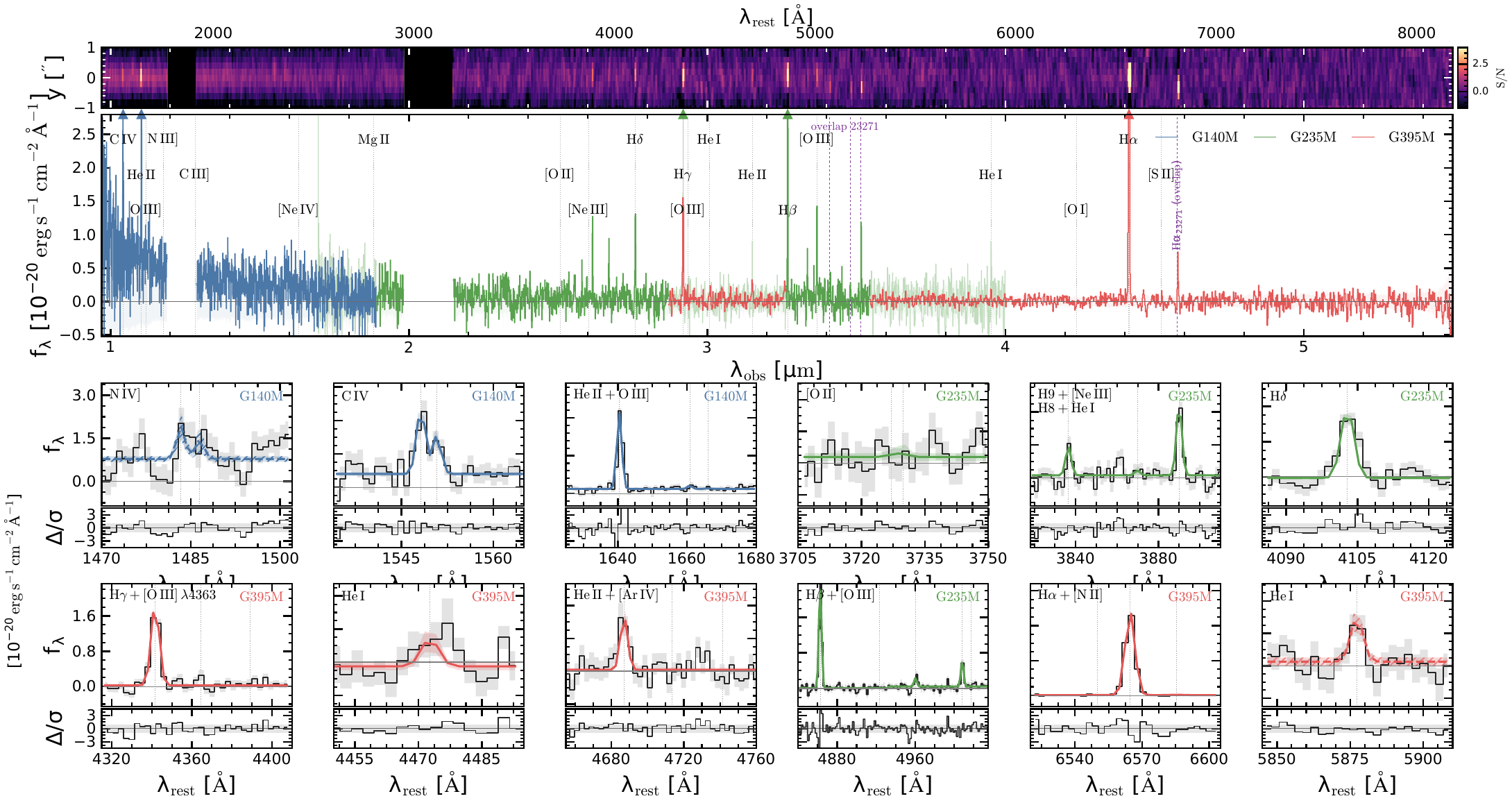}
  \caption{Fiducial \NIRSpec spectra of AMORE6-B. \textit{Top:} a display-only mosaic of the source-centred, rectified two-dimensional spectra, shown as signal-to-noise and lightly smoothed in the spatial and spectral directions. At each wavelength we show the locally most sensitive valid grating, so the extended G235M spectrum fills the G395M detector gap containing \hb and \oiii; the fits use the unsmoothed spectra on their native grids. \textit{Middle:} the profile-weighted G140M, G235M, and G395M one-dimensional bootstrap stacks on their native wavelength grids. Each grating is shown at full saturation where it is locally adopted, including G235M across the G395M detector gap, and in a lighter shade where another grating is more sensitive. Grey dotted lines identify selected transitions at the fitted nebular redshift, including C\,\textsc{iii}] in the G140M detector gap; triangles mark strong lines clipped by the display range. Purple markers identify the H$\beta$+[O\,\textsc{iii}] and \ha lines from the overlapping spectrum of UNCOVER~23271. \textit{Bottom:} selected rest-frame line complexes. Thin vertical lines mark the expected wavelengths of the fitted transitions. Black steps and grey bands show the data and $1\sigma$ uncertainties, while the coloured curves and bands show the median and 16th--84th percentiles of the joint covariance-aware posterior predictive model; the lower panels show the signed residuals. Marginal complexes retained for inspection are shown with dashed models and hatched posterior bands. C\,\textsc{iv} is retained as a tentative, repeat-sensitive summed-doublet flux, whereas N\,\textsc{iv}] remains an upper limit; He\,\textsc{i}\,$\lambda5876$ is also tentative. Other components adopted as upper limits after the nod/repeat robustness tests are suppressed from the emission-only display; all signed measurements and limits are retained in Table~\ref{tab:line_fluxes}. The plotted spectra retain the point-source path-loss calibration applied by the pipeline. The morphology-aware line and continuum corrections are applied separately to the tabulated fluxes and equivalent widths and therefore cannot be represented by a single rescaling of this figure.}
  \label{fig:total_spectrum}
\end{figure*}

\paragraph{Spectral overlaps.}
Several dispersed spectra intersect the detector rows occupied by AMORE6. The dominant contribution comes from the slitlet containing UNCOVER~23271 and UNCOVER~23399; after accounting for the nodded subtraction, the estimated net continuum contributions are approximately 5.5, 5.6, and 5.3\,nJy in G140M, G235M, and G395M, respectively. Their potentially relevant emission lines are marked in Fig.~\ref{fig:total_spectrum}. We do not model these overlaps explicitly because they are either spectrally separated from the lines used in our analysis or absorbed by the independently fitted local continua. \NIRSpec equivalent widths are therefore reported with this caveat, while our fiducial equivalent widths use the \NIRSpec line fluxes and the continuum inferred from the \NIRCam photometry.

\paragraph{Extraction, combination, and uncertainties.}
Instead of the boxcar spectra of the pipeline, which we keep as a check, we extract the one-dimensional spectra of AMORE6-B from the rectified two-dimensional spectra of the individual integrations with a profile-weighted extraction \citep{Horne1986} over the five spatial pixels within $\pm0.5''$ of the trace. The spatial profile is measured from the strongest lines (\heii$\lambda1640$ in G140M, H$\delta$ and H$\gamma$ in G235M, and H$\gamma$ and \ha in G395M) and, for each integration, is rebuilt excluding its parent exposure, so that the same noise does not define both the weights and the flux. Pixels flagged by the pipeline are omitted, and residual outliers are removed with an $8\sigma$ clipping. No exposure was rejected. Since the resampling correlates adjacent pixels and the pipeline variance may not describe the noise of deep stacks, we do not use the pipeline uncertainties, but derive the covariance between pixels from 2000 bootstrap resamplings of the individual integrations, stratified by nod \citep{Curti2025,Scholtz2026DR4}. A bootstrap of complete nod patterns, which also captures the nod-to-nod residuals of the background, is kept as a more conservative control: it changes the \hb flux by 4\%, and this difference is propagated in the analyses that depend on the Balmer decrement. 
We do not stack the gratings where they overlap, but fit the lines in each grating with its own line-spread function.

\subsubsection{Astrometry, slit losses, and effective line-spread function}
\label{sec:slitloss}

The UNCOVER mosaics used to design the MSA configuration agree with the GLASS-JWST and \textit{Gaia}~DR3 astrometry to within $\simeq0.01''$ around AMORE6, and the MSA target position lies within $0.035''$ of the peak of AMORE6-B in F356W and F444W. 
The final MSA metafiles independently encode the same AMORE6-B position in the three nods, $(x,y)=(0.662,0.570)$, $(0.668,0.581)$, and $(0.656,0.558)$ in shutter-pitch coordinates. Relative to the shutter centre, these positions correspond to an across-shutter displacement of only $|u|=0.042$--$0.045''$ for a half-width of $0.10''$. 
 As an independent check, the velocity offset between H$\gamma$ and \ha, which is sensitive to the across-shutter position through the wavelength zero-point correction, is $+0.6\pm10.0$\,km\,s$^{-1}$ and places the Balmer emission $+0.002\pm0.033''$ from the metafile position. We therefore use the final metafile coordinates for the wavelength calibration, the pipeline path-loss correction, and all subsequent slit-loss and line-spread-function calculations.

The pipeline path-loss step assumes a point source and uses a flight-derived map of throughput as a function of position within the shutter and wavelength \citep{Scholtz2026DR4}. The realised position of AMORE6-B lies within the calibrated central 0.35 shutter pitches, but the source is resolved and its continuum and line emission might have different spatial distributions. We use the intrinsic morphologies of Section~\ref{sec:morphology}: a compact plus an extended component for the short-wavelength continuum, and more extended single Gaussians ($\sigma\simeq0.10$--$0.11''$) for the nebular emission. We weight each intrinsic morphology with the flight point-source path-loss map in the CRDS reference, which includes the PSF, shutter geometry, and slit diffraction, and is the same reference adopted for JADES DR4, accurate to a few per cent for sources within 0.3 shutter pitches of the centre, and explicitly propagate the three nods, self-subtraction, and boxcar extraction. An independent Fourier-optics calculation with \texttt{msafit} \citep{deGraaff2024}, evaluated at the final metafile positions, agrees with the flight-pathloss calculation to 0.4--6.3\% for the continuum morphologies and to 4.2--7.9\% for the two nebular morphologies; we conservatively include an 8\% model uncertainty for the line slit-loss correction.

We derive separate smooth correction curves for continuum and lines over the full wavelength range of the three gratings, both defined relative to the position-dependent point-source path loss already applied by the pipeline. Each curve is obtained by evaluating a fixed intrinsic morphology at every wavelength with the local PSF and path-loss map. The continuum curve uses the line-free F115W morphology. The line curve uses the \hb+\oiii and \ha maps of Section~\ref{sec:morphology}. We adopt the \hb+\oiii morphology for all lines blueward of \hb, including the rest-frame UV lines, for which no line map is available, the \ha morphology redward of \ha, and a linear interpolation in wavelength between the two. 

The two curves and the resulting equivalent-width correction are shown in Fig.~\ref{fig:slitloss_lsf} (Appendix~\ref{app:lensing}). For reference, at the flight-metafile position and for the 5-pixel boxcar extraction, the continuum corrections are 1.604, 1.565, and 1.523 at \heii$\lambda1640$, \hb, and \ha, respectively, while the corresponding line corrections are 1.446, 1.462, and 1.534. Consequently, an equivalent width measured from the point-source-corrected spectrum is multiplied by 0.902, 0.934, and 1.007 at these three lines, respectively. Differential slit losses modify even the widest-baseline nebular-line ratios by $\lesssim6\%$ (for example \heii$\lambda1640$/\ha), but are essential for the equivalent widths because the compact UV continuum lies closer to the shutter edge than the extended line emission. 
We retain a conservative $\pm0.02''$ across-shutter registration envelope; this dominates the uncertainty for the compact continuum. 
The line fluxes of Table~\ref{tab:line_fluxes} are given in the image plane, corrected for slit losses but not for magnification; the absolute quantities derived from them are divided by the whole-image magnification $\mu_\mathrm{B}$ (Section~\ref{sec:lensmodel}), while the differential magnification between the compact UV continuum and the more extended line emission is discussed in Appendix~\ref{app:lensing}. 
Finally, we compare the corrected spectral continuum with the \NIRCam photometry. In G140M, where the continuum is well detected, it is lower than the photometric one by a factor of $\simeq1.7$ (Section~\ref{sec:continuum}), more than the $\simeq0.1$\,dex scatter found by \citet{Scholtz2026DR4} for the JADES sample, while in G235M and G395M it is too faint for a useful comparison. We therefore measure the equivalent widths against the photometric continuum.

The same \texttt{msafit} forward models provide a morphology-dependent line-spread function (Fig.~\ref{fig:slitloss_lsf}). AMORE6-B lies between the two standard limiting cases: its LSF is broader than that of a centred point source but narrower than the nominal STScI LSF for a uniformly illuminated shutter. For the adopted nebular morphology, the Gaussian-equivalent resolving powers are $R=1181$ at \heii$\lambda1640$ ($\lambda_\mathrm{obs}=1.103\,\mu$m) in G140M, $R=2036$ at \hb (3.270\,$\mu$m) and $R=2101$ at \oiii$\lambda5007$ (3.368\,$\mu$m) in G235M, and $R=1714$ at \ha (4.415\,$\mu$m) in G395M, corresponding to instrumental FWHM values of 254, 147, 143, and 175\,km\,s$^{-1}$, respectively. These values average over the sub-pixel phase at which a line falls on the detector: for an extended source the dispersion-direction profile is only $\simeq0.6$\,pixel wide ($\sigma$) and close to a top hat, so its sampled width depends on this phase by up to $\pm8$\%. \hb and \oiii$\lambda5007$ fall in the G395M detector gap and are observed only in G235M, redward of its nominal wavelength range and of the 3.2\,$\mu$m limit of the PSF library, so that their LSF is a short extrapolation. The LSF is grating-specific even at the same wavelength (e.g.\ at \heii$\lambda4686$, $R=1957$ in G235M and $R=1209$ in G395M). We convolve the intrinsic line profiles with the appropriate wavelength-dependent LSF in each grating and allow a multiplicative LSF-scale nuisance parameter in the kinematic fit, with a prior width of 8\% that reflects the unknown pixel phase. 

\section{Spectroscopic measurements}
\label{sec:measurements}

\subsection{Emission-line fitting}
\label{sec:linefit}

We fitted the one-dimensional spectra of the three gratings jointly with a forward-modelling approach based on the \textsc{unite} package\footnote{Available at \href{https://github.com/TheSkyentist/unite}{https://github.com/TheSkyentist/unite}.}
\citep{Hviding+2025,Hviding2026}, retaining the native wavelength grid of each grating.  
The rectified two-dimensional spectra were used to construct and validate the extraction profiles, contamination masks, spatial centroids, and aperture corrections.

We first extracted every rectified integration separately using empirical spatial profiles measured from the strong nebular lines, and then formed one inverse-variance-weighted stack per grating.  
To account for the correlations between neighbouring spectral pixels introduced by rectification and stacking, we measured the noise and its covariance directly from 2000 bootstrap realisations of the individual integrations, following the approach of \citet{Curti2025} and \citet{Scholtz2026DR4}. The bootstrap preserved the sampling of the three nod positions, and the resulting covariance was included in the fit in local wavelength windows around each line complex. 
As a conservative check, we repeated the analysis by resampling complete three-nod patterns rather than individual integrations. 
The line complexes shared a common redshift and velocity width, while their fluxes and local continua were fitted independently. 

For each wavelength range of interest the model consists of a constant local continuum and one or more emission lines convolved with the grating- and wavelength-dependent, morphology-aware \NIRSpec line-spread function described in Section~\ref{sec:reduction}.  
A single intrinsic velocity dispersion, $\sigma_v$, and a common velocity centroid were adopted for the non-resonant nebular lines.  
Lines observed in more than one grating were fitted simultaneously with shared kinematics but independent continua, covariance matrices, and amplitudes.  
We fitted the flux of each repeated line independently in each grating, rather than applying a single relative flux correction between gratings. This provides a direct check of their native flux calibration and combines the evidence for the presence of a line without requiring its measured flux to be identical in the different gratings.
For line ratios we adopt the measurement from the grating with the most reliable local calibration and sensitivity; in particular, H$\gamma$/H$\beta$ is measured internally from G235M, whereas H$\alpha$ is measured on the native G395M scale.

The fluxes of transitions arising from the same upper level were tied to their theoretical ratios, including \oiii$\lambda5007/\lambda4959$ and [N\,\textsc{ii}]$\lambda6584/\lambda6548$, whereas the density-sensitive \oii and \sii doublet ratios were left free within their physically allowed ranges \citep[e.g.,][]{Sanders+2016,Kewley2019}. 
For fully blended features, e.g. the \neiii+He\,I+H$\epsilon$ complex, all components shared the same kinematics but had independent fluxes except for the fixed doublet ratios.  
We found no evidence for an additional broad or velocity-offset nebular component. We repeated the fit with a second Gaussian having a common velocity offset, width, and fractional contribution across the securely detected lines. This model increases the Bayesian information criterion by $\Delta\mathrm{BIC}=\mathrm{BIC}_{\rm two}-\mathrm{BIC}_{\rm one}=+2.5$, while changing the integrated line fluxes by less than 3\%. We therefore use the single-component model throughout.

The systemic redshift was measured from a joint fit to the detected Balmer and non-resonant forbidden lines, giving $z_{\rm sys}=5.723539^{+0.000015}_{-0.000018}$. After removal of the instrumental broadening, the intrinsic line width is $\sigma_v=55.6\pm1.5$\,km\,s$^{-1}$ (statistical uncertainty), corresponding to an intrinsic FWHM of $131\pm4$\,km\,s$^{-1}$. Varying the instrumental line-spread function by its adopted 8\% uncertainty changes $\sigma_v$ by $^{+7}_{-9}$\,km\,s$^{-1}$. The \NIRSpec redshift is blueshifted by $\Delta v=-79\pm22$\,km\,s$^{-1}$ relative to that inferred from the WFSS \hb detection. Conversely, relative to the nebular systemic redshift, the joint centroid of the \MUSE \lya line is redshifted by $\Delta v_{\rm Ly\alpha}=+79\pm2$\,km\,s$^{-1}$, where the quoted uncertainty is statistical and does not include the relative absolute wavelength calibration of the two instruments.

\subsection{Detections, upper limits, and equivalent widths}
\label{sec:continuum}

The measured line fluxes and rest-frame equivalent widths are reported in Table~\ref{tab:line_fluxes}, and the principal fitted spectral windows are shown in Fig.~\ref{fig:total_spectrum}.  
We report the posterior median as the line flux and the 16th--84th percentile interval as its statistical uncertainty.
Lines with ${\rm S/N}\geq5$ are considered detected and those with $3\leq{\rm S/N}<5$ tentative. For lines with ${\rm S/N}<3$, and for those not reproduced in the individual nods, we quote $3\sigma$ upper limits, $\max(0,F)+3\sigma$, from the posterior of the joint fit, which includes the correlated noise and the covariance with the neighbouring lines.

We detect the series of Balmer lines down to H10 with high significance. 
For reference, the path- and slit-loss-corrected \hb flux is $F_{\rm H\beta}=20.26\pm0.38\times10^{-19}$\,erg\,s$^{-1}$\,cm$^{-2}$, in agreement with the NIRCam/WFSS measurement of AMORE6-B within $1\sigma$.
We also detect \heii$\lambda4686$ and He\,\textsc{i}$\lambda5876$, whereas the \oii$\lambda\lambda3727,3730$ doublet, \neiii$\lambda3869$, the auroral \oiii$\lambda4363$, [O\,\textsc{i}]$\lambda6300$, [N\,\textsc{ii}]$\lambda6584$, and \sii$\lambda\lambda6716,6731$ are not detected.
The only metal line detected at high-significance is \oiii$\lambda5007$, at $10\sigma$: we measure $F_{5007}=5.05\pm0.50\times10^{-19}$\,erg\,s$^{-1}$\,cm$^{-2}$, yielding $R_3\equiv$\,\oiii$\lambda5007$/\hb$=0.249\pm0.025$, an exceptionally low value; the implications are discussed in Sects.~\ref{sec:empirical_oh} and \ref{sec:sed}.

In the rest-frame UV, \heii$\lambda1640$ is detected at $35\sigma$ and \civ$\lambda\lambda1548,1551$ at $8\sigma$, whereas O\,\textsc{iii}]$\lambda\lambda1661,1666$ is not detected.  
Unfortunately, C\,\textsc{iii}]$\lambda\lambda1907,1909$ falls in the G140M detector gap and provides no useful \NIRSpec constraint. However, archival VLT/X-shooter observations from ESO programme 115.29G6.001 (PI: T.~Morishita) yield a conservative $3\sigma$ upper limit of $F_{\rm CIII]}<7.5\times10^{-18}$\,erg\,s$^{-1}$\,cm$^{-2}$. 

\paragraph{Continuum and equivalent widths.}
The continuum under the optical emission lines is measured from the photometry of Table~\ref{tab:photometry}, in the medium bands F250M, F300M, F360M, F410M, F430M, F460M, and F480M, which sample the rest frame between 3700 and 7100\,\AA, after subtracting the contribution of the lines measured with \NIRSpec from each filter. A power law in $f_\nu$ describes these points well, with $f_\nu\propto\lambda^{-0.2}$, and gives the continuum at the wavelength of each line. In the rest-frame UV, a power-law fit to F115W, F150W, F162M, and F182M (1700--2700\,\AA; F090W contains \lya), after subtracting the \heii$\lambda1640$ flux from F115W, gives $\beta_{\rm UV}=-3.01\pm0.23$. Both measurements are independent of stellar-population models; the spectral energy distribution is modelled in Section~\ref{sec:sed}.

We compute the equivalent widths against this continuum, which is on the same total-flux scale as the path-loss-corrected line fluxes and has a higher signal-to-noise ratio than the continuum from the NIRSpec spectrum. We obtain $\mathrm{EW}_0(\mathrm{H}\alpha)=2430^{+650}_{-460}$\,\AA, $\mathrm{EW}_0(\mathrm{H}\beta)=478^{+70}_{-62}$\,\AA, $\mathrm{EW}_0(\mathrm{H}\gamma)=171^{+29}_{-25}$\,\AA, $\mathrm{EW}_0(\mathrm{He\,II}\,\lambda4686)=35\pm8$\,\AA, and $\mathrm{EW}_0(\mathrm{He\,II}\,\lambda1640)=15.6\pm1.8$\,\AA; the uncertainties include those of the photometry, of the line fluxes, and of the path-loss correction of the lines, which does not cancel against the photometric continuum. For comparison, Table~\ref{tab:line_fluxes} also lists the equivalent widths against the local continuum of the \NIRSpec spectrum. In G140M the continuum is well detected, but the \NIRSpec estimate of $\mathrm{EW}_0(\mathrm{He\,II}\,\lambda1640)$, $26.6^{+2.9}_{-2.4}$\,\AA, is 1.7 times the photometric one, i.e.\ the G140M continuum is lower than the photometric UV continuum. In G235M and G395M the continuum is barely detected: for most lines it is positive in less than 95\% of the posterior draws, and where it is formally significant, around \oiii$\lambda\lambda4959,5007$ and \ha, it is 2--3 times higher than the photometric continuum, which gives equivalent widths 2--3 times lower. Magnification cancels in the line ratios and equivalent widths, and is not applied to the values in Table~\ref{tab:line_fluxes}.

{\tabletypesize{\footnotesize}
\setlength{\tabcolsep}{3pt}
\begin{deluxetable*}{lcccccll}
  \tablecaption{Emission-line fluxes and rest-frame equivalent widths of AMORE6-B.\label{tab:line_fluxes}}
  \tablehead{
    \colhead{Line} & \colhead{$\lambda_{\rm rest}$\tablenotemark{a}} & \colhead{$\lambda_{\rm obs}$\tablenotemark{b}} &
    \colhead{$F$\tablenotemark{c}} & \colhead{$C_{\rm path}$\tablenotemark{d}} & \colhead{EW$_0^{\rm phot}$\tablenotemark{f}} &
    \colhead{EW$_0^{\rm spec}$\tablenotemark{h}} & \colhead{Notes} \\
    \colhead{} & \colhead{(\AA)} & \colhead{($\mu$m)} &
    \colhead{($10^{-19}$\,erg\,s$^{-1}$\,cm$^{-2}$)} & \colhead{} & \colhead{(\AA)} & \colhead{(\AA)} & \colhead{}
  }
  \startdata
  \sidehead{\textit{G140M/F100LP}}
  N\,\textsc{iv}]$\lambda\lambda$1483,1486 & 1483.32, 1486.50 & 0.9973, 0.9995 & $<4.7$ & $1.45\pm0.14$ & $<4.4$ & \nodata & doublet sum \\
  \civ$\lambda\lambda$1548,1551 & 1548.19, 1550.77 & 1.0409, 1.0427 & $5.57\pm0.70$ & $1.45\pm0.14$ & $5.9\pm1.0$ & \nodata & doublet sum \\
  \heii$\lambda$1640 & 1640.42 & 1.1029 & $12.19\pm0.35$ & $1.45\pm0.14$ & $15.6\pm1.8$ & $26.6^{+2.9}_{-2.4}$ &  \\
  O\,\textsc{iii}]$\lambda\lambda$1661,1666 & 1660.81, 1666.15 & 1.1167, 1.1202 & $<2.1$ & $1.45\pm0.14$ & $<2.8$ & \nodata & doublet sum \\
  N\,\textsc{iii}]$\lambda$1750 & 1749.67 & 1.1764 & $<1.8$ & $1.45\pm0.14$ & $<2.7$ & \nodata & multiplet sum; systematic limit\tablenotemark{e} \\
  \ciii$\lambda\lambda$1907,1909 & 1906.68, 1908.73 & 1.2826, 1.2840 & \nodata & \nodata & \nodata & \nodata & detector gap \\
  C\,\textsc{ii}]$\lambda$2326 & 2326.00 & 1.5639 & $<1.0$ & $1.46\pm0.11$ & $<3.7$ & \nodata &  \\
  {[}Ne\,\textsc{iv}]$\lambda\lambda$2422,2425 & 2422.56, 2425.14 & 1.6288, 1.6306 & $<1.4$ & $1.46\pm0.11$ & $<5.8$ & \nodata & doublet sum \\
  \sidehead{\textit{G235M/F170LP}}
  Mg\,\textsc{ii}$\lambda\lambda$2796,2803 & 2796.35, 2803.53 & 1.8801, 1.8850 & $<2.8$ & $1.47\pm0.11$ & $<18$ & \nodata & doublet sum \\
  \heii$\lambda$3203 & 3204.03 & 2.1542 & $<1.1$ & $1.47\pm0.11$ & $<16$ & \nodata &  \\
  {[}Ne\,\textsc{v}]$\lambda$3426 & 3426.84 & 2.3041 & $<0.85$ & $1.47\pm0.11$ & $<13$ & \nodata &  \\
  \oii$\lambda\lambda$3727,3730 & 3727.09, 3729.88 & 2.5059, 2.5078 & $<1.2$ & $1.47\pm0.12$ & $<23$ & \nodata & doublet sum \\
  H10 & 3798.98 & 2.5543 & $0.89\pm0.25$ & $1.47\pm0.12$ & $17.6\pm6.0$ & $26.4^{+21.8}_{-10.6}$ & tentative \\
  H9 & 3836.48 & 2.5795 & $1.16\pm0.26$ & $1.47\pm0.12$ & $23.5\pm7.0$ & $34.6^{+27.1}_{-13.1}$ & tentative \\
  \neiii$\lambda$3869 & 3869.86 & 2.6019 & $<0.82$ & $1.47\pm0.12$ & $<17$ & \nodata &  \\
  H8 + He\,\textsc{i}$\lambda$3889 & 3890.00 & 2.6155 & $2.89\pm0.19$ & $1.47\pm0.12$ & $60\pm12$ & $85.5^{+61.2}_{-26.1}$ & blend \\
  H$\epsilon$ + He\,\textsc{i}$\lambda$3965 & 3970.50 & 2.6696 & $1.39\pm0.25$ & $1.47\pm0.12$ & $30.3\pm7.6$ & $56.9^{+108.0}_{-27.3}$\tablenotemark{i} & blend \\
  He\,\textsc{i}$\lambda$4026 & 4027.33 & 2.7078 & $<0.86$ & $1.47\pm0.12$ & $<19$ & \nodata &  \\
  H$\delta$ & 4102.89 & 2.7586 & $3.62\pm0.28$ & $1.47\pm0.12$ & $85\pm16$ & $272^{+742}_{-156}$\tablenotemark{i} &  \\
  H$\gamma$ & 4341.68 & 2.9191 & $6.38\pm0.27$ & $1.47\pm0.12$ & $171^{+29}_{-25}$ & $190^{+267}_{-76}$\tablenotemark{i} &  \\
  \oiii$\lambda$4363 & 4364.44 & 2.9344 & $<1.2$ & $1.47\pm0.12$ & $<34$ & \nodata & systematic limit\tablenotemark{e} \\
  \hb\tablenotemark{g} & 4862.68 & 3.2694 & $13.85\pm0.26$ & $1.46\pm0.12$ & $478^{+70}_{-62}$ & $568^{+1061}_{-256}$\tablenotemark{i} &  \\
  \oiii$\lambda$4959\tablenotemark{g} & 4960.30 & 3.3351 & $1.16\pm0.11$ & $1.47\pm0.12$ & $41.6\pm6.7$ & $14.8^{+5.9}_{-3.7}$ & tied to $\lambda$5007 \\
  \oiii$\lambda$5007\tablenotemark{g} & 5008.24 & 3.3673 & $3.44\pm0.34$ & $1.47\pm0.12$ & $127^{+22}_{-19}$ & $44.1^{+17.6}_{-10.9}$ &  \\
  \sidehead{\textit{G395M/F290LP}}
  He\,\textsc{i}$\lambda$4471 & 4472.73 & 3.0073 & $<1.4$ & $1.46\pm0.12$ & $<39$ & \nodata &  \\
  \heii$\lambda$4686 & 4687.02 & 3.1513 & $1.11\pm0.20$ & $1.46\pm0.12$ & $34.9\pm8.1$ & $127^{+287}_{-63}$\tablenotemark{i} &  \\
  He\,\textsc{i}$\lambda$5876 & 5877.25 & 3.9516 & $0.74\pm0.20$ & $1.50\pm0.11$ & $39\pm12$ & $43.9^{+86.1}_{-23.3}$\tablenotemark{i} & tentative \\
  {[}O\,\textsc{i}]$\lambda$6300 & 6302.04 & 4.2372 & $<0.49$ & $1.52\pm0.11$ & $<31$ & \nodata &  \\
  {[}N\,\textsc{ii}]$\lambda$6548 & 6549.86 & 4.4038 & $<0.22$ & $1.53\pm0.11$ & $<15$ & \nodata &  \\
  \ha & 6564.61 & 4.4137 & $34.85\pm0.33$ & $1.53\pm0.11$ & $2425^{+645}_{-458}$ & $1044^{+857}_{-341}$ &  \\
  {[}N\,\textsc{ii}]$\lambda$6584 & 6585.27 & 4.4276 & $<0.65$ & $1.53\pm0.11$ & $<46$ & \nodata &  \\
  He\,\textsc{i}$\lambda$6678 & 6679.99 & 4.4913 & $<0.96$ & $1.53\pm0.11$ & $<69$ & \nodata &  \\
  \sii$\lambda$6716 & 6718.29 & 4.5171 & $<0.67$ & $1.53\pm0.11$ & $<49$ & \nodata &  \\
  \sii$\lambda$6731 & 6732.67 & 4.5267 & $<1.1$ & $1.53\pm0.11$ & $<82$ & \nodata &  \\
  He\,\textsc{i}$\lambda$7065 & 7067.14 & 4.7516 & $<0.81$ & $1.54\pm0.13$ & $<66$ & \nodata &  \\
  {[}Ar\,\textsc{iii}]$\lambda$7136 & 7137.77 & 4.7991 & $<0.74$ & $1.54\pm0.13$ & $<62$ & \nodata &  \\
  \oii$\lambda\lambda$7320,7330 & 7321.94, 7331.69 & 4.9229, 4.9295 & $<1.6$ & $1.54\pm0.13$ & $<143$ & \nodata & doublet sum
  \enddata
  \tablecomments{Fluxes from the joint fit of the \NIRSpec spectra (Section~\ref{sec:measurements}), not corrected for magnification or attenuation. Detections have ${\rm S/N}\geq5$ and tentative detections $3\leq{\rm S/N}<5$; non-detections are given as $3\sigma$ upper limits, $\max(0,F)+3\sigma$.}
  \tablenotetext{a}{Vacuum wavelengths.}
  \tablenotetext{b}{At the nebular redshift $z=5.72354\pm0.00002$ (Section~\ref{sec:measurements}).}
  \tablenotetext{c}{Measured flux, with the path-loss calibration of the \JWST pipeline for a point source.}
  \tablenotetext{d}{Path-loss correction for the morphology of AMORE6-B relative to the point-source calibration: the total flux is $C_{\rm path}F$. The uncertainty (fit and model terms) is not included in that of $F$, but is propagated to the equivalent widths and to the quantities derived from absolute fluxes. It arises from the morphology shared by all lines and varies smoothly with wavelength, so that it changes the ratios of optical lines by $\lesssim1\%$ and \heii$\lambda1640$/\hb by $\lesssim3\%$.}
  \tablenotetext{e}{Not reproduced in the individual nods and adopted as an upper limit.}
  \tablenotetext{f}{Of the total flux, against power-law fits to the \NIRCam photometry: F115W, F150W, F162M, and F182M below 3000\,\AA, the medium bands from F250M to F480M above (Section~\ref{sec:continuum}).}
  \tablenotetext{g}{Measured beyond the nominal G235M range (Table~\ref{tab:observations}); \hb\ lies at the NRS1 edge in G395M, and \oiii$\lambda\lambda$4959,5007 fall in the G395M detector gap.}
  \tablenotetext{h}{Against the local continuum of the \NIRSpec spectrum in the same grating, with the line and the continuum each corrected for their path loss; given only for lines detected at ${\rm S/N}\geq3$. In G235M and G395M the continuum is barely detected; where it is formally significant, at \oiii$\lambda\lambda$4959,5007 and \ha, it is 2--3 times higher than the photometric continuum (Section~\ref{sec:continuum}).}
  \tablenotetext{i}{Local continuum positive in less than 95\% of the posterior draws: the equivalent width is limited by the continuum and not reliable.}
\end{deluxetable*}}

\section{Physical conditions of the ionized gas}
\label{sec:conditions}

Before comparing the spectrum with photoionization models, we derive the properties of the gas that follow from the hydrogen and helium lines and from line ratios alone. These measurements rely only on atomic data and on the photometry, and they define the conditions under which the models of Section~\ref{sec:models} are compared with the data.

\subsection{Attenuation, temperature, density, and helium abundance}
\label{sec:helium}
\label{sec:attenuation}
\label{sec:tene}

\paragraph{Direct constraints.}
The main Balmer decrements are \ha/\hb$=2.63\pm0.06$, H$\gamma$/\hb$=0.461\pm0.022$, and H$\delta$/\hb$=0.263\pm0.021$. At the electron temperatures expected for gas of very low metallicity, $T_{\rm e}\simeq2$--$3\times10^{4}$\,K, the Case-B values for the above ratios are $\simeq2.7$, 0.47, and 0.26, respectively \citep{Storey1995}, so that the three ratios already indicate little or no attenuation. 
However, the usual diagnostics of temperature and density are not available. The auroral \oiii$\lambda4363$ line is not detected, with \oiii$\lambda4363/\lambda5007<0.36$ at $3\sigma$, a limit that does not constrain the temperature of gas in which \oiii$\lambda5007$ is itself weak; moreover, \oii$\lambda\lambda3727,3730$ and \sii$\lambda\lambda6716,6731$ are not detected, and the N\,\textsc{iv}]$\lambda\lambda1483,1486$ doublet, the only density diagnostic of the high-ionization gas in the G140M range, is detected at less than $3\sigma$.
The attenuation, temperature, and density must therefore be constrained together with the helium abundance from the hydrogen and helium recombination lines, whose ratios depend on all four.

\paragraph{Model.}
We fit the hydrogen and helium lines simultaneously, following the approach developed for the primordial helium abundance by \citet{Aver2021jcap} and references therein. The data are the fluxes (or upper limits) of H10, H9, H$\delta$, H$\gamma$, \hb, and \ha; of the blends H8+He\,\textsc{i}$\lambda3889$ and H$\epsilon$+He\,\textsc{i}$\lambda3965$ (from which \neiii$\lambda3968$ is fitted separately; Section~\ref{sec:measurements}); of He\,\textsc{i}$\lambda\lambda$4026, 4388, 4471, 4922, 5876, 6678, and 7065; and of \heii$\lambda\lambda$1640, 4686. H$\gamma$ enters twice, as the two independent measurements in G235M and G395M, for a total of 18 measurements. The predicted flux of each line is
\begin{equation}
\begin{split}
F_{\lambda} = {} & F_{0}\,C_{g}\,10^{-0.4A_{V}[k(\lambda)-k({\rm H}\beta)]/R_{V}}\\
& \times\sum_{j}y_{j}\,\frac{\epsilon_{j}(T_{\rm e},n_{\rm e})}{\epsilon_{{\rm H}\beta}(T_{\rm e},n_{\rm e})}\,f_{\tau,j}(\tau_{3889},T_{\rm e},n_{\rm e}),
\end{split}
\label{eq:helium_model}
\end{equation}
where the sum runs over the components of each feature, $y_{j}$ is 1 for hydrogen, $y^{+}$ for He\,\textsc{i}, and $y^{++}$ for He\,\textsc{ii}, $\epsilon_{j}$ are the emissivities, $f_{\tau,j}$ the radiative-transfer correction of the He\,\textsc{i} triplets, and $F_{0}$ the intrinsic \hb flux. The H\,\textsc{i} and He\,\textsc{ii} emissivities are from \citet{Storey1995} and the He\,\textsc{i} emissivities from \citet{Porter2012,Porter2013}, extended above $2.5\times10^{4}$\,K with the temperature dependence of \citet{DelZanna2022}, all evaluated with \textsc{PyNeb} \citep{Luridiana2015}; the optical-depth corrections are from \citet{Benjamin2002}. For consistency with the SED fit of Section~\ref{sec:sed}, we adopt the SMC bar extinction curve of \citet{Gordon2003} ($R_{V}=2.74$). Since the measurement fit applies no relative flux correction between gratings (Section~\ref{sec:measurements}), the lines measured in G395M are multiplied by a factor $C_{g}=C_{395}$, with a Gaussian prior $1\pm0.15$, while $C_{g}=1$ for G235M. The two H$\gamma$ measurements share the same intrinsic flux, so that their ratio fixes $C_{395}$; without the G395M measurement, $C_{395}$ would be constrained only by its prior and would be degenerate with $A_V$ through \ha/\hb, since \ha is observed only in G395M and \hb only in G235M. \heii$\lambda1640$ is the only line taken from G140M, whose flux scale relative to G235M is not constrained by any line in common: the ratio of the G140M and G235M continua where the two gratings overlap, $1.6^{+1.5}_{-0.7}$, is consistent with unity but does not constrain it. We therefore add in quadrature to its uncertainty a calibration term of 15\% of the predicted flux, equivalent to marginalizing over a factor $C_{140}$ with a Gaussian prior $1\pm0.15$. We include \heii$\lambda1640$ for two reasons: its ratio to \heii$\lambda4686$ is about seven times more sensitive to the attenuation than \ha/\hb, and the UV line ratios of the diagnostic diagrams (Section~\ref{sec:models}) are corrected with the attenuation derived here, which should therefore be constrained by a UV line as well. Lines measured with ${\rm S/N}\geq3$ enter with Gaussian likelihoods. For the six He\,\textsc{i} lines with ${\rm S/N}<3$, which are given as upper limits in Table~\ref{tab:line_fluxes}, we use a one-sided likelihood, which penalizes the model only when it exceeds $\max(0,F)$ and thus treats these measurements as upper limits rather than as negative fluxes. The free parameters are $y^{+}$, $y^{++}$, $T_{\rm e}$, $n_{\rm e}$, $\tau_{3889}$, $A_{V}$, $F_{0}$, and $C_{395}$, with uniform priors $1.5\leq T_{\rm e}/10^{4}\,{\rm K}\leq2.5$, $1\leq\log(n_{\rm e}/{\rm cm^{-3}})\leq3$, $0\leq A_{V}\leq1.5$, and $0\leq\tau_{3889}\leq10$. We sample the posterior with \textsc{emcee} \citep{ForemanMackey2013}. The recombination lines constrain the temperature only weakly: with a uniform prior over $1\leq T_{\rm e}/10^{4}\,{\rm K}\leq4$ the posterior rises towards the upper edge, where the atomic data are extrapolated, without improving the fit ($\Delta\chi^{2}=1$). We therefore restrict the prior to the range of the photoionization models of Section~\ref{sec:models} that reproduce $R_3$ and \heii$\lambda4686$/\hb, $T_{\rm e}(\mathrm{O\,III})\simeq1.8$--$2.5\times10^{4}$\,K.

\begin{figure*}
  \centering
  \includegraphics[width=\textwidth]{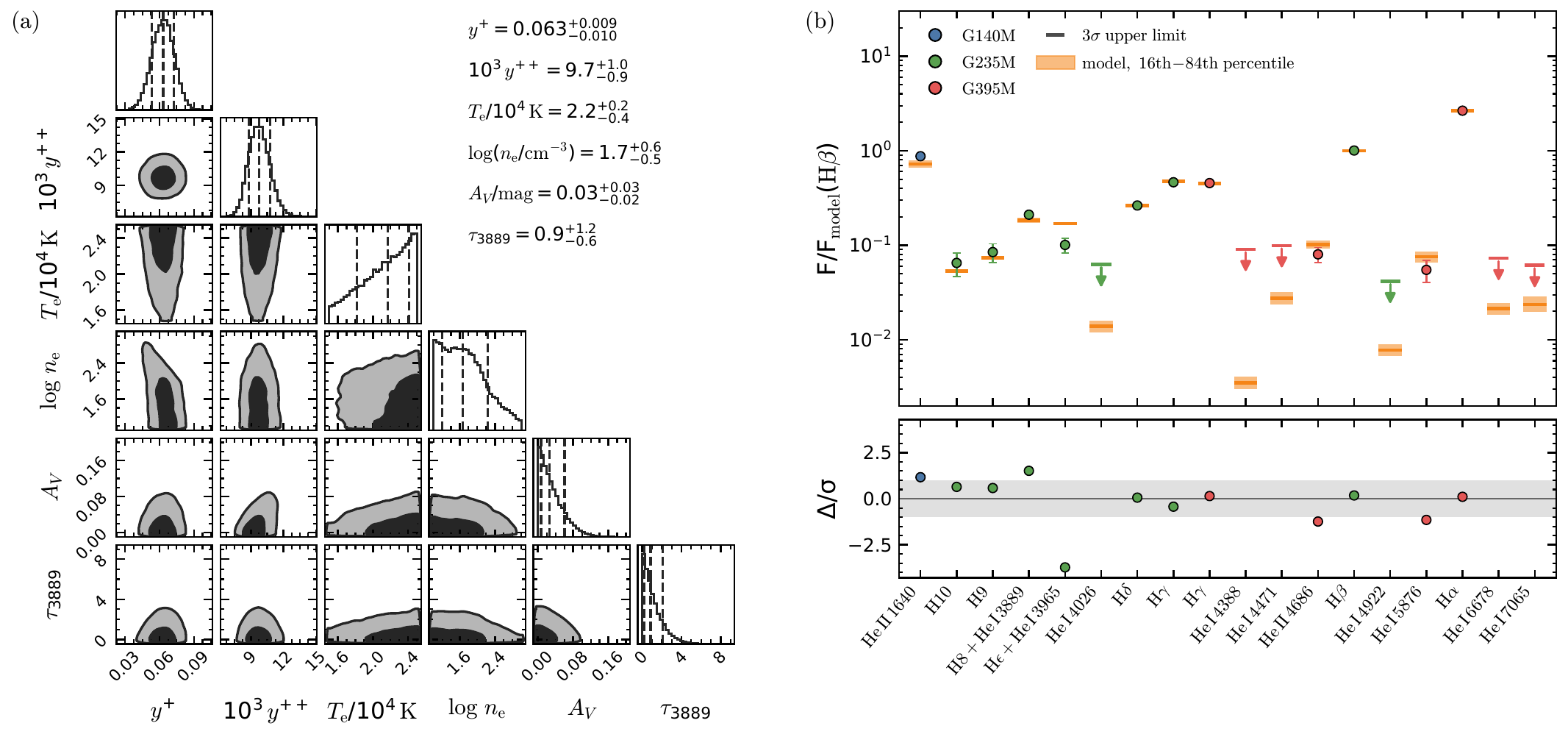}
  \caption{Joint fit of the hydrogen and helium lines of AMORE6-B (Section~\ref{sec:helium}). \textit{(a)} Posterior distribution of the abundances of singly and doubly ionized helium, $y^{+}$ and $y^{++}$, the electron temperature, the electron density ($n_{\rm e}$ in cm$^{-3}$), the attenuation $A_V$ in mag (SMC bar curve), and the optical depth of He\,\textsc{i}$\lambda3889$. Contours enclose 39 and 86\% of the posterior (1 and $2\sigma$ in two dimensions), and dashed lines in the histograms mark the 16th, 50th, and 84th percentiles, which are listed at the top right. The temperature and density posteriors are bounded by the uniform priors, $1.5\leq T_{\rm e}/10^{4}\,{\rm K}\leq2.5$ and $1\leq\log(n_{\rm e}/{\rm cm^{-3}})\leq3$. \textit{(b)} Measured fluxes relative to the intrinsic \hb flux of the model, $F_{0}$, ordered by wavelength and coloured by grating (blue: G140M; green: G235M; red: G395M); He\,\textsc{i} lines measured below $3\sigma$ are shown as $3\sigma$ upper limits, $\max(0,F)+3\sigma$, and enter the fit through a one-sided likelihood. Orange boxes and lines show the 16th--84th percentile range and median of the posterior-predictive model, including the attenuation and, for G395M, the relative flux scale $C_{395}$; H$\gamma$ is measured independently in both gratings. The lower panel shows the residuals normalized by the combined uncertainty of the measurement (including, for \heii$\lambda1640$, the 15\% calibration term, which is not included in its error bar) and of the model, with the grey band marking $\pm1\sigma$. Fluxes are corrected for path losses but not for magnification.}
  \label{fig:helium}
\end{figure*}

\paragraph{Results.}
The model reproduces the 18 fluxes with $\chi^{2}=23$ (Fig.~\ref{fig:helium}b). The only significant deviation is the H$\epsilon$+He\,\textsc{i}$\lambda3965$ blend, 40\% below the model ($3.7\sigma$); all the other measurements are reproduced within $1.5\sigma$, and the Balmer lines from \ha to H$\delta$ within $0.5\sigma$. The two He\,\textsc{ii} lines bracket the model, $\lambda4686$ 21\% below and $\lambda1640$ 21\% above it (both $1.2\sigma$): their observed ratio, $\lambda1640/\lambda4686=10.9\pm2.0$, exceeds the Case-B value, 7.0--7.6 at $T_{\rm e}=1.5$--$2.5\times10^{4}$\,K \citep{Storey1995}, and therefore leaves no room for attenuation by itself. The posterior is shown in Fig.~\ref{fig:helium}a. The attenuation is $A_V=0.03^{+0.03}_{-0.02}$\,mag ($<0.10$\,mag at the 97.5th percentile), 
which changes the ratios of optical lines by at most 1\% and \heii$\lambda1640$/\hb by 8\% (0.03\,dex). Without \heii$\lambda1640$ the fit gives $A_V=0.09^{+0.07}_{-0.06}$\,mag, consistent within $1\sigma$ (Appendix~\ref{app:helium}). The diagnostic diagrams of Section~\ref{sec:models} show the line ratios corrected for the attenuation of the fiducial fit. The correction is smaller than the measurement uncertainty for every ratio, including those that involve \heii$\lambda1640$ once its calibration uncertainty is accounted for; the ratios quoted in the text are the observed ones. The attenuation is instead varied within its posterior in the comparison of the emission lines with the models (Section~\ref{sec:inference}), and in the SED fit (Section~\ref{sec:sed}), and it is corrected for in $\xi_{\rm ion}$ (Section~\ref{sec:xiion}). The temperature is $T_{\rm e}=2.2^{+0.2}_{-0.4}\times10^{4}$\,K, within the adopted prior; at this temperature \oiii$\lambda4363/\lambda5007\simeq0.04$, well below the observed limit, and so is O\,\textsc{iii}]$\lambda1666$/\oiii$\lambda5007\simeq0.1$. 
The density is $\log(n_{\rm e}/{\rm cm^{-3}})<2.8$ (97.5th percentile), and enters the photoionization models as a free parameter (Section~\ref{sec:models}). The optical depth is small, $\tau_{3889}=0.9^{+1.2}_{-0.6}$, and $C_{395}=0.95\pm0.02$, consistent with the agreement of the two native H$\gamma$ measurements. The abundances are $y^{+}=0.063\pm0.010$ and $y^{++}=0.0097\pm0.0010$. The neutral-helium correction is negligible for an ionizing spectrum as hard as that of AMORE6-B, so that $y=0.073\pm0.010$ and the helium mass fraction is $Y=4y(1-Z)/(1+4y)=0.226^{+0.022}_{-0.024}$. Fixing $T_{\rm e}$, restricting $n_{\rm e}$, changing the treatment of the relative flux scale of the gratings, or leaving out \heii$\lambda1640$ changes $Y$ by at most 0.004, and allowing for collisional excitation of H\,\textsc{i} raises it by 0.015, within its uncertainty; these tests, and the larger change obtained with the resampling-control fluxes, are described in Appendix~\ref{app:helium}.

\paragraph{Comparison with the primordial Helium abundance.}
Two aspects of this measurement are remarkable. 
First, 13\% of the helium is doubly ionized ($y^{++}/y=0.13\pm0.02$), whereas in the local galaxies used to determine the primordial abundance $y^{++}$ is in general below 1\% of $y^{+}$, except in a few high-excitation H\,\textsc{ii} regions \citep{Izotov2014}. Second, AMORE6-B extends the measurements of helium in metal-poor gas to an oxygen abundance 1\,dex below that of the most metal-poor galaxies in those samples, which reach $12+\log(\mathrm{O/H})\simeq7$ \citep{Izotov2014,Hsyu2020,Aver2021jcap,Matsumoto2022}. Over this range the increase of $Y$ with O/H is negligible, so that at the abundance of AMORE6 the helium mass fraction should equal the primordial value. Our measurement is consistent with the primordial abundances derived from these samples, $Y_{\rm p}=0.2453\pm0.0034$ \citep{Aver2021jcap} and $0.2436^{+0.0039}_{-0.0040}$ \citep{Hsyu2020}, from which it differs by $0.9\sigma$ and $0.8\sigma$, respectively. It rests, however, on few, faint He\,\textsc{i} lines, of which only He\,\textsc{i}$\lambda5876$ is measured at more than $3\sigma$ outside the blends with the Balmer lines, on the \hb normalization (Appendix~\ref{app:helium}), and on a temperature that is not directly measured. Its uncertainty is several times larger than those of individual local galaxies, but at the same time it does not require the extrapolation to zero metallicity on which $Y_{\rm p}$ relies; a measurement of the same lines with a precision of $\simeq0.005$ in $Y$ would constrain $Y_{\rm p}$ directly in a system with $Z\lesssim10^{-2}\,Z_\odot$, delivering a unique test on the primordial Helium abundance $\approx$1~Gyr after the Big Bang.

\subsection{Ionization parameter}
\label{sec:ionization}

Neither \oii$\lambda\lambda3727,3730$ nor \neiii$\lambda3869$ is detected, and O32$\,\equiv\,$\oiii$\lambda5007$/\oii$\lambda\lambda3727,3730>2.9$ at $3\sigma$. At the low abundances of interest, O32 depends almost only on $U$, and for the stellar ionizing spectra of Section~\ref{sec:seds} this limit requires $\log U\gtrsim-3$ at any gas metallicity. The ratio of \civ to the X-shooter limit on \ciii (Section~\ref{sec:measurements}), \civ/\ciii$\,>0.11$, gives the same lower limit, $\log U\gtrsim-3$, in all the model families; it does not depend on C/O, since both lines are from carbon, but relies on the relative flux scales of the two instruments. Oxygen is therefore mostly O$^{2+}$, and $R_3$ traces the oxygen abundance with a small ionization correction. The strong \heii lines imply, however, that part of the oxygen is O$^{3+}$, whose ionization potential (54.9\,eV) is close to that of He$^{+}$ (54.4\,eV). With the usual approximation O$^{3+}$/(O$^{+}$+O$^{2+})\simeq$\,He$^{2+}$/He$^{+}$ \citep{Izotov2006}, the helium abundances of Section~\ref{sec:helium} give 0.15, a correction of $+0.06$\,dex to O/H, small compared with the calibration uncertainty of Section~\ref{sec:empirical_oh}. O\,\textsc{iv}]$\lambda\lambda1401$--1407 is not detected in the X-shooter spectrum ($<5.7\times10^{-18}$\,erg\,s$^{-1}$\,cm$^{-2}$ at $3\sigma$), but this limit allows O$^{3+}$/O$^{2+}$ up to $\simeq60$ at $T_{\rm e}=2.2\times10^{4}$\,K and does not constrain it. The ionization parameter is constrained jointly with the abundance in Section~\ref{sec:inference}.

\subsection{Ionizing-photon production efficiency}
\label{sec:xiion}

The ratio of the Balmer-line luminosity to the UV continuum gives the production rate of hydrogen-ionizing photons per unit UV luminosity, $\xi_{\rm ion}=Q({\rm H}^{0})/L_{\nu}(1500\,\mbox{\AA})$, independently of the magnification. From \ha, with the Case-B emissivity and recombination coefficient at $T_{\rm e}=2\times10^{4}$\,K, close to the lower limit set by the helium lines (Section~\ref{sec:helium}), and the UV continuum extrapolated to 1500\,\AA\ from the power law of Section~\ref{sec:continuum}, we obtain $\log(\xi_{\rm ion}/{\rm Hz\,erg^{-1}})=25.52^{+0.06}_{-0.07}$ after correcting both quantities for the attenuation of Section~\ref{sec:attenuation}, and $25.58\pm0.05$ without the correction; the uncertainties include the path-loss correction of \ha, and the attenuation dominates the first. These values assume that no ionizing photons escape. They are 0.03\,dex lower with the conversion of \citet{Leitherer1995}, which holds at $T_{\rm e}=10^{4}$\,K and is adopted by most of the studies quoted below, 0.03\,dex higher at $T_{\rm e}=3\times10^{4}$\,K, close to the median of the helium posterior, and 0.07\,dex higher if the UV continuum is taken from the fiducial SED model of Section~\ref{sec:sed} instead of the power law. The corrected value is slightly above the median of the spectroscopic samples of star-forming galaxies at $z>4.6$, $25.38\pm0.38$ \citep{Pahl2025}, within the range of other samples at $z>4$ \citep{Llerena2025,Begley2025}, and consistent with the average of faint \lya emitters at $z>5.8$, $\simeq25.57$ \citep{Saxena2024}. It is, however, 0.3--0.4\,dex below the values reported for galaxies of similar UV luminosity: $25.80\pm0.14$ for eight lensed galaxies with $-17\lesssim\muv\lesssim-15$ behind Abell~2744 \citep{Atek2024}, and $\simeq25.9$ from the relation of \citet{Simmonds2024} at the redshift and $\muv$ of AMORE6-B. In the fiducial model of Section~\ref{sec:sed}, the Pop~III.1 stars have $\log\xi_{\rm ion}\simeq26.5$ on their stellar continuum (those younger than 2\,Myr produce 91\% of the ionizing photons and 82\% of the UV continuum), but 86\% of their ionizing photons escape, so that the Balmer lines trace only the remaining 14\%. The measured value then corresponds to an intrinsic production efficiency $\log\xi_{\rm ion}\simeq25.52-\log(1-f_{\rm esc})\simeq26.4$, consistent with that of the stars. The single Pop~III.1 burst of Section~\ref{sec:sed} gives the same picture: $\log\xi_{\rm ion}\simeq26.6$ for the stars, $f_{\rm esc}=0.90$, and an intrinsic efficiency of 26.5. The measurement, which assumes no escape, is thus a lower limit on the intrinsic efficiency, and its moderate value is a consequence of the leakage rather than of a soft ionizing spectrum. In a radiation-bounded nebula the comparison would instead require the total continuum, stellar and nebular: for a young metal-free population the nebular continuum supplies 60--80\% of $L_\nu(1500\,\mbox{\AA})$, which lowers $\xi_{\rm ion}$ by 0.4--0.7\,dex with respect to the stellar value.

\paragraph{He$^{+}$-ionizing photons.}
In the same way, the \heii recombination lines count the photons above 54.4\,eV absorbed by He$^{+}$, $Q({\rm He}^{+})=L(\heii)\,\alpha_{\rm B}({\rm He}^{+})/\epsilon(\heii)$. The ratio of a \heii line to \ha thus gives the fraction of hard photons in the ionizing spectrum, $Q({\rm He}^{+})/Q({\rm H}^{0})$, independently of the magnification and of the UV continuum. We use, for both ions, the Case-B emissivities of \citet{Storey1995} computed with PyNeb \citep{Luridiana2015}, the recombination coefficients of \citet{Draine2011} (scaled to He$^{+}$ as for a hydrogenic ion), and $T_{\rm e}=2\times10^{4}$\,K in both zones. With these, \heii$\lambda4686$ gives $\log Q({\rm He}^{+})/Q({\rm H}^{0})=-1.33^{+0.07}_{-0.09}$ ($-1.37$ at $10^{4}$\,K): 5\% of the hydrogen-ionizing photons have energies above 54.4\,eV. The attenuation changes this value by 0.01\,dex, and \heii$\lambda4686$/\hb gives the same result. \heii$\lambda1640$ gives $-1.11^{+0.06}_{-0.04}$, 0.2\,dex higher, as expected from the excess of the observed $\lambda1640/\lambda4686$ over the Case-B value (Section~\ref{sec:helium}). We adopt the value from \heii$\lambda4686$, which is measured on the same flux scale as the Balmer lines.

The measured ratio can be compared directly with the ionizing spectra of the model families used in Sections~\ref{sec:models} and \ref{sec:sed}, which are described in Section~\ref{sec:seds} and summarized in Table~\ref{tab:sed_families}; here we quote only their $Q({\rm He}^{+})/Q({\rm H}^{0})$, integrated over the stellar spectra. The extremely metal-poor Pop~II populations fall far below it. At $\Zstar=10^{-5}$ and 1\,Myr, $\log Q({\rm He}^{+})/Q({\rm H}^{0})=-4.9$ for the BPASS v2.2.1 binary populations \citep{Eldridge2017} and $-4.0$ for the \textsc{pySTARBURST99} populations with very massive stars up to 300\,$M_\odot$ \citep{Hawcroft2025}, and neither exceeds $-2.7$ at any age up to 3\,Myr. Only metal-free populations younger than about 2\,Myr approach the measured value. The \textsc{pySTARBURST99} very-massive-star population at $\Zstar=0$ gives $-1.7$ at 1--2\,Myr. The Pop~III spectra of \textsc{Yggdrasil} \citep{Zackrisson2011} give, at 1\,Myr, $-1.9$ for a \citet{Kroupa2001} IMF over 0.1--100\,$M_\odot$ \citep{Schaerer2002}, $-1.6$ for the log-normal IMF with a characteristic mass of 10\,$M_\odot$ (Pop~III.2; \citealt{Raiter2010}), and $-1.2$ for a Salpeter IMF over 50--500\,$M_\odot$ (Pop~III.1; \citealt{Schaerer2002}). Pop~III.1 is the only one that exceeds the measured value. All of these fall below $-3$ after 3\,Myr.

For comparison, the integrated \heii$\lambda4686$/\hb of the most metal-poor \heii-emitting galaxies in the local Universe is 0.02--0.04 in I\,Zw\,18 and 0.026 in SBS\,0335$-$052E, i.e.\ $Q({\rm He}^{+})/Q({\rm H}^{0})\lesssim10^{-1.7}$, and reaches 0.06 only in individual knots of SBS\,0335$-$052E \citep{Kehrig2015,Kehrig2018,Schaerer2019}. 
The integrated ratio of AMORE6-B thus exceeds even those of the hardest regions known locally. The measured value is close to the stellar ratio of the fiducial model of Section~\ref{sec:sed}, $-1.22$, whose Pop~III.1 stars produce all the \heii and Balmer emission, and to that of the single Pop~III.1 burst at 1\,Myr, $-1.24$. 
Two assumptions enter this comparison. Photons above 54.4\,eV absorbed by H$^{0}$ and He$^{0}$ rather than by He$^{+}$ are not counted, so that $Q({\rm He}^{+})$ is a lower limit on the emitted rate. More importantly, the escape of ionizing photons must not depend on their energy. This holds for the picket-fence geometry of the SED fits of Section~\ref{sec:sed}, in which the photons escape through channels free of gas and all the nebular emission scales by $1-f_{\rm esc}$, so that the measured ratio is that of the stars. In a density-bounded nebula, instead, the He$^{++}$ zone, which lies inside the H$^{+}$ zone, would remain ionization bounded while hydrogen-ionizing photons escape, and the measured ratio would exceed the intrinsic one by up to $1/(1-f_{\rm esc})$, i.e.\ 0.85\,dex for the escape fraction of the fiducial model.

\subsection{An indirect estimate of the oxygen abundance}
\label{sec:empirical_oh}

\paragraph{The $R_3$ ratio.}
The \NIRSpec spectrum detects \oiii$\lambda5007$ at $10\sigma$ and gives $R_3=$\,\oiii$\lambda5007$/\hb$\,=0.249\pm0.025$, i.e.\ $\log R_3=-0.60\pm0.04$ (Table~\ref{tab:line_fluxes}). Both lines are measured in the same G235M spectrum, so that the ratio does not depend on the relative calibration of the gratings, and their path-loss corrections differ by less than 1\%. They fall at 3.27 and 3.37\,$\mu$m, beyond the nominal range of G235M, where the flux calibration is extrapolated; the ratio is affected only by the change of the response over 0.1\,$\mu$m, and H$\gamma$/\hb, with H$\gamma$ inside the calibrated range, agrees with the Case-B value (Section~\ref{sec:helium}). The main systematic is the \hb flux, which changes by 4\% when complete nod patterns are resampled instead of individual integrations (Section~\ref{sec:measurements}); this changes $\log R_3$ by 0.02\,dex and is negligible against the calibration uncertainty discussed below. Since O32$\,>2.9$ and $\log U\gtrsim-3$ (Section~\ref{sec:ionization}), oxygen is mostly O$^{++}$ and $R_3$ traces the oxygen abundance with a small ionization correction. The measurement is consistent with, and well below, the WFSS limits of $R_3<0.43$ for AMORE6-B and $<0.33$ for the stack of the two images \citep{Morishita2026}.

\begin{figure*}
  \centering
  \includegraphics[width=0.49\textwidth]{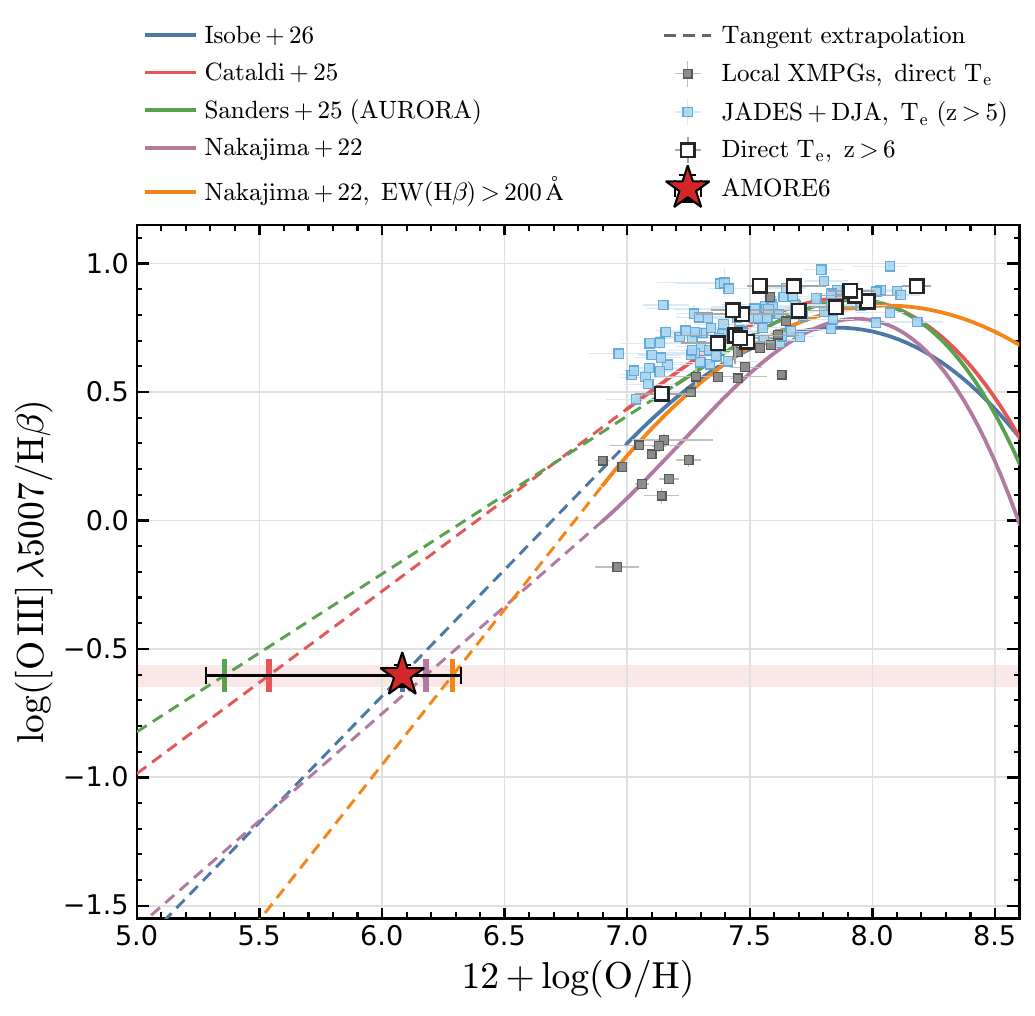}\hfill
  \includegraphics[width=0.49\textwidth]{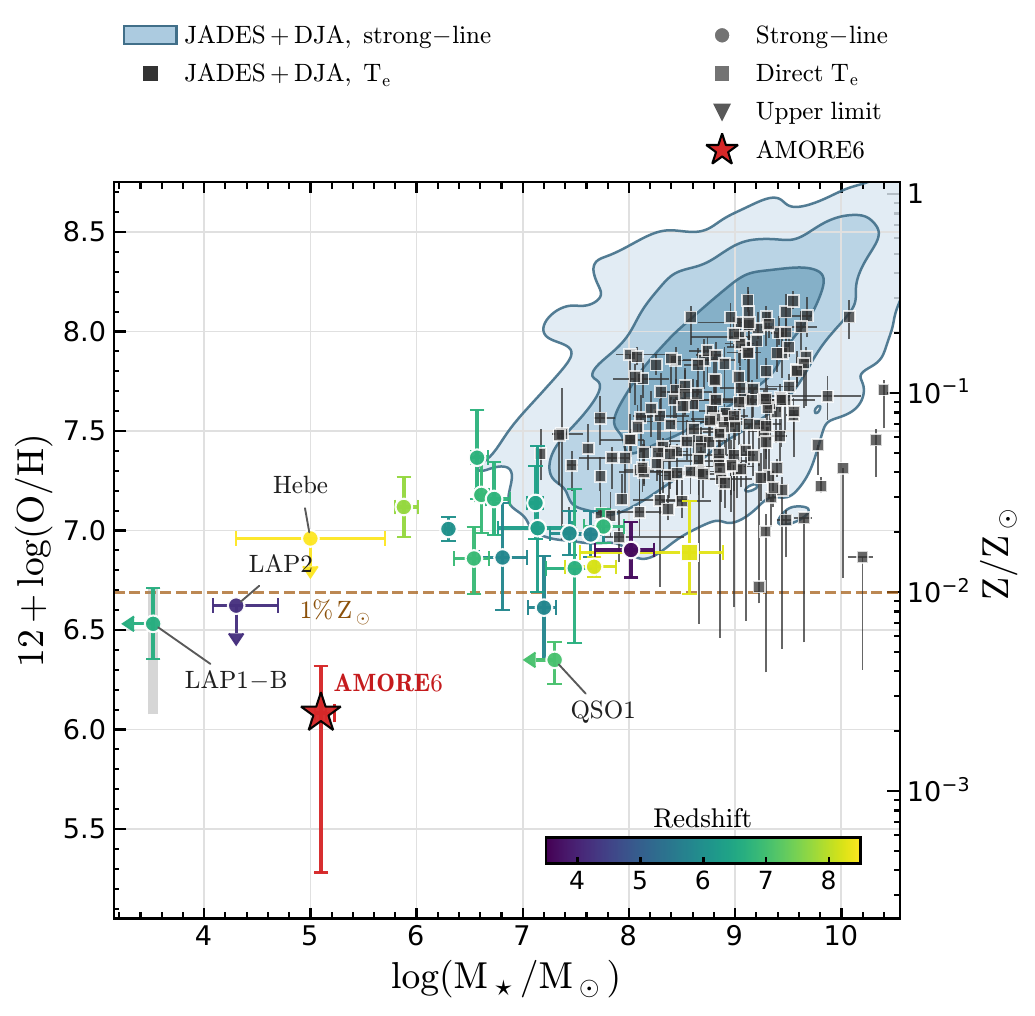}
  \caption{Oxygen abundance of AMORE6. \textit{Left:} $R_3=$\,\oiii$\lambda5007$/\hb against 12+log(O/H). Solid curves are the strong-line calibrations of \citet{Isobe2026mzr}, \citet{Cataldi2025}, \citet{Sanders2025aurora}, and \citet{Nakajima2022empress} (whole sample and EW(\hb)$>200$\,\AA) over their calibrated range, and dashed lines their extrapolation along the tangent at the lowest calibrated abundance. Grey squares are local extremely metal-poor galaxies and H\,\textsc{ii} regions with direct-$T_\mathrm{e}$ abundances \citep{Skillman2013,Berg2016,Izotov2018,Annibali2019,Berg2019,Kojima2020,Aver2021,Isobe2022}; open squares are direct-$T_\mathrm{e}$ measurements at $z>6$ \citep{MarquesChaves2024,Topping2024,Schaerer2024,Topping2025,Curti2025,Calabro2024,Rai2026}; light-blue squares are a compilation of $z>5$ galaxies with $T_\mathrm{e}$-based abundances from both JADES and DJA latest data releases with. The shaded band is the 16th--84th percentile range of $R_3$ of AMORE6-B, the coloured ticks mark the abundance given by each calibration, and the red star is their median, with a bar spanning all of them: $12+\log(\mathrm{O/H})=6.08^{+0.24}_{-0.80}$. \textit{Right:} mass--metallicity plane at $z>3$. Contours enclose 50, 80, and 95\% of the strong-line metallicities for a sample combining both JADES DR4 \citep{CurtisLake2026,Scholtz2026DR4} and the DAWN JWST Archive \citep[DJA;][]{Heintz2025primal,deGraaff2025rubies}, derived as in \citet{Curti2024,Isobe2026mzr} with the calibration of \citet{Isobe2026mzr}; grey squares are the available $T_\mathrm{e}$-based metallicities from the same sample. Coloured symbols are individually studied extremely metal-poor systems, coloured by redshift: LAP1-B \citep{Nakajima2025}, LAP2 \citep{Vanzella2026}, SAPPHIRES \citep{Hsiao2025}, UNCOVER \citep{Chemerynska2024}, EXCELS-63107 \citep{Cullen2025}, A370-z8-LAE \citep{Willott2025}, CAPERS-39810 \citep{CaiYu2026}, Hebe \citep{Maiolino2026,Ubler2026}, and QSO1 \citep{Maiolino2026BHnearpristine}. Their strong-line abundances are recomputed from $R_3$ with the calibration of \citet{Isobe2026mzr} (for the UNCOVER galaxies, from the $R_3$ implied by the published abundances; Hebe and QSO1 are given on that scale by the authors), which is also the median of the five calibrations of the left panel; the grey bar on LAP1-B spans the abundances given by the five calibrations; EXCELS-63107 has a direct-$T_\mathrm{e}$ abundance. AMORE6 is shown at the abundance of the left panel and at the stellar mass of the fiducial SED model; the dashed bar extends to the highest mass of the SED scenarios that are as likely as the fiducial one (Section~\ref{sec:sed}). The right axis gives $Z/Z_\odot$ for a solar $12+\log(\mathrm{O/H})=8.69$ \citep{Asplund2021}.}
  \label{fig:mzr_emp}
\end{figure*}

\paragraph{Strong-line calibrations below their calibrated range.}
Fig.~\ref{fig:mzr_emp} (left) compares this value with the $R_3$ calibrations derived from direct-$T_{\rm e}$ abundances of galaxies at high redshift \citep{Isobe2026mzr,Cataldi2025,Sanders2025aurora} and of local metal-poor galaxies \citep{Nakajima2022empress}, and with the galaxies that have both a direct abundance and a published $R_3$. AMORE6-B has the lowest $R_3$ of any of them. It lies 1.3--1.5\,dex below the turnover of the calibrations ($\log R_3\simeq0.75$--$0.87$ at $12+\log(\mathrm{O/H})\simeq7.9$), more than 1\,dex below every direct-method measurement at $z>5$, and 0.4\,dex below the most extreme local region, DDO~68 Reg-7, with $\log R_3=-0.18$ at $12+\log(\mathrm{O/H})=6.96$ \citep{Annibali2019}. No calibration therefore covers AMORE6-B, and the abundance requires an extrapolation below the lowest calibrating abundance, $12+\log(\mathrm{O/H})=6.9$--$7.2$, which we take along the tangent of each calibration at that point.

The five extrapolations give $12+\log(\mathrm{O/H})$ between 5.36 \citep{Sanders2025aurora} and 6.29 \citep[][galaxies with EW(\hb)$>200$\,\AA, as AMORE6-B]{Nakajima2022empress}. The spread of 0.9\,dex is not a measurement error, since the statistical uncertainty of each value is $\leq0.08$\,dex. It reflects the different slopes of the calibrations at their low-metallicity edge, from $\mathrm{d}\log R_3/\mathrm{d}\log(\mathrm{O/H})\simeq0.6$ for AURORA to 1.2 for the high-EW EMPRESS galaxies, which, over extrapolations of 0.6--1.1\,dex in $\log R_3$ below the calibrated range, become a difference of almost 1\,dex in abundance. A slope close to unity is the expected limit: when oxygen no longer dominates the cooling, the electron temperature stops rising as the abundance decreases, and at fixed ionization parameter $R_3$ becomes proportional to O/H. The calibrations with slopes of 0.8--1.2 \citep{Isobe2026mzr,Nakajima2022empress} give 6.08--6.29, while the shallower ones, set by galaxies in which oxygen still regulates $T_{\rm e}$, give 5.4--5.5. We adopt the median of the five values with an interval that includes all of them and their statistical uncertainties, $12+\log(\mathrm{O/H})=6.08^{+0.24}_{-0.80}$, i.e.\ 0.04--0.4\% of the solar value. This interval quantifies the calibration uncertainty only, since the calibrations also assume the ionization parameter and density typical of their calibrating samples; with these as free parameters, the SED fit gives $12+\log(\mathrm{O/H})=6.34\pm0.12$ (Section~\ref{sec:sed}), and the comparison of all the lines with the models favours $\log(\Zgas/Z_\odot)$ between $-3$ and $-2.5$ (Section~\ref{sec:inference}). For comparison, \citet{Morishita2026} derived $12+\log(\mathrm{O/H})<5.97$ ($3\sigma$) from the WFSS limit on $R_3$ with photoionization models at fixed ionization parameter and density, a value that falls within our interval.

\paragraph{Position in the mass--metallicity plane.}
Fig.~\ref{fig:mzr_emp} (right) places AMORE6 against the strong-line and direct-method abundances of galaxies at $z>3$ and of the individually studied extremely metal-poor systems. We adopt the stellar mass of the fiducial SED model, the intrinsic mass still in stars, $\log(M_\star/M_\odot)=5.10\pm0.06$ (Section~\ref{sec:sed}); the scenarios that are as likely as the fiducial one (Table~\ref{tab:sed_scenarios}) give up to $\log(M_\star/M_\odot)=5.2$. AMORE6 lies two orders of magnitude in mass below the bulk of the survey galaxies, and the relation fitted to them at $7<\log(M_\star/M_\odot)<9$ (slope 0.30) predicts $12+\log(\mathrm{O/H})\simeq6.74$ at its mass, 0.7\,dex above the measured value. In mass it falls between LAP1-B and LAP2 ($\log(M_\star/M_\odot)\lesssim4.3$, $12+\log(\mathrm{O/H})=6.5$--6.6 on the same scale; \citealt{Nakajima2025,Vanzella2026}) and the low-mass UNCOVER galaxies ($\log(M_\star/M_\odot)\geq5.9$, 6.9--7.4), and its abundance is 0.4--0.5\,dex below the former and 0.8--1.3\,dex below the latter. Over the five calibrations of Fig.~\ref{fig:mzr_emp} (left), LAP1-B spans 6.1--6.7, i.e.\ the comparison holds on a common calibration but not across calibrations. Hebe has $12+\log(\mathrm{O/H})<6.96$ at $\log(M_\star/M_\odot)=4.3$--5.7 \citep{Maiolino2026,Ubler2026}, and QSO1, which hosts an accreting black hole, $6.35^{+0.09}_{-0.12}$ at $\log(M_\star/M_\odot)<7.3$ \citep{Maiolino2026BHnearpristine}, both on the same scale. AMORE6 has the lowest abundance of the compilation; only QSO1 reaches a comparable value, at a higher mass. A system this far below the relation fits the picture of Section~\ref{sec:enrichment}, in which the metals released so far have enriched only the gas around the young stars.

\subsection{Carbon-to-oxygen ratio}
\label{sec:co}

The detection of \civ$\lambda\lambda1548,1551$ at $8\sigma$ gives a direct estimate of C/O. We compute the ionic abundances with \textsc{PyNeb} \citep{Luridiana2015} at $n_{\rm e}=100$\,cm$^{-3}$ and at the temperature of the fit to the hydrogen and helium lines, $T_{\rm e}=2.2^{+0.2}_{-0.4}\times10^{4}$\,K (Section~\ref{sec:helium}), after correcting for the attenuation of Section~\ref{sec:attenuation}. To avoid the flux scale of G140M relative to the other gratings, \civ is referred to \heii$\lambda1640$ in the same spectrum (\civ/\heii$\lambda1640=0.46\pm0.06$) and placed on the \hb scale with the Case-B ratio \heii$\lambda1640/\lambda4686$ and the measured \heii$\lambda4686$/\hb. Together with \oiii$\lambda5007$/\hb, this gives $\log(\mathrm{C^{3+}/O^{2+}})=-0.51\pm0.11$. The ratio of \civ to the undetected O\,\textsc{iii}]$\lambda\lambda1661,1666$, which is almost independent of temperature, requires $\log(\mathrm{C^{3+}/O^{2+}})>-1.0$ ($3\sigma$), and the X-shooter limit on \ciii constrains C$^{2+}$ only weakly. O$^{+}$ is negligible. In the photoionization models of Section~\ref{sec:models} that reproduce $R_3$ and \heii$\lambda4686$/\hb, C$^{3+}$/O$^{2+}$ falls short of C/O by 0.19\,dex (0.10--0.27\,dex), the fraction of carbon in C$^{2+}$. We obtain $\log(\mathrm{C/O})=-0.30^{+0.23}_{-0.18}$, where the uncertainty includes those of $T_{\rm e}$, to which the collisionally excited UV lines are very sensitive ($+0.2$ at $1.5\times10^{4}$\,K and $-0.47$ at $2.5\times10^{4}$\,K), and of the ionization correction. The direct G140M flux scale would raise the value by 0.2\,dex, and resonant scattering of \civ makes it a lower limit. An independent estimate with \textsc{HII-CHI-mistry-UV} \citep{PerezMontero2017}, which compares the UV carbon and oxygen lines with a grid of photoionization models, gives $\log(\mathrm{C/O})>-0.5$ at 99\% confidence; it is also a lower limit, because \ciii, and with it C$^{2+}$, is not detected, and is consistent with the value above. C/O is thus about 0.5\,dex above the plateau of metal-poor galaxies, $\log(\mathrm{C/O})\simeq-0.8$, adopted in our models \citep{Nicholls2017}, and consistent with the solar value, $\log(\mathrm{C/O})_\odot=-0.23$ \citep{Asplund2021}; it would be at the plateau only for $T_{\rm e}\gtrsim3\times10^{4}$\,K, above the range of the models that reproduce the other lines. As shown in Fig.~\ref{fig:co}, this C/O is at or above the upper end of the yields of Pop~II core-collapse supernovae, and within the region of the NEFERTITI galaxies enriched by Pop~III stars \citep{Rusta2025}, which at the O/H of AMORE6 have $\log(\mathrm{C/O})$ between $-0.57$ and $-0.03$ (16th--84th percentiles). Galaxies enriched by Pop~II stars rarely reach such a low O/H in these simulations and, when they do, have $\log(\mathrm{C/O})\simeq-0.42$, at the lower end of the measured range.
\iftrue
  \begin{figure*}
  \centering
  \includegraphics[width=\columnwidth]{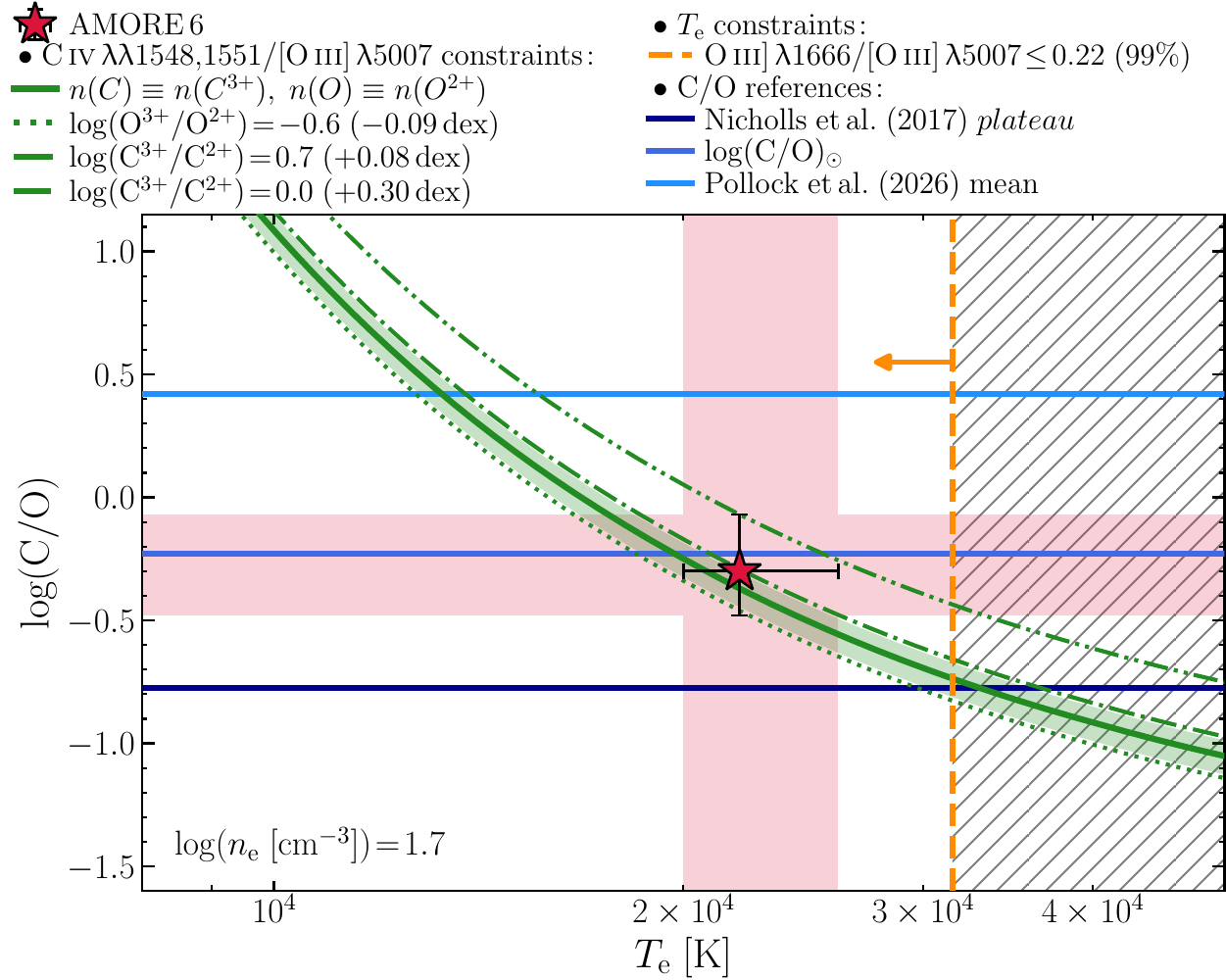}
  \includegraphics[width=\columnwidth]{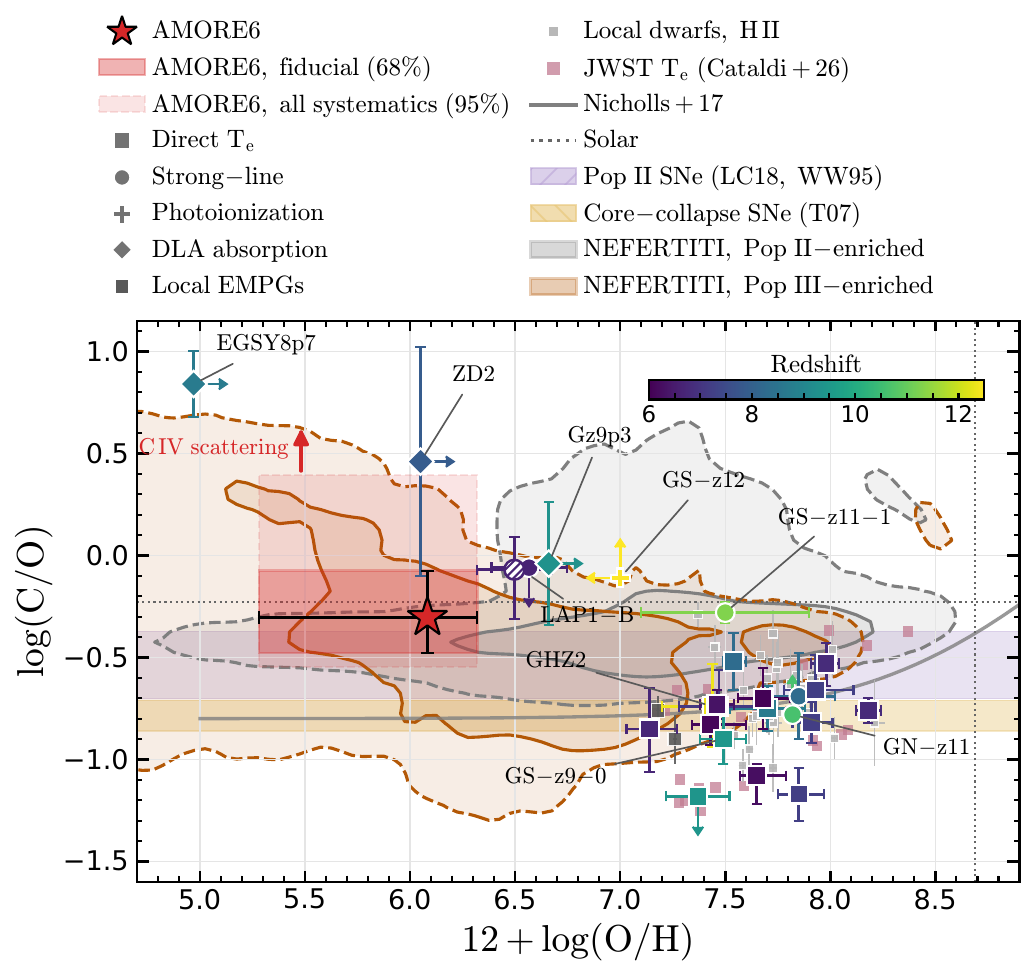}
  \caption{C/O abundance in AMORE6. \textit{Top panel:} the constraints from the model
  (red star and shaded cross) agree with direct constraints from \textsc{PyNeb}: the
  green curves are from the observed \civ$\lambda\lambda1548,51$ to \oiii$\lambda5007$
  flux ratio, without ICF (solid) and showing the effect of individual ICF corrections,
  as labelled. The horizontal lines show reference abundances from the literature.
  \textit{Bottom panel:} C/O against O/H. AMORE6 (red star) is shown at the strong-line abundance of Section~\ref{sec:empirical_oh}; the dark box is the 68\% range of the direct C/O estimate, and the light box the 95\% range including the temperature over $1.5$--$2.5\times10^{4}$\,K and the flux scale of G140M. The arrow marks that resonant scattering of \civ makes the value a lower limit. Coloured symbols are galaxies at $z>6$, coloured by redshift, with the shape giving the method of the O/H as in Fig.~\ref{fig:mzr_emp} (squares: direct $T_\mathrm{e}$; circles: strong lines; crosses: photoionization models) \citep{Cameron2023gnz11,DEugenio2024gsz12,Castellano2024ghz2,Calabro2024,MarquesChaves2024,Schaerer2024,NavarroCarrera2024,Isobe2023cn,Topping2024,Topping2025,Curti2025,Rai2026,Pollock2026}, including GS-z11-1 \citep{Scholtz2025gsz11}. LAP1-B is shown at the abundance of Fig.~\ref{fig:mzr_emp} with both the C/O measured by \citet{Nakajima2025} (hatched) and the upper limit of \citet{Scholtz2026}, slightly offset in O/H for clarity, and GS-z12 with the low-metallicity solution and the C/O lower limit of the original analysis of \citet{DEugenio2024gsz12}. Diamonds are the abundances of the damped \lya absorbers of three galaxies at $z=7.9$--9.3 \citep{Pollock2026}, for which the O/H are lower limits. Grey squares are local metal-poor galaxies and H\,\textsc{ii} regions, pink squares the JWST galaxies with direct abundances of \citet{Cataldi2026}, and the grey curve the local trend of \citet{Nicholls2017}. The hatched bands give the C/O of the yields of Pop~II core-collapse supernovae \citep{WoosleyWeaver1995,LimongiChieffi2018} and of the 13--50\,$M_\odot$ core-collapse supernovae of \citet{Tominaga2007}, converted with $\log(\mathrm{C/O})_\odot=-0.26$. The orange and grey regions enclose 68\% (solid contours) and 95\% (dashed) of the NEFERTITI galaxies \citep{Rusta2025} enriched by Pop~III and by Pop~II stars, respectively.}
  \label{fig:co}
  \end{figure*}
\fi

\section{Modeling framework}
\label{sec:models}

We now compare the spectral properties of AMORE6-B with predictions from a set of different model families.
Such a comparison rests on three ingredients: a set of photoionization models that predicts every line under one common set of assumptions about the gas; a set of ionizing spectra that spans physically distinct scenarios proposed for AMORE6-like systems; and a comparison between models and data that uses the full information content of the spectrum, including the non-detections, rather than a single line ratio or specific diagnostic diagram. We describe the three in turn, and defer the technical details of the grid and of the inference to Appendix~\ref{app:model_grids}.

\subsection{Photoionization model grid}
\label{sec:grid}

All the model predictions used in this paper come from \Cloudy \citep[version 25;][]{Gunasekera2025} with the same nebular setup for every ionizing spectrum, so that the differences between families arise from the ionizing spectrum alone. The nebula is an ionization-bounded cloud of constant density, without dust, illuminated by the central source. The oxygen abundance of the gas, $\Zgas$, is independent of the metallicity of the stars, $\Zstar$, so that a metal-free stellar population can ionize gas already enriched by the first supernovae, as expected during the transition from Pop~III to Pop~II star formation \citep{Rusta2025}. The grid spans $12+\log(\mathrm{O/H})$ from 5.0 to 8.7, in addition to pristine gas, $-3.5\leq\log U\leq0$, and $10^{2}\leq n_{\rm H}\leq10^{7}$\,cm$^{-3}$. The density range extends beyond the critical density of \oiii$\lambda5007$ ($n_{\rm crit}\simeq7\times10^{5}$\,cm$^{-3}$), so that the collisional suppression of the forbidden lines invoked to reconcile a low $R_3$ with a higher abundance (Section~\ref{sec:empirical_oh}) is part of the parameter space explored by the model. Carbon and nitrogen follow the abundance pattern of \citet{Nicholls2017}, which at the abundances of interest reduces to the primary plateau, $\log(\mathrm{C/O})\simeq-0.8$ and $\log(\mathrm{N/O})\simeq-1.7$, and the other elements scale with oxygen. For each model we adopt the fluxes of Table~\ref{tab:line_fluxes} relative to \hb and their equivalent widths against the stellar and nebular continuum.

We note that the fiducial models discussed here and included in the analysis of Section~\ref{sec:inference} are radiation bounded. A density-bounded nebula that leaks ionizing photons loses preferentially the low-ionization lines and can lower $R_3$ at fixed abundance. For each ionizing spectrum we therefore also computed clouds truncated where 20, 50, and 80\% of the hydrogen-ionizing photons escape. Their effect is shown as a vector in the diagnostic diagrams and discussed in Section~\ref{sec:leakage}. 
The grid limits, the stopping criteria, and the atomic data are listed in Appendix~\ref{app:model_grids}.

\subsection{Families of ionizing spectra}
\label{sec:seds}

The same \Cloudy grids are computed for the families of ionizing spectra listed in Table~\ref{tab:sed_families}, as representative, physically-motivated but distinct explanations for a system with strong Balmer and \heii emission and weak metal lines such as AMORE6.
Within each family, the gas metallicity and the ionization parameter are free.

\begin{table*}
  \caption{Families of ionizing spectra.\label{tab:sed_families}}
  \centering
  \small
  \begin{tabular}{p{2.7cm}p{5.6cm}p{2.9cm}p{5.6cm}}
    \toprule
    Family & Spectra & Gas & Distinctive prediction \\
    \midrule
    Pop~III, self-consistent & NEFERTITI galaxies \citep{Rusta2025} at successive evolutionary stages & enriched by Pop~III supernovae, as in the simulations & strong \heii with metal lines from the first supernovae; abundance pattern of Pop~III yields \\
    Pop~III, population synthesis & \citet{Schaerer2002} and \citet{Raiter2010} SSPs through \textsc{Yggdrasil} \citep{Zackrisson2011}: Pop~III.1 (Salpeter IMF over 50--500\,$M_\odot$), Pop~III.2 (log-normal IMF), and a Kroupa IMF over 0.1--100\,$M_\odot$; 1--10\,Myr & pristine to $12+\log(\mathrm{O/H})=7$, abundances scaled & \heii$\lambda4686$/\hb from $\simeq0.02$ (Kroupa IMF) to $\simeq0.08$ (Salpeter IMF over 50--500\,$M_\odot$) at 1\,Myr; no \heii after $\simeq3$\,Myr \\
    Extremely metal-poor Pop~II & BPASS v2.2 binaries \citep{Eldridge2017}, $\Zstar=10^{-5}$ (1\,Myr) and $10^{-4}$ (1--10\,Myr) & as above & \heii$\lambda4686$/\hb$\lesssim0.003$; \civ and O\,\textsc{iii}] only for $12+\log(\mathrm{O/H})\gtrsim6.5$ \\
    Very massive stars & \textsc{pySTARBURST99} \citep{Hawcroft2025} populations with stars up to 300\,$M_\odot$ at $\Zstar=0$ (1--3\,Myr) and $10^{-5}$ (1\,Myr) & as above & \heii$\lambda4686$/\hb$\simeq0.03$ at $\Zstar=0$ for 2.5\,Myr, independent of $U$ and $\Zgas$; $\simeq0.004$ at $\Zstar=10^{-5}$ \\
    X-ray binaries & BPASS $\Zstar=10^{-4}$ population with an X-ray binary component, a multicolour accretion disc with inner temperature $kT_{\rm in}=1$\,keV (stellar-mass binaries) or 0.2\,keV (ultraluminous sources), which supplies a fraction $f_Q=0.01$--1 of the hydrogen-ionizing photons & as above & \heii increasing with $f_Q$, with little effect on the metal lines \\
    Massive black holes (AGN/DCBH) & \Cloudy AGN continua in the configuration of \citet{NakajimaMaiolino2022}: thermal bump at $T_{\rm bb}=5\times10^{4}$, $10^{5}$, and $2\times10^{5}$\,K (the disc temperatures of $10^{5}$--$10^{6}\,M_\odot$ black holes) with an optical-to-X-ray slope $\alpha_{\rm ox}=-1.2$, $-1.6$, and $-2.0$, alone and added to the BPASS $\Zstar=10^{-4}$ population & as above & \heii$\lambda4686$/\hb from $\simeq0.005$ ($\alpha_{\rm ox}=-2.0$) to $\simeq0.15$ ($\alpha_{\rm ox}=-1.2$); very high-ionization lines ([Ne\,\textsc{iv}], [Ne\,\textsc{v}]), low-ionization lines from the extended partially ionized zone \\
    \bottomrule
  \end{tabular}
  \begin{flushleft}\footnotesize All families share the gas grid of Section~\ref{sec:grid}, and all appear both in the diagnostic diagrams (Fig.~\ref{fig:model_comparison}) and in the comparison of Section~\ref{sec:inference} (Fig.~\ref{fig:line_evidence}).\end{flushleft}
\end{table*}

The first family defines the Pop~III scenario in its most complete form: the locally-calibrated NEFERTITI model follows the formation of Pop~III stars and the enrichment of their surroundings by the first supernovae using high-resolution N-body simulations of the Milky Way assembly \citep{Koutsouridou2023, Koutsouridou2024, Koutsouridou2026}. NEFERTITI provides nebular spectra of Pop~III galaxies at successive evolutionary stages, from a self-polluted phase in which all the stellar mass is in Pop~III stars but the gas has been self-enriched by Pop~III SNe, through the hybrid (PopIII+PopII) phases, in which the mass fraction of Pop~III stars with respect to the total decreases \citep{Rusta2025,Rusta2026}. Each galaxy is computed at $\log U=-3$, $-2$, $-1$, $-0.5$, and 0. Following \citet{Rusta2025}, here we use the models at $\log U\geq-2$. These models allow us to test whether the spectrum of AMORE6 can be reconciled by combining the hard spectrum of metal-free stars with the metal lines expected from gas that the first supernovae have already polluted. 

The second family attempts to isolate the role of the stellar initial mass function. It consists of the metal-free, instantaneous-burst spectra of \textsc{Yggdrasil} \citep{Zackrisson2011}, stars only, for three IMFs: a Salpeter slope over 50--500\,$M_\odot$ (Pop~III.1) and a \citet{Kroupa2001} IMF over 0.1--100\,$M_\odot$, both from the single stellar populations of \citet{Schaerer2002}, and a log-normal IMF with a characteristic mass of 10\,$M_\odot$ over 1--500\,$M_\odot$ (Pop~III.2; \citealt{Raiter2010}). The ages span 1--10\,Myr, before and after the most massive stars have died. 

The third family represents the main alternative for a very young, extremely metal-poor (though not fully pristine) system, namely an ordinary binary population at $\Zstar=10^{-5}$ and $10^{-4}$. We use the instantaneous-burst spectra of BPASS v2.2.1 \citep{Eldridge2017,StanwayEldridge2018}, which include binary evolution, with an IMF slope of $-1.35$ above 0.5\,$M_\odot$ and an upper mass of 300\,$M_\odot$, at 1\,Myr for $\Zstar=10^{-5}$ and 1--10\,Myr for $\Zstar=10^{-4}$.

The fourth family tests whether the inclusion of very massive stars (VMS), which reach effective temperatures and \heii-ionizing luminosities comparable to those of Pop~III stars even at non-zero metallicity \citep{Schaerer2025}, can reproduce the spectrum without invoking metal-free stars. We use instantaneous bursts computed with \textsc{pySTARBURST99} \citep{Hawcroft2025}, with a \citet{Kroupa2001} IMF extended to 300\,$M_\odot$, the non-rotating \textsc{Genec} tracks at $\Zstar=0$ \citep{Murphy2021} and $10^{-5}$, including the very-massive-star models of \citet{Martinet2023}, and the \textsc{Fastwind} atmospheres of \citet{Hawcroft2025}, at 1--3\,Myr for $\Zstar=0$ and 1\,Myr for $\Zstar=10^{-5}$. At $\Zstar=10^{-5}$ the very massive stars are assigned the metal-free atmospheres, since the atmosphere grid at that metallicity does not reach their temperatures.

The fifth family adds to the BPASS population at $\Zstar=10^{-4}$ of the third family the emission of X-ray binaries, which at low metallicity provide a hard component that can contribute to the \heii emission \citep{Schaerer2019,Senchyna2020,Lecroq2024}. 
We describe their spectrum as a multicolour accretion disc \citep{Mitsuda1984}, with an inner temperature of 1\,keV, typical of stellar-mass black holes, or of 0.2\,keV, typical of ultraluminous sources and of black holes of $\sim10^{3}$--$10^{4}\,M_\odot$. The stars and the X-ray source illuminate separate clouds with the same $\Zgas$ and $U$, and the two nebular spectra are co-added; the strength of the X-ray component is set by the fraction of the hydrogen-ionizing photons it supplies, $f_Q$ (from 1 to 100\%), rather than by $L_X$/SFR, since the X-ray luminosity of binaries at these metallicities is poorly known \citep{Lehmer2021}. 
We note that the adopted BPASS spectra include the effects of binary evolution on the stars but not the emission from accretion onto compact remnants. 
In the comparison of Section~\ref{sec:inference}, $f_Q$ and the age of the stars are independent, so that X-ray binaries can accompany even a 1\,Myr old population. This favours the scenario, since X-ray binaries appear only once the most massive stars have collapsed into compact remnants, after $\simeq4$\,Myr; the SED fits, which impose this delay, might instead disfavour it (see Section~\ref{sec:alternatives}). 
For the soft disc, \Cloudy does not converge at $\log U>-2.5$, and these models are limited to lower ionization parameters; for the hard disc, the models at $-1.5\leq\log U\leq-0.5$, which do not converge, are interpolated between $\log U=-2$ and 0. 

The sixth family includes accretion onto a massive black hole, either a low-mass AGN or a direct-collapse black hole (DCBH) formed in a pristine halo, which can also produce strong \heii with weak metal lines \citep{NakajimaMaiolino2022}. X-ray binaries are also accretion-powered, but the two families differ in the mass of the compact object and hence in the spectrum: the disc temperature scales as $M_{\rm BH}^{-1/4}$, so that the discs of stellar-mass and intermediate-mass black holes peak in the X-rays, whereas those of black holes of $10^{5}$--$10^{6}\,M_\odot$ peak in the extreme ultraviolet, supplying far more He$^{+}$-ionizing photons together with a bright ultraviolet and optical continuum. We adopt the configuration of \citet{NakajimaMaiolino2022}: the \Cloudy multi-component AGN continuum, in which a thermal ``big bump'' of temperature $T_{\rm bb}$ is joined to an X-ray power law by the optical-to-X-ray slope $\alpha_{\rm ox}$, with $T_{\rm bb}=5\times10^{4}$, $10^{5}$, and $2\times10^{5}$\,K, the disc temperatures of black holes of $10^{5}$--$10^{6}\,M_\odot$, and $\alpha_{\rm ox}=-1.2$, $-1.6$, and $-2.0$ (see also Appendix~\ref{app:model_grids}). 
Each is taken alone and, in the same way as the X-ray binaries, added to the BPASS population at $\Zstar=10^{-4}$ with $f_Q$ free, since an accreting source alone does not reproduce the Balmer equivalent widths (Section~\ref{sec:ionizing_source}). In the SED fits (Section~\ref{sec:sed}) the disc continuum can also be hidden from view, as in a type-2 AGN, so that only its nebular emission is observed. Here we note that, unlike \citet{NakajimaMaiolino2022}, who stop their clouds at a neutral column of $10^{21}$\,cm$^{-2}$, our models are radiation bounded and include, beyond that column, an X-ray-heated, partially ionized zone. This zone strengthens mainly the low-ionization lines and leaves the ratios between high-ionization lines unchanged; nonetheless, the nine continua were computed with both criteria (Appendix~\ref{app:model_grids}), and our conclusions mainly rest on the equivalent widths of the Balmer and \heii lines (Section~\ref{sec:ionizing_source}), which depend little on this choice: removing \oii, [N\,\textsc{ii}], and [S\,\textsc{ii}] from the comparison of Section~\ref{sec:inference} changes the evidence of these models by less than 0.1.

\subsection{Emission line diagnostic diagrams}
\label{sec:ionizing_source}

\begin{figure*}
  \centering
  \includegraphics[width=\textwidth]{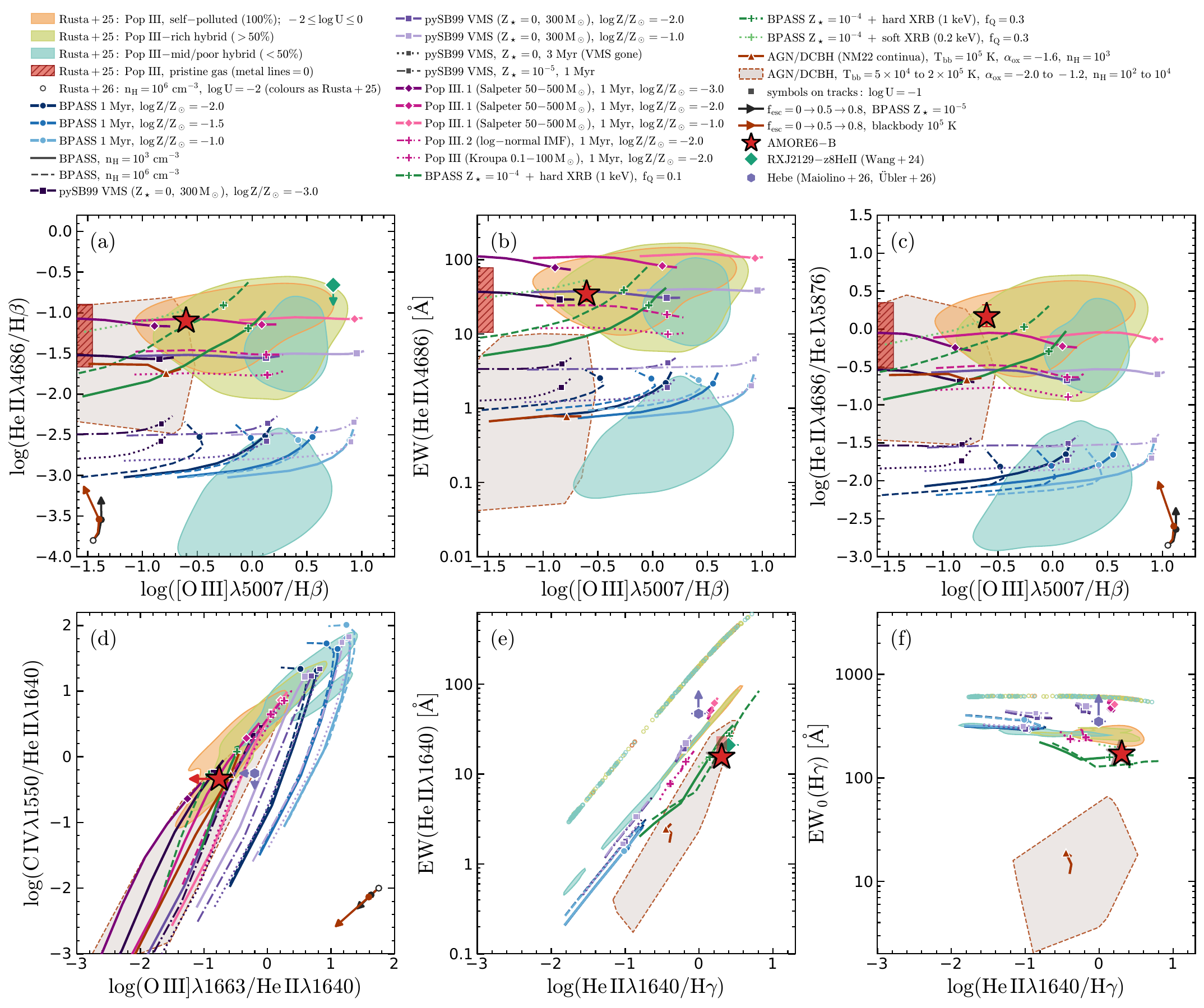}
  \caption{Diagnostic diagrams for AMORE6-B (red star). \textit{(a)--(c)} Hardness of the ionizing spectrum: \heii$\lambda4686$/\hb, EW(\heii$\lambda4686$), and \heii$\lambda4686$/He\,\textsc{i}$\lambda5876$ against $R_3$; \textit{(d)} \civ/\heii$\lambda1640$ against O\,\textsc{iii}]$\lambda1663$/\heii$\lambda1640$; \textit{(e)}, \textit{(f)} the planes based on H$\gamma$ used for Hebe \citep{Ubler2026,Maiolino2026,Rusta2026}: EW(\heii$\lambda1640$) and the rest-frame EW of H$\gamma$ against \heii$\lambda1640$/H$\gamma$. Filled regions enclose 68\% of the NEFERTITI Pop~III galaxies of \citet{Rusta2025} at $-2\leq\log U\leq0$ and $n_{\rm H}=10^{3}$\,cm$^{-3}$, grouped by evolutionary stage; the hatched strip marks the pristine phase, drawn as the self-polluted models with every metal line set to zero and put at the edge of the axes. Blue tracks are BPASS $\Zstar=10^{-5}$, 1\,Myr models at three gas metallicities, for $n_{\rm H}=10^{3}$ (solid) and $10^{6}$\,cm$^{-3}$ (dashed); purple tracks are the very-massive-star models at three gas metallicities, for $\Zstar=0$ at 1\,Myr (solid) and 3\,Myr (dotted) and for $\Zstar=10^{-5}$ at 1\,Myr (dash-dotted), all at $n_{\rm H}=10^{3}$\,cm$^{-3}$; magenta tracks are the metal-free Pop~III.1 population (Salpeter IMF over 50--500\,$M_\odot$) at 1\,Myr, at the same three gas metallicities and $n_{\rm H}=10^{3}$\,cm$^{-3}$, and the magenta dashed and dotted tracks are the Pop~III.2 (log-normal IMF) and Kroupa-IMF Pop~III populations at 1\,Myr and $\log(\Zgas/Z_\odot)=-2$; green tracks are the BPASS $\Zstar=10^{-4}$, 1\,Myr population with X-ray binaries that supply 10\% (solid) and 30\% (dashed) of the hydrogen-ionizing photons for $kT_{\rm in}=1$\,keV, and 30\% for $kT_{\rm in}=0.2$\,keV (dotted, $\log U\leq-2.5$), at $\log(\Zgas/Z_\odot)=-2$; symbols on the tracks mark $\log U=-1$. The shaded band with dashed edge encloses the massive-black-hole models in the \citet{NakajimaMaiolino2022} configuration at $12+\log(\mathrm{O/H})=6.0$ (all nine continua, $-3\leq\log U\leq-0.5$, $10^{2}\leq n_{\rm H}\leq10^{4}$\,cm$^{-3}$), and the brown track is the fiducial continuum ($T_{\rm bb}=10^{5}$\,K, $\alpha_{\rm ox}=-1.6$, $n_{\rm H}=10^{3}$\,cm$^{-3}$; the triangle marks $\log U=-1$). Arrows show the effect of an escape fraction of ionizing photons of 0.5 (dot) and 0.8 (head) for a BPASS population (black) and for a $10^{5}$\,K blackbody (brown). In panels (e) and (f), open circles are the NEFERTITI models at $n_{\rm H}=10^{6}$\,cm$^{-3}$ and $\log U=-2$ of \citet{Rusta2026}, coloured by stage as the filled regions, with the pristine phase in dark red. Where a grid gives only EW(\hb), EW(H$\gamma$) is computed as EW(\hb)$\times$(H$\gamma$/\hb)$\times0.80$, where 0.80 is the ratio of the continuum flux densities at \hb and H$\gamma$ in the models that give both (0.73--0.87), and EW(\heii$\lambda4686$) as EW(\hb)$\times$(\heii$\lambda4686$/\hb)$\times0.93$ for the same continuum slope. The AMORE6-B line ratios are from the joint fit of the \NIRSpec spectra, with arrows for $3\sigma$ limits, and are corrected for the attenuation of Section~\ref{sec:attenuation}, whose uncertainty is included in the error bars; in panels (e) and (f) the faint star marks the observed \heii$\lambda1640$/H$\gamma$, before the correction; the equivalent widths of \heii$\lambda4686$, H$\gamma$, and \heii$\lambda1640$ are on the total-flux scale, over power-law fits to the photometric continuum (Section~\ref{sec:continuum}). The hexagon is Hebe (total \heii of \citealt{Maiolino2026} over the H$\gamma$ of \citealt{Ubler2026}); other symbols are extremely metal-poor and \heii-emitting galaxies from the literature.}
  \label{fig:model_comparison}
\end{figure*}

Fig.~\ref{fig:model_comparison} places the line ratios, equivalent widths, and upper limits of AMORE6-B in six diagnostic planes, together with the loci of the model families. The equivalent widths are measured against the total continuum, which includes any older stellar population, and are therefore lower limits on those of the young population alone.
The top row mainly isolates the hardness of the ionizing spectrum through \heii$\lambda4686$/\hb, EW(\heii$\lambda4686$), and \heii$\lambda4686$/He\,\textsc{i}$\lambda5876$ against $R_3$; these ratios depend on the ratio of He$^{+}$- to H-ionizing photons and only weakly on $U$ and $\Zgas$ within each family. The bottom row shows the UV plane of \civ and O\,\textsc{iii}]$\lambda1663$ relative to \heii$\lambda1640$ proposed by \cite{Rusta2025}, which separates stellar spectra from those of massive black holes and limits the carbon and oxygen content of the gas, and the two planes based on \heii$\lambda1640$ and H$\gamma$ adopted in the analysis of Hebe, the \heii emitter near GN-z11 \citep{Maiolino2026,Ubler2026,Rusta2026}. 
We note here that since \ciii$\lambda\lambda1907,1909$ falls in the G140M detector gap, the UV plane is built on O\,\textsc{iii}]$\lambda1663$ and \civ instead of the \ciii-based criteria of \citet{Rusta2025}. The X-shooter limit, \ciii/\heii$\lambda1640<1.7$, is too shallow to constrain the models beyond the other diagnostics (Fig.~\ref{fig:model_comparison_extra}g).

The remaining planes, $R_3$ against O32, [N\,\textsc{ii}]/\ha, and \neiii$\lambda3869$/\oii, those of the undetected [Ne\,\textsc{iv}]$\lambda2424$ and [Ne\,\textsc{v}]$\lambda3426$, \neiii/H$\gamma$ against \heii$\lambda1640$/H$\gamma$, and the \ciii plane, are shown in Fig.~\ref{fig:model_comparison_extra}; Figs.~\ref{fig:model_grids_appendix_main} and \ref{fig:model_grids_appendix_extra} show every model of every family in all the planes.

\paragraph{Hardness of the ionizing spectrum.}
AMORE6-B has \heii$\lambda4686$/\hb$=0.080\pm0.015$ at $R_3=0.25$ (Fig.~\ref{fig:model_comparison}a). This value lies within the region of the self-polluted NEFERTITI galaxies, whose 16th--84th percentiles span 0.022--0.13, and in the upper part of that of the Pop~III-rich hybrids. Among the population-synthesis models it is reached only by metal-free stars with a top-heavy IMF at ages below $\simeq2$\,Myr (0.08 at 1\,Myr for a Salpeter IMF over 50--500\,$M_\odot$). It is 0.4\,dex above the very-massive-star population at $\Zstar=0$, 1.4\,dex above the very massive stars at $\Zstar=10^{-5}$, and 1.7\,dex above the BPASS populations at any $U$ and gas metallicity; the Pop~III.2 and Kroupa-IMF populations give 0.03 and 0.02. Leakage of ionizing photons raises \heii$\lambda4686$/\hb of the BPASS p
opulation by only $\simeq0.3$\,dex at an escape fraction of 0.8; only matter-bounded clouds that leak almost all their ionizing photons reach the observed ratio, at the price of an ionizing-photon budget that the continuum excludes (Section~\ref{sec:leakage}). In a radiation-bounded nebula, a BPASS population reaches the observed ratio only with an additional hard source: X-ray binaries that supply 30\% of the hydrogen-ionizing photons give \heii$\lambda4686$/\hb$\,=0.06$--0.11 at $R_3\simeq0.25$, for both disc temperatures. The ratio \heii$\lambda4686$/He\,\textsc{i}$\lambda5876=1.5\pm0.5$ (Fig.~\ref{fig:model_comparison}c) leads similar conclusions though with a lower significance because He\,\textsc{i}$\lambda5876$ is detected at $4\sigma$. The equivalent width of \heii$\lambda4686$, $35\pm8$\,\AA\ (Fig.~\ref{fig:model_comparison}b), is within the range of the self-polluted galaxies (10--78\,\AA) and at the lower end of those of the very-massive-star models at $\Zstar=0$ (35--60\,\AA), while the BPASS and massive-black-hole models give less than 4 and 10\,\AA, respectively; with X-ray binaries supplying 30\% of the hydrogen-ionizing photons, the BPASS population reaches 20--50\,\AA\ at the observed $R_3$. 

\paragraph{Metal lines.}
With $R_3=0.25$ and O32$\,>2.9$ (Fig.~\ref{fig:model_comparison_extra}a), the very-massive-star models at $\Zstar=0$ require $\log(\Zgas/Z_\odot)$ between $-3.0$ (at $\log U=-0.5$) and $-2.0$ (at $\log U=-3$), i.e.\ $12+\log(\mathrm{O/H})\simeq5.7$--6.7. \civ is detected, with \civ/\heii$\lambda1640=0.46\pm0.06$, and O\,\textsc{iii}]$\lambda1663$/\heii$\lambda1640<0.17$ (Fig.~\ref{fig:model_comparison}d). These values lie at the low end of the NEFERTITI galaxies: \civ/\heii$\lambda1640\leq0.58$ together with the O\,\textsc{iii}] limit is met by 28\% of the self-polluted galaxies and 14\% of the Pop~III-rich hybrids, against 6\% of the Pop~III-mid hybrids and none of the Pop~III-poor ones, which predict \civ/\heii$\gtrsim1$. For the Pop~III.1 and very-massive-star populations at the C/O plateau of the grid, the observed \civ/\heii and $R_3$ are reproduced together only at $\log(\Zgas/Z_\odot)\simeq-3$ and $\log U\simeq-1$ to $-0.5$; at the lower ionization parameter preferred by the optical lines, \civ requires a higher C/O (Sections~\ref{sec:co} and \ref{sec:sed}). The radiation-bounded BPASS models reproduce neither ratio, because of their weak \heii. 
[N\,\textsc{ii}]/\ha$<0.019$ (Fig.~\ref{fig:model_comparison_extra}b) is above the predictions of all the stellar families and excludes part of the AGN grid of \citet{Feltre2016} (Fig.~\ref{fig:model_grids_appendix_extra}). \neiii$\lambda3869$/H$\gamma<0.13$ (Fig.~\ref{fig:model_comparison_extra}f) is met by 78\% of the self-polluted galaxies and by the very-massive-star models at $\log(\Zgas/Z_\odot)\leq-2.5$, or at $-2$ for $\log U\leq-2$, and excludes most of the star-forming grid of \citet{Gutkin2016} and the BPASS models at $\log(\Zgas/Z_\odot)\geq-1.25$; it is about three times deeper than the limit for Hebe.

\paragraph{Equivalent widths of H$\gamma$ and \heii$\lambda1640$.}
The H$\gamma$ planes separate stellar and massive black holes (Fig.~\ref{fig:model_comparison}f). Without a host stellar population, the massive-black-hole models have EW(H$\gamma)\lesssim70$\,\AA, against $171^{+29}_{-25}$\,\AA\ in AMORE6-B, while the stellar families predict 250--300\,\AA\ at $n_{\rm H}=10^{3}$\,cm$^{-3}$ and the BPASS population with X-ray binaries 130--200\,\AA. The massive-black-hole models reach the observed EW(\heii$\lambda1640$) without a host population and with the hardest continuum (Fig.~\ref{fig:model_comparison}e), but then underpredict EW(\heii$\lambda4686$) and EW(H$\gamma$) by factors of 3--5: their continuum is much flatter between 1640 and 4686\,\AA\ than that of AMORE6-B, whose equivalent widths require $f_\lambda(1640)/f_\lambda(4686)\simeq24$. A host population would lower all the equivalent widths further. The NEFERTITI models at $n_{\rm H}=10^{6}$\,cm$^{-3}$ \citep{Rusta2026} predict a \heii/H$\gamma$ similar to that of the low-density ones, but equivalent widths of H$\gamma$ and \heii$\lambda1640$ higher by a factor of 2--7, since at this density two-photon emission is collisionally suppressed and the nebular continuum is weaker. AMORE6-B lies on the low-density locus in both equivalent-width planes (Fig.~\ref{fig:model_comparison}e,f). \heii$\lambda1640$/H$\gamma=2.0^{+0.2}_{-0.1}$ is twice that of Hebe. 

In summary, at a first glance the diagnostic diagrams favour metal-free stellar populations with a top-heavy IMF or in the self-polluted phase of the NEFERTITI galaxies, and metal-poor populations with a hard X-ray component that supplies a significant fraction of the ionizing photons; very massive stars at $\Zstar=0$ and the less top-heavy Pop~III IMFs fall short of the \heii$\lambda4686$/\hb ratio by 0.4--0.7\,dex. Binary populations at $\Zstar\geq10^{-5}$ alone are excluded by all the \heii diagnostics, and massive black holes by the equivalent widths of the Balmer lines.

\subsection{Ranking the ionizing sources with the emission lines}
\label{sec:inference}

Although very powerful, classical diagnostic diagrams such as those of Fig.~\ref{fig:model_comparison} compare only two line ratios at a time. To leverage all the information delivered by JWST/NIRSpec, we here assess, for each family of Table~\ref{tab:sed_families}, how well its models reproduce the whole emission-line spectrum. 
For every model we compare the predicted fluxes of all the lines of Table~\ref{tab:line_fluxes}, relative to \hb, with the measured ones, together with the equivalent widths of \hb, \heii$\lambda4686$, and \heii$\lambda1640$. The lines that are not detected enter, like the detected ones, with their measured fluxes and uncertainties rather than as upper limits, so that the models that predict them brighter than the data allow are disfavoured. Three further uncertainties are added to those of the measurements: the attenuation, which is varied within the posterior of Section~\ref{sec:attenuation}; the flux scale of G140M relative to the other gratings, which is varied by 15\%; and an uncertainty of 0.1\,dex on every model prediction, which accounts for the simplifications of the photoionization models. The likelihood of each model is then $\mathcal{L}\propto\exp(-\chi^{2}/2)$, where $\chi^{2}$ sums over all the lines and equivalent widths the squared differences between measured and predicted values, divided by the combined uncertainties, with the predicted fluxes scaled to the \hb normalisation that best matches the data.

Averaging the likelihood over all the models of a family gives the evidence of that family, $\mathcal{Z}$. A family scores well if a large part of its models reproduce the data, and poorly if none do or if only a narrow, finely tuned subset does. The difference $\Delta\ln\mathcal{Z}$ between two families measures how strongly the data prefer one over the other: following \citet{Trotta2008}, $|\Delta\ln\mathcal{Z}|<1$ is inconclusive, and differences larger than 1, 2.5, and 5 indicate a weak, moderate, and strong preference. All the families of Table~\ref{tab:sed_families} enter the comparison; for the grids computed here we use $n_{\rm H}=10^{3}$\,cm$^{-3}$ and $-3.5\leq\log(\Zgas/Z_\odot)\leq-2$, the range common to all of them, and $-3.5\leq\log U\leq0$ for every family, massive black holes included, except the soft X-ray source, whose models do not converge above $\log U=-2.5$. The evidence depends on these ranges, and the families are neither exhaustive nor of known prior probability: we use $\Delta\ln\mathcal{Z}$ to rank the families, not to assign them probabilities. The technical details are given in Appendix~\ref{app:model_grids}.

\begin{figure}
  \centering
  \includegraphics[width=\columnwidth]{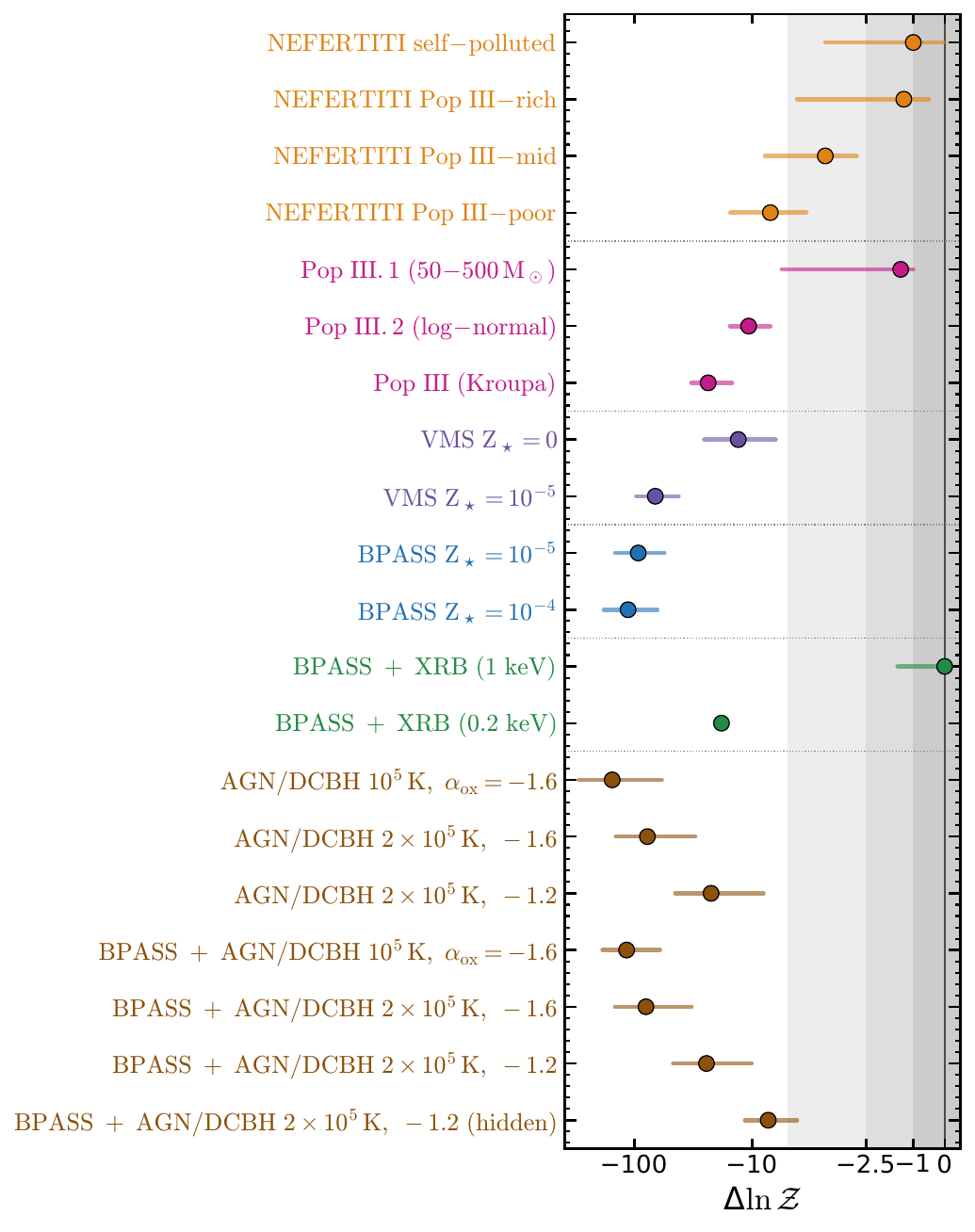}
  \caption{Ranking of the ionizing sources by the emission lines of AMORE6-B alone: evidence $\mathcal{Z}$ of each family of Table~\ref{tab:sed_families} relative to the best one, as computed from all the emission lines fluxes together with the equivalent widths of \hb, \heii$\lambda4686$, and \heii$\lambda1640$ (Section~\ref{sec:inference}). Circles assume an uncertainty of 0.1\,dex on the model predictions; the bars span the values for 0.05 and 0.2\,dex. The grey bands mark the inconclusive ($|\Delta\ln\mathcal{Z}|<1$), weak (1--2.5), and moderate (2.5--5) ranges of the scale of \citet{Trotta2008}; beyond 5 the preference is strong. The horizontal axis is linear between $-5$ and 0 and logarithmic beyond.}
  \label{fig:line_evidence}
\end{figure}

The ranking is shown in Fig.~\ref{fig:line_evidence}. Four families are statistically equivalent ($|\Delta\ln\mathcal{Z}|\leq1.4$): the self-polluted and the Pop~III-rich NEFERTITI galaxies, the Pop~III.1 population, and a BPASS population with X-ray binaries of $kT_{\rm in}=1$\,keV that supply $\simeq20\%$ of the hydrogen-ionizing photons. The best models of these families share a young age ($\simeq1$\,Myr), a high ionization parameter ($\log U\simeq-1$ to 0), and $\log(\Zgas/Z_\odot)$ between $-3$ and $-2.5$. The Pop~III-mid hybrids are disfavoured moderately ($\Delta\ln\mathcal{Z}=-3.8$). All the other families are disfavoured strongly: the Pop~III-poor hybrids ($-7.0$), the Pop~III.2 population ($-10.7$), the very massive stars at $\Zstar=0$ ($-13.1$), the BPASS population with the soft X-ray source ($-18$, although these models are limited to $\log U\leq-2.5$), the Pop~III population with a Kroupa IMF ($-23.5$), and the massive black holes (AGN or DCBH), alone or with a host population ($\leq-22$, or $-7.2$ with a host population and the disc hidden from view). Metal-poor stellar populations without a hard additional source, BPASS or very massive stars at $\Zstar\geq10^{-5}$, are excluded ($\Delta\ln\mathcal{Z}<-60$). The model uncertainty changes the numbers but not the grouping: for 0.05\,dex the X-ray binary combination is preferred over the other three leading families by 4--6, and for 0.2\,dex the self-polluted galaxies rank first, with all four within 1.6. Using only the lines of Fig.~\ref{fig:model_comparison}, the order of the families is unchanged, and $\Delta\ln\mathcal{Z}$ changes by at most 1.5, except for the soft X-ray source ($+3.9$). The emission lines alone therefore cannot separate metal-free stars from a metal-poor population with a hard X-ray source of high luminosity; further information come from the full-SED fitting (Section~\ref{sec:sed}) and from an estimate of the X-ray luminosity that such a source would require (Section~\ref{sec:alternatives}).

\section{Full spectral fitting}
\label{sec:sed}

We model the spectral energy distribution of AMORE6-B with \textsc{bagpipes} \citep{Carnall2018}, fitting the total photometry of Table~\ref{tab:photometry} in the 18 \NIRCam bands from F115W to F480M (F070W and F090W are blueward of, or contain, \lya, and are excluded) together with the fluxes of fourteen UV and optical lines, from \civ to \ha, including the non-detections of O\,\textsc{iii}]$\lambda1663$, \oii, \neiii, [N\,\textsc{ii}], and \sii, which enter at zero flux with the uncertainty of the measurement. The line fluxes carry a calibration uncertainty of 10\%, raised to 20\% for the G140M lines (\civ, \heii$\lambda1640$, O\,\textsc{iii}]$\lambda1663$), whose flux scale relative to the other gratings cannot be measured, and a free factor accounts for the relative scale of spectroscopy and photometry. Since the photoionization grid assumes the C/O plateau of \citet{Nicholls2017}, a free factor (0.1--30) multiplies the \civ flux; at these metallicities carbon contributes little to the cooling, so we assume that \civ scales almost linearly with C/O. The nebular continuum and lines of each population are computed with the setup of Section~\ref{sec:grid} for stellar ages up to 50\,Myr, $-3.5\leq\log(\Zgas/Z_\odot)\leq-2$, and $-3.5\leq\log U\leq0$; $\Zgas$ and $U$ are free and independent of the stellar metallicity, and the attenuation follows the SMC curve of \citet{Gordon2003} with $0\leq A_V\leq1$. A fraction $f_{\rm esc}$ of the ionizing photons, free between 0 and 0.95, escapes without being absorbed, which scales all the nebular emission by $1-f_{\rm esc}$. Posteriors and evidences are computed with \textsc{nautilus} \citep{Lange2023}, and masses are corrected for $\mu=78$ (Section~\ref{sec:lensmodel}).

We fit the seven scenarios reported in Table~\ref{tab:sed_scenarios}; the fiducial fit is shown in Fig.~\ref{fig:sed}, and the other six in Appendix~\ref{app:sed_scenarios}. The first five are those that the emission lines alone do not exclude (Section~\ref{sec:inference}); the last two, for comparison, belong to the massive-black-hole family: its most favourable configuration (AGN or DCBH; $T_{\rm bb}=2\times10^{5}$\,K, $\alpha_{\rm ox}=-1.2$) added to the BPASS population, with the disc visible or hidden (Section~\ref{sec:alternatives}). 

In three of the first five scenarios the \heii is produced by metal-free stars with a Salpeter IMF over 50--500\,$M_\odot$ (Pop~III.1), the only stellar population that reaches the observed \heii$\lambda4686$/\hb: these are the star-formation history of a NEFERTITI galaxy, a single Pop~III.1 burst coupled with an older Pop~III.2 population of free age and mass, and a single Pop~III.1 burst coupled with a hard X-ray source. In the other two, the hard X-ray source produces the \heii, and a Pop~III.2 or a BPASS $\Zstar=10^{-4}$ population produces the continuum and most of the Balmer emission. The X-ray source has the spectrum of the X-ray binaries of Section~\ref{sec:seds}, a multicolour disc with $kT_{\rm in}=1$\,keV, and a free luminosity and it is meant to reproduce the X-ray binaries or ultraluminous X-ray sources formed by the stellar population that accompanies it.

\begin{table*}
  \caption{SED fits of AMORE6-B.\label{tab:sed_scenarios}}
  \centering
  \footnotesize
  \setlength{\tabcolsep}{2.5pt}
  \begin{tabular}{lccccccl}
    \toprule
    Scenario & $\Delta\ln\mathcal{Z}$ & $f_{\rm esc}$ & $\log U$ & $\log(\Zgas/Z_\odot)$ & $\log(\mathrm{C/O})$ & $\log M/M_\odot$ & \heii source \\
    \midrule
    \textcolor{famNEF}{Pop~III.1, NEFERTITI SFH (fiducial)} & 0 & $0.86\pm0.02$ & $-2.28^{+0.28}_{-0.22}$ & $-2.35^{+0.11}_{-0.12}$ & $+0.2\pm0.3$ & $5.13^{+0.05}_{-0.06}$ & Pop~III.1 (100\%) \\
    \textcolor{famPopIII}{Pop~III.1 burst + older Pop~III.2}$^{a}$ & $+0.4$ & $0.90^{+0.01}_{-0.02}$ & $-2.28^{+0.31}_{-0.22}$ & $-2.34^{+0.10}_{-0.11}$ & $+0.1\pm0.3$ & $5.20^{+0.06}_{-0.07}$ & Pop~III.1 (100\%) \\
    \textcolor{famPopIII}{Pop~III.1 burst + HXS} & $0.0$ & $0.90\pm0.01$ & $-2.26^{+0.34}_{-0.25}$ & $-2.34^{+0.10}_{-0.11}$ & $+0.1\pm0.3$ & $5.19\pm0.06$ & Pop~III.1 ($\geq99\%$) \\
    \textcolor{famXRB}{Pop~III.2 + HXS} & $-3.0$ & $0.37^{+0.18}_{-0.21}$ & $-0.39^{+0.21}_{-0.17}$ & $-2.56^{+0.13}_{-0.11}$ & $-0.7^{+0.1}_{-0.2}$ & $6.36^{+0.13}_{-0.17}$ & HXS (99\%) \\
    \textcolor{famXRB}{BPASS + XRB} & $-6.9$ & $0.67^{+0.06}_{-0.07}$ & $-0.33^{+0.24}_{-0.36}$ & $-2.68^{+0.13}_{-0.10}$ & $-0.7^{+0.1}_{-0.2}$ & $5.46^{+0.09}_{-0.10}$ & XRB (97\%) \\
    \textcolor{famAGN}{BPASS + AGN/DCBH, disc visible} & $-22.8$ & $0.61^{+0.05}_{-0.07}$ & $-0.22^{+0.15}_{-0.23}$ & $-2.80^{+0.09}_{-0.08}$ & $-0.7\pm0.2$ & $5.55^{+0.07}_{-0.06}$ & BPASS (99\%) \\
    \textcolor{famAGN}{BPASS + AGN/DCBH, disc hidden} & $-3.3$ & $0.92^{+0.02}_{-0.04}$ & $-2.07^{+0.28}_{-0.25}$ & $-2.36^{+0.09}_{-0.10}$ & $+0.1^{+0.2}_{-0.3}$ & $5.51\pm0.06$ & AGN/DCBH (99.8\%) \\
    \bottomrule
  \end{tabular}
  \begin{flushleft}\footnotesize Medians and 16th--84th percentiles of the fits with \civ, O\,\textsc{iii}]$\lambda1663$, and a free C/O. $\Delta\ln\mathcal{Z}$ is relative to the fiducial model. $M$ is the intrinsic mass formed by the young (or only) stellar population; $^{a}$ Older population: $\log M/M_\odot<5.5$ (84th percentile). HXS: hard X-ray source; XRB: X-ray binaries; both are a multicolour disc with $kT_{\rm in}=1$\,keV. AGN/DCBH: the continuum of \citet{NakajimaMaiolino2022} with $T_{\rm bb}=2\times10^{5}$\,K and $\alpha_{\rm ox}=-1.2$; with the disc visible the fit switches it off. BPASS has $\Zstar=10^{-4}$. The last column gives the source of \heii$\lambda4686$ and its share. Scenario names are coloured by the source of \heii as the families of Fig.~\ref{fig:line_evidence}.\end{flushleft}
\end{table*}

The star formation history of one of the Pop~III galaxies in NEFERTITI \citep{Koutsouridou2026} at $z\simeq6$ (Pop~III star formation over the last 4\,Myr, four times faster in the last 2\,Myr than before; only its normalisation is free) reproduces the photometry and the lines with $\chi^{2}=21$ for the 18 bands, and $\chi^{2}=19$ for the fourteen lines at maximum likelihood (Fig.~\ref{fig:sed}). This is our fiducial model. 
All the stellar mass, $1.3\pm0.2\times10^{5}\,M_\odot$, formed in the last 4\,Myr, 80\% of it in the last 2\,Myr, and no older population is required. These stars produce all the \heii and Balmer emission and the whole continuum. The gas has $\log(\Zgas/Z_\odot)=-2.35\pm0.12$, i.e.\ $12+\log(\mathrm{O/H})=6.34\pm0.12$, and $\log U=-2.3\pm0.25$, consistent with the abundance derived from $R_3$ in Section~\ref{sec:empirical_oh}, and the attenuation, $A_V=0.03^{+0.04}_{-0.02}$\,mag, agrees with that of the hydrogen and helium lines (Section~\ref{sec:attenuation}). The fit finds a very large escape fraction, $f_{\rm esc}=0.86\pm0.02$, as primarily driven by the continuum: in a radiation-bounded nebula the nebular continuum of these stars would dominate the rest-frame UV and make it redder than the observed $\beta_{\rm UV}\simeq-3$. 
Two helium lines remain slightly discrepant: He\,\textsc{i}$\lambda5876$/\hb is over-predicted (0.10 against $0.055\pm0.016$, $2.8\sigma$) and \heii$\lambda1640$ under-predicted ($2.3\sigma$), while \heii$\lambda4686$/\hb, 0.056 in the model, is $1.6\sigma$ below the observed value. \civ is reproduced, and O\,\textsc{iii}]$\lambda1663$ is below its limit, only with a C/O above the plateau: $\log(\mathrm{C/O})=+0.2\pm0.3$, about 1\,dex above it and also above the solar value. The ionization parameter, which is set by the optical lines, does not change; at the plateau C/O, \civ would require $\log U\simeq-1.7$. The direct estimate of Section~\ref{sec:co}, $\log(\mathrm{C/O})=-0.30^{+0.23}_{-0.18}$, is consistent with this value within $1.5\sigma$: both place C/O above the plateau, and the direct estimate, which assumes the ionization correction of models at $\log U\simeq-2$ to $-1.5$, would be higher with that of the fit.

The single Pop~III.1 burst is as likely as the fiducial model: a burst of $1.6\times10^{5}\,M_\odot$, younger than 1.1\,Myr, with $f_{\rm esc}=0.90$ and the same gas properties. An older population is not required. Since Pop~III.1 stars older than 3.6\,Myr have all died, an earlier generation can contribute to the light only with a less top-heavy IMF, and we model it as a Pop~III.2 burst of free age and mass: its mass is $5\times10^{3}\,M_\odot$, with a 84th percentile of $3\times10^{5}\,M_\odot$. Adding a hard X-ray source to the young burst leaves the fit unchanged: the source produces at most 1\% of the \heii (84th percentile), for an intrinsic $L_X(0.5$--$8\,{\rm keV})<4\times10^{41}$\,erg\,s$^{-1}$.

The \heii can also be produced without young metal-free stars, by the X-ray binaries or ultraluminous sources of an older population. In this scenario, $2\times10^{6}\,M_\odot$ of Pop~III.2 stars formed 15\,Myr ago produce the continuum and 73\% of the Balmer emission, and an X-ray source of $L_X=1.8\times10^{42}$\,erg\,s$^{-1}$ the \heii, in gas at $\log U\simeq-0.4$, where \civ is reproduced with the plateau C/O; it is moderately disfavoured ($\Delta\ln\mathcal{Z}=-3.0$). The same configuration with a BPASS $\Zstar=10^{-4}$ population is disfavoured ($\Delta\ln\mathcal{Z}=-6.9$). Finally, the massive black hole added to the BPASS population is strongly disfavoured with its disc visible ($-23$), whose continuum would dilute the equivalent widths, and moderately with its disc hidden ($-3.3$); the latter requires an obscured source of $\simeq2\times10^{43}$\,erg\,s$^{-1}$, and the emission lines alone disfavour it more strongly ($\Delta\ln\mathcal{Z}\simeq-7$; Section~\ref{sec:inference}). We discuss these alternatives in Section~\ref{sec:alternatives}.

\begin{figure*}
  \centering
  \includegraphics[width=\textwidth]{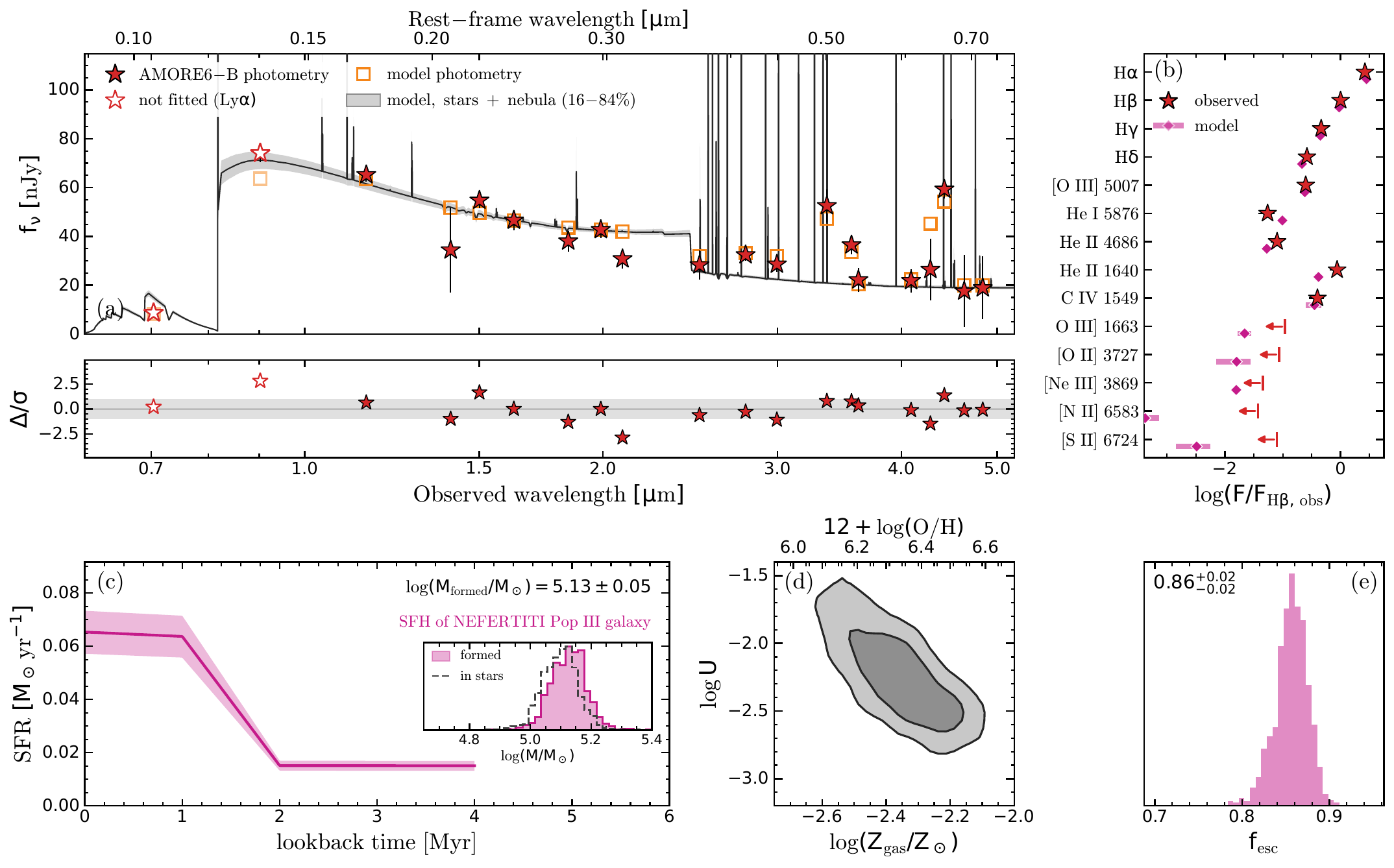}
  \caption{Fiducial model of AMORE6-B: Pop~III.1 stars (Salpeter IMF over 50--500\,$M_\odot$) formed with the star-formation history of a NEFERTITI galaxy forming Pop~III stars at $z\approx 6$ \citep{Koutsouridou2026}, fitted to the photometry and the emission lines with a free escape fraction of ionizing photons. \textit{(a)} Observed photometry (stars; F070W and F090W, which contain \lya, are not fitted), posterior median (black line) and 16--84th percentile range (grey band) of the model spectrum, and model photometry (squares); the lower panel shows the residuals in units of the photometric uncertainty. \textit{(b)} Line fluxes relative to the observed \hb: measurements (stars), $3\sigma$ upper limits (arrows), and the model (diamonds and 16--84th percentile bars). \textit{(c)} Star-formation history, with the median and 16--84th percentile range of its normalisation; the inset shows the posterior of the mass formed and of the mass still in stars. \textit{(d)} Gas metallicity and ionization parameter (68 and 95\% contours). \textit{(e)} Escape fraction. Masses and star-formation rates are corrected for a magnification $\mu=78$.}
  \label{fig:sed}
\end{figure*}

\section{Discussion}
\label{sec:discussion}

\subsection{AMORE6 as a self-polluted Population~III system}
\label{sec:selfpolluted}

Based on the multiple lines of evidence emerging from the analysis presented in the previous sections, we suggest that the properties of AMORE6 are best reproduced by a self-polluted PopIII-galaxy scenario: a population of massive and young ($1-4$~Myr old) metal-free stars is ionizing the gas recently self-enriched by the very-first generation of supernovae evolving in a few, $\approx (2-4)$~Myrs. %
First, the diagnostic diagrams and the likelihood test based on all the emission lines (Sections~\ref{sec:ionizing_source} and \ref{sec:inference}) favour metal-free stars with a top-heavy IMF, as in the self-polluted and Pop~III-rich phases of the NEFERTITI galaxies \citep{Rusta2025}, leaving open only a hard X-ray or accretion source added to a metal-poor population as a viable alternative. 
Then, the SED fitting (Section~\ref{sec:sed}) assigns the \heii to Pop~III.1 stars, unless a hard X-ray source is accompanied by a very massive older Pop~III.2 population ($2\times10^{6}\,M_\odot$, several times the largest Pop~III starbursts in simulations of this epoch; \citealt{Jeong2026}), and disfavours X-ray binaries ($\Delta\ln\mathcal{Z}=-6.9$), which would also require stars too young to host compact objects and an X-ray luminosity above the \textit{Chandra} limit (Section~\ref{sec:alternatives}). 
However, this scenario must also account for the metallicity of the gas: the abundance we measure in fact has to be reconciled with the metals that the stars can have produced.

\paragraph{Metal budget.}
\label{sec:enrichment}
Whether the metals in the gas can originate from the stars we observe depends on the timescale of their evolution. In our fiducial star-formation history, the stars formed 2--4\,Myr ago include very massive stars with progenitor masses of 140--260\,$M_\odot$, which end their lives as energetic pair-instability supernovae (PISNe, $E\approx10^{52}$--$10^{53}$\,erg). These stars explode after $\simeq2$--2.5\,Myr, returning a substantial fraction of their mass as metals \citep{HegerWoosley2002}. Given the Pop~III.1 IMF, which contains no stars below 50\,$M_\odot$, core-collapse supernovae (CCSNe) from 8--40\,$M_\odot$ progenitors \citep{Heger2010} do not contribute. Instead, stars in the 40--140\,$M_\odot$ and above 260\,$M_\odot$ ranges are expected to collapse directly into black holes. Adopting the lifetimes of metal-free stellar models from \citet{Schaerer2002}, we estimate that the $\approx30$ PISN progenitors that have already exploded have released $\approx3\times 10^{3}\,M_\odot$ of metals. By comparison, the ionized gas, whose mass is inferred from the H$\beta$ luminosity assuming $n_{\rm e}=10^{2}$--$10^{3}$\,cm$^{-3}$, contains only 6--60\,$M_\odot$ of metals at the observed metallicity.

Assuming a closed-box model, with no SN-driven outflows, we can estimate the gas mass over which the newly produced metals would need to be diluted to reproduce the observed abundance:
$M_{\rm dil}=M_Z/Z_{\rm gas}\simeq2\times10^{7}\,M_\odot$.
This is approximately 150 times the stellar mass. The inferred dilution mass lies between the mass of the ionized gas estimated above and the dynamical mass. The latter is $M_{\rm dyn}=5\sigma_v^{2}R/G\simeq1$--$3\times10^{8}\,M_\odot$, for $\sigma_v=55.6$\,km\,s$^{-1}$ and half-light radii of $\simeq35$\,pc for the stellar component (Table~\ref{tab:amore6_literature}) and $\simeq80$\,pc for the Balmer emission (Section~\ref{sec:morphology}, scaled by $\mu^{-1/2}$). This dynamical mass also includes the contribution of dark matter.

The observed stellar population has therefore produced 50--500 times more metals than are currently contained in the ionized gas. Our data do not allow us to establish the fate of the remaining metals. However, PISNe are up to two orders of magnitude more energetic than typical CCSNe and are expected to drive powerful outflows. Simulations have shown that even a single PISN exploding in a high-redshift minihalo (with a dark matter mass of $\approx10^{6}\,M_\odot$) can completely evacuate its gas \citep[e.g.,][]{Bromm2003}. Although the estimated dark matter mass of AMORE6 is two orders of magnitude higher than that of such minihaloes, and is consistent with expectations for pristine star-forming galaxies after the end of reionization, a substantial fraction of the gas and metals may have been expelled by PISN-driven outflows or may reside in a hot, diffuse phase. Neither component contributes significantly to the nebular emission lines used to measure O/H. The observed metallicity may therefore trace only the small reservoir of cool, neutral gas that remains in the galaxy.

Finally, the combination of metal-enriched PISN-driven outflows and the relatively massive dark matter halo hosting AMORE6 may lead to incomplete mixing of the gas, allowing some regions to remain pristine. This would resolve the apparent tension with the critical metallicity for the transition to Pop~II star formation ($10^{-6}$--$10^{-3.5}\,Z_\odot$; \citealt{KlessenGlover2023}): the youngest stars, which power the \heii\ emission, could have formed from pristine gas, while the gas they subsequently ionize had already been enriched by the first PISNe.

The C/O ratio provides a further, albeit more tentative, piece of evidence. The values above the plateau, indicated by both the direct estimate and the fiducial fit (Sections~\ref{sec:co} and \ref{sec:sed}), fall in the region occupied by NEFERTITI galaxies enriched by Pop~III stars (Fig.~\ref{fig:co}), and are therefore consistent with this scenario. However, the present data do not allow us to distinguish between the progenitor masses and explosion energies of the supernovae responsible for the carbon enrichment \citep[see, e.g.,][]{Koutsouridou2023, Vanni2023}. In the alternative scenario involving a hard X-ray source and an older Pop~III.2 population, with a higher ionization parameter, C/O remains consistent with the plateau and therefore provides no discriminatory power.

For the single burst scenario, the young stars have not yet released any metals, which must then come from an earlier generation. The older Pop~III.2 population of the fit releases $5\times10^{2}\,M_\odot$ of metals (16th--84th percentiles $9$--$1.8\times10^{4}\,M_\odot$), i.e.\ $M_{\rm dil}\sim7\times10^{6}\,M_\odot$, enough for the ionized gas, but are poorly constrained. This scenario also requires incomplete mixing as well, now over the $\simeq40$\,Myr between the two bursts, during which the gas that fuels the more recent burst must remain pristine. 

\subsection{Constraints on the initial mass function of Population~III stars}
\label{sec:imf}

The mass distribution of the first stars is one of the main unknowns of early galaxy formation. Because metal-free gas cools inefficiently, simulations of primordial star formation predict typical masses of tens to hundreds of solar masses \citep{Bromm2002,BrommLarson2004,Bromm2013}, but the fragmentation of protostellar discs and the radiative feedback of the accreting protostars spread the final masses over more than two orders of magnitude \citep{StacyBromm2013,Hirano2014,KlessenGlover2023}. So far, observational constraints have come only from near field cosmology. The absence of surviving metal-free stars in the Galactic halo and in ultra-faint dwarf galaxies sets the lower mass limit of the Pop~III IMF, $>\,0.8 M_\odot$ \citep{Hartwig2015,Rossi2021}. 
The abundance patterns of extremely metal-poor stars and the fraction of carbon-enhanced stars observed, suggest a most-likely characteristic mass for the Pop~III IMF of $\approx 10-20$ \citep{Ishigaki2018,Koutsouridou2023}.
Finally, the lack of stars that carry the chemical signature of pair-instability supernovae excludes the flattest, most top-heavy IMFs \citep{Koutsouridou2024}. \citet{Rusta2026} showed that Hebe, a pristine \heii emitter at very high redshift, adds an independent constraint: in fact, the lower limit on \heii$\lambda1640$/H$\gamma$ excludes the least top-heavy IMFs, complementing the near-field limits.

We here apply the same approach to AMORE6-B. As in \citet{Rusta2026}, we describe the Pop~III IMF with the form of \citet{Larson1998}, $dN/dm\propto m^{-x}\exp(-M_{\rm ch}/m)$ between 0.8 and 1000\,$M_\odot$, where $x$ is the slope at high masses ($x=2.35$ for a Salpeter IMF) and $M_{\rm ch}$ the characteristic mass. We then compute the nebular emission of a single, fully sampled burst at ages of 1 and 2\,Myr, with the same photoionization setup as in \citet{Rusta2026}, exploring the PopIII IMF by varying $M_{ch}$ and $x$. 
The ratio of \heii to Balmer emission depends on the IMF, through the fraction of the mass in the hottest stars, but not on the mass of the burst, and therefore constrains $x$ and $M_{\rm ch}$ jointly. For AMORE6-B the relevant ratio is that of the burst alone. In the fiducial model of Section~\ref{sec:sed} the Pop~III.1 stars produce all of the \heii and Balmer emission, and the escape of ionizing photons lowers both by the same factor, so that the ratio of the stars is the observed one, $0.080\pm0.015$. Because both \heii and the Balmer lines are detected, the ratio is a measurement rather than a limit. The IMFs that produce too much \heii are therefore excluded, as well as those that produce too little.

\begin{figure}
  \centering
  \includegraphics[width=\columnwidth]{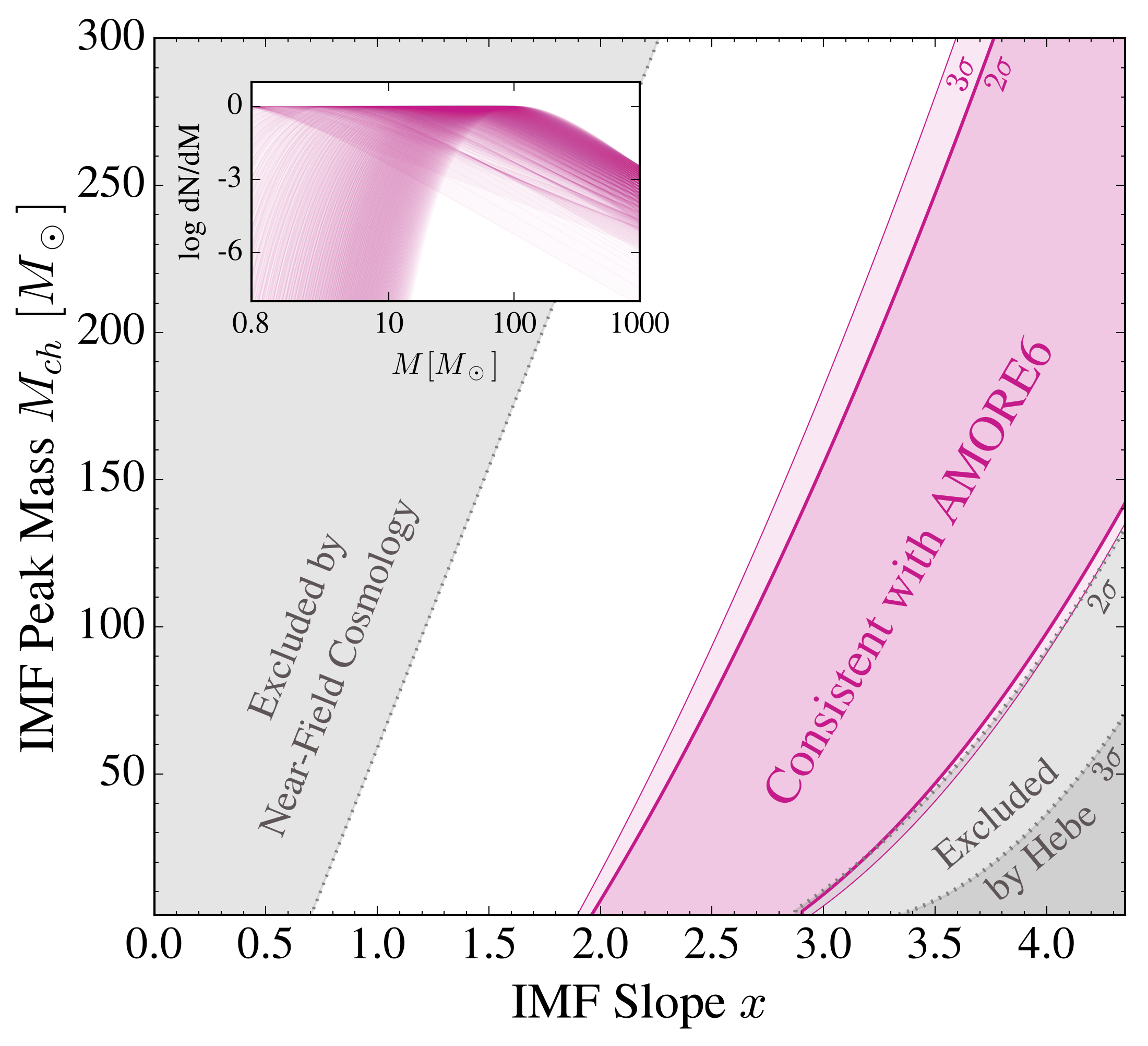}
  \caption{Constraints on the IMF of Pop~III stars, described as $dN/dm\propto m^{-x}\exp(-M_{\rm ch}/m)$ \citep{Larson1998} between 0.8 and 1000\,$M_\odot$, in the plane of slope $x$ and characteristic mass $M_{\rm ch}$, following \citet{Rusta2026}. The pink band encloses the IMFs whose \heii-to-Balmer ratio is consistent with that of AMORE6-B; grey regions are excluded by near-field cosmology \citep{Koutsouridou2024} and by the lower limit on \heii$\lambda1640$/H$\gamma$ in the C1 component of Hebe \citep{Rusta2026}. Both AMORE6-B and Hebe constraints are shown for 2 and 3$\sigma$ confidence levels, with the ranges spanning stellar ages of 1-2 Myr. The top inset shows the Pop~III IMFs consistent with the observed line ratios of AMORE6-B.}
  \label{fig:imf}
\end{figure}

Fig.~\ref{fig:imf} shows the result.  
In the diagram linking the characteristic mass to the slope, AMORE6-B is consistent with the region of approximately $90x-268\lesssim M_{\rm ch}/M_\odot\lesssim172x-330$ ($3\sigma$). For a Salpeter slope, the band requires $M_{\rm ch}\lesssim70\,M_\odot$ ($3\sigma$); flatter IMFs are allowed only with lower characteristic masses, and steeper IMFs only with higher ones. 
The band lies within the region allowed by the near-field constraints and by Hebe at $>2\sigma$, while being narrower than either: AMORE6 alone excludes both the least top-heavy IMFs and the most top-heavy ones, which the near field excludes only for the flattest slopes. 

Differential effects can impact the shape of the edges of the band. 
On the one hand, leakage of ionizing photons from a matter-bounded nebula (Section~\ref{sec:leakage}), a contribution of X-ray binaries (Section~\ref{sec:alternatives}), and an age of 2\,Myr instead of 1\,Myr, close to the peak of the \heii emission \citep{Schaerer2002}, would all raise \heii/\hb for a given IMF and shift the band towards less top-heavy IMFs; on the other hand, an older population contributing to the Balmer lines would shift it the other way, though the SED fits do not require one (Section~\ref{sec:sed}). The exclusion of the most top-heavy IMFs is therefore robust, whereas that of the least top-heavy ones holds only for a young, radiation-bounded nebula ionized by the stars alone. Finally, the \heii emission depends more on the maximum stellar mass than on the minimum one \citep{Rusta2026}, and our models of Sec. \ref{sec:sed} stop at 500\,$M_\odot$ rather than 1000\,$M_\odot$.
The band also assumes a fully sampled IMF, which however is a good approximation for the inferred stellar mass of $1.3\times10^{5}\,M_\odot$ \citep{rossi26}.

Unlike the line ratio, the \heii luminosity scales with the mass of the burst. The fiducial mass, $1.3\times10^{5}\,M_\odot$ for a Salpeter slope over 50--500\,$M_\odot$ 
is within the $2\times10^{4}$--$6\times10^{5}\,M_\odot$ range inferred for the C1 component of Hebe \citep{Rusta2026}. %
Simulations limit Pop~III starbursts in this epoch to less than $10^{6}\,M_\odot$, even under a strong Lyman--Werner background, because the first supernovae enrich the gas internally \citep{Jeong2026}; and UV-bright Pop~III systems at $z\simeq6$ require either Pop~III star formation in massive atomic-cooling haloes or a very bursty star-formation history \citep{Venditti2025}. AMORE6-B needs neither: its Pop~III stars, $1.3\times10^{5}\,M_\odot$ formed within 4\,Myr, are below this limit, and no older population is required (Section~\ref{sec:sed}).

\subsection{Alternative scenarios}
\label{sec:alternatives}

The alternatives to metal-free stars differ mainly in the mass of the accreting black hole, which sets the temperature of the disc ($T\propto M_{\rm BH}^{-1/4}$; Section~\ref{sec:seds}): stellar-mass and intermediate-mass black holes, in X-ray binaries and ultraluminous X-ray sources, have hot discs that peak in the X-rays, whereas massive black holes of $10^{5}$--$10^{6}\,M_\odot$ (AGN or DCBH) have cooler discs that peak in the extreme ultraviolet.

\paragraph{X-ray binaries and ultraluminous X-ray sources.}
In the comparison of the emission lines (Section~\ref{sec:inference}), the hard disc ($kT_{\rm in}=1$\,keV) added to a BPASS population is as likely as the Pop~III models, whereas the soft disc of intermediate-mass black holes ($kT_{\rm in}=0.2$\,keV) is disfavoured ($\Delta\ln\mathcal{Z}=-18$, although these models are limited to $\log U\leq-2.5$). The SED fits test the hard disc with three stellar populations (Section~\ref{sec:sed}, Table~\ref{tab:sed_scenarios}).
Added to the Pop~III.1 stars, it is not needed and produces at most 1\% of the \heii. Conversely, if X-ray binaries supplied part of the \heii alongside the Pop~III.1 stars, the hardness required of the stars would decrease (Section~\ref{sec:imf}).
With a BPASS $\Zstar=10^{-4}$ population, it reproduces the emission lines but not the photometry ($\Delta\ln\mathcal{Z}=-6.9$). This fit places the stars at an age of 1\,Myr, before any compact object has formed, so that the stars, the X-ray source, and the metals of the gas would all need an earlier generation, and it requires $L_X(0.5$--$8\,{\rm keV})\simeq3\times10^{42}$\,erg\,s$^{-1}$ for 97\% of the \heii, orders of magnitude above the X-ray output of metal-poor star-forming galaxies \citep{Lehmer2021}. The predicted flux, $\simeq1.7\times10^{-16}$\,erg\,s$^{-1}$\,cm$^{-2}$ in the observed 0.5--7\,keV band, is above the typical sensitivity, of order $10^{-16}$\,erg\,s$^{-1}$\,cm$^{-2}$, of the 2.1\,Ms \textit{Chandra} catalogue of the field, in which AMORE6 is not detected \citep{Wang2026clues}: the X-ray data already disfavour this scenario, although a limit measured at the position of AMORE6 is needed to exclude it.
With an older Pop~III.2 population, the X-ray source can instead replace the Pop~III.1 stars as the source of the \heii, with a moderately less likely fit ($\Delta\ln\mathcal{Z}=-3.0$), at three conditions: $L_X\simeq1.8\times10^{42}$\,erg\,s$^{-1}$, the luminosity of about a hundred bright ultraluminous X-ray sources (a single black hole would need $\gtrsim10^{4}\,M_\odot$ at the Eddington rate, and would then have a cooler disc than the one assumed); gas at $\log U\simeq-0.4$; and $2\times10^{6}\,M_\odot$ of Pop~III.2 stars formed 15\,Myr ago, which produce the continuum and 73\% of the \hb emission. The predicted X-ray flux, $\simeq9\times10^{-17}$\,erg\,s$^{-1}$\,cm$^{-2}$, is close to the sensitivity of the \textit{Chandra} catalogue, so that the forced photometry at the position of AMORE6 is needed to test this case as well. However, in this case, the mass of metal-free stars is several times larger than the maximum expected from Pop~III starbursts in simulations of this epoch \citep{Jeong2026}, and their supernovae would have released $1.5\times10^{5}\,M_\odot$ of metals, which would require $M_{\rm dil}\simeq4\times10^{9}\,M_\odot$, 15--40 times the dynamical mass (Section~\ref{sec:enrichment}): at least 92--97\% of the metals must have been expelled from a system with $\sigma_v=55$\,km\,s$^{-1}$, while the remaining gas stayed pristine enough to form Pop~III.2 stars for $\sim10$\,Myr.

\paragraph{Massive black holes.}
Accretion onto a massive black hole alone is excluded by the equivalent widths of the Balmer lines and of \heii$\lambda4686$ (Section~\ref{sec:ionizing_source}), and requires a host population. The hardest configuration of \citet{NakajimaMaiolino2022} ($T_{\rm bb}=2\times10^{5}$\,K, $\alpha_{\rm ox}=-1.2$) added to a BPASS population is disfavoured by the SED fits (Table~\ref{tab:sed_scenarios}): $\Delta\ln\mathcal{Z}=-23$ with the disc visible, whose continuum would dilute the equivalent widths, so that the fit switches it off, and $-3.3$ with the disc hidden, which requires an accreting source of $\simeq2\times10^{43}$\,erg\,s$^{-1}$, i.e.\ a black hole of $\gtrsim10^{5}\,M_\odot$ at the Eddington rate. The predicted X-ray flux, $\simeq7\times10^{-17}$\,erg\,s$^{-1}$\,cm$^{-2}$ in the observed 0.5--7\,keV band, is below the \textit{Chandra} sensitivity at this position \citep{Wang2026clues}, so the existing X-ray data do not test this scenario. Other constraints do not exclude an obscured black hole either: [Ne\,\textsc{iv}]$\lambda2424$ and [Ne\,\textsc{v}]$\lambda3426$ are not detected, N\,\textsc{v}/\lya$\,<0.022$ (\MUSE), the narrow lines trace the potential of the host, and there is no significant broad \ha: $F_{\rm broad}/F_{\rm narrow}<0.13$--0.20 at $3\sigma$ for FWHM 1000--3000\,km\,s$^{-1}$, the width expected for a black hole of $10^{5}$--$10^{6}\,M_\odot$ at this luminosity \citep{GreeneHo2005}.

\subsection{Leakage of ionizing photons and matter-bounded clouds}
\label{sec:leakage}

Several independent observables suggest that AMORE6-B is leaking most of its ionizing photons. The SED fit requires $f_{\rm esc}=0.86\pm0.02$ in the fiducial model, within 4\,Myr of the onset of star formation, and $f_{\rm esc}=0.90$ for a single burst at 1\,Myr. The escape fraction is set by the UV continuum: in a radiation-bounded nebula the nebular continuum of these stars would dominate the rest-frame UV and redden it, and the observed $\beta_{\rm UV}\simeq-3$ (Section~\ref{sec:continuum}) is that of a nebular continuum weakened by the escape. 

\begin{figure}
  \centering
  \includegraphics[width=\columnwidth]{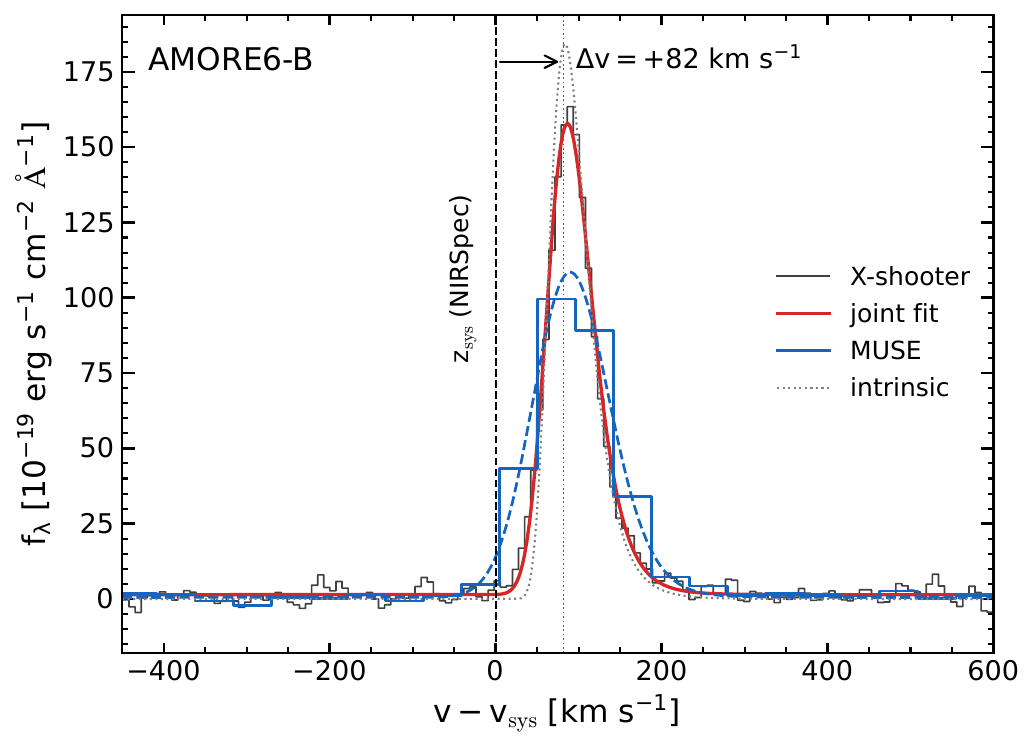}
  \caption{\lya of AMORE6-B from X-shooter (black; slit spectrum) and \MUSE (blue; total flux), in velocity relative to the systemic redshift measured with \NIRSpec (dashed vertical line). The red and blue dashed curves are the joint fit convolved with the line-spread functions of X-shooter and \MUSE, and the dotted curve is the intrinsic profile; the arrow marks the offset of the peak.}
  \label{fig:lya}
\end{figure}

The \lya emission leads to the same picture. Fig.~\ref{fig:lya} shows its profile from archival \MUSE and X-shooter spectra (Section~\ref{sec:continuum}), fitted jointly in the two images with an asymmetric Gaussian \citep{Shibuya2014} convolved with the line-spread function of each instrument. The line is narrow (intrinsic FWHM $55\pm4$\,km\,s$^{-1}$) and asymmetric towards the red, with no blue peak (blue-to-red flux ratio $<0.02$ at $3\sigma$ between $-800$ and $-100$\,km\,s$^{-1}$, although the IGM suppresses most of the blue side), and peaks at only $+82\pm1$\,km\,s$^{-1}$ from the systemic redshift ($+80$\,km\,s$^{-1}$ in AMORE6-A); such small offsets are found in Lyman-continuum leakers \citep{Verhamme2015,Verhamme2017}. For a static slab, the peak offset corresponds to $N_{\rm HI}\simeq(5$--$7)\times10^{18}$\,cm$^{-2}$ for $T=1$--$2\times10^{4}$\,K \citep{Neufeld1990,Dijkstra2006} (though we note that $a\tau_0\simeq70$--200 is at the edge of the validity of the approximation). Such a column is opaque to the Lyman continuum ($\tau_{912}\simeq30$): in the `picket-fence picture' adopted by the SED fitting, the ionizing photons escape through channels of much lower column, and the \lya photons that cross them emerge close to the systemic velocity, where the IGM absorbs most of them. 

In the \MUSE cube the line is more extended than the UV continuum (FWHM of $0.71''$ against $0.55''$), as expected from resonant scattering. The total \lya flux, $(3.45\pm0.03)\times10^{-17}$\,erg\,s$^{-1}$\,cm$^{-2}$, gives \lya/\hb$\,=17.0^{+1.6}_{-1.3}$, i.e.\ $f_{\rm esc}({\rm Ly}\alpha)=0.71\pm0.07$ for Case-B recombination, or $0.43\pm0.04$ if collisional excitation is included as in the \Cloudy nebula of the fiducial model (\lya/\hb$\,\simeq38$); both are lower limits, since the IGM still absorbs part of the emission close to the systemic velocity. Finally, the continuum is detected blueward of \lya, in F070W ($8.7\pm1.7$\,nJy; Table~\ref{tab:photometry}), where only the flux transmitted by the \lya forest reaches us, so that the line of sight is not blocked by dense neutral gas around the source. Together, these observables provide a strong support to the high escape fraction inferred by our modeling, whose value, however, rests mainly on the photometric UV slope (Section~\ref{sec:sed}), which \lya and the F070W flux support but do not measure.

Such a high escape fraction at an age below 4\,Myr is not expected for present-day-like leakers. In the MEGATRON simulations \citep{Choustikov2026}, escape starts only after $\gtrsim2.5$--10\,Myr, once stellar feedback has opened channels through the birth cloud; the leakers have low densities ($n_{\rm e}\simeq30$--100\,cm$^{-3}$), the strongest ones have weak emission lines, and escape fractions estimated from the UV slope are too high for stars younger than 5\,Myr. AMORE6-B has the low density of these leakers ($\log(n_{\rm e}/{\rm cm^{-3}})<2.8$; Section~\ref{sec:helium}) but combines a high escape fraction with strong lines and a young age. The difference may lie in the powering source: the ionizing output of $\sim10^{5}\,M_\odot$ of Pop~III.1 stars in a low-mass halo may ionize the birth cloud almost at once, so that a Str\"omgren sphere bounded by neutral gas hardly forms, and the photons escape through gas that is ionized but too tenuous to reprocess them. In the picket-fence description of the fit, the nebula covers only $\simeq15\%$ of the sky seen from the stars.

Finally, the fit lowers all the nebular emission by the same factor and leaves the line ratios unchanged, whereas a density-bounded nebula would raise \heii/\hb and lower \oii and $R_3$ (Fig.~\ref{fig:model_comparison}), so that the stars would need a softer spectrum (Section~\ref{sec:imf}). This degeneracy between the hardness of the ionizing spectrum and the geometry of the nebula is illustrated by a fit of the line fluxes with HOMERUN \citep{Marconi2024,Moreschini2026}, which combines single-cloud \Cloudy models, radiation bounded and matter bounded, with free non-negative weights (Fig.~\ref{fig:homerun}, Appendix~\ref{app:model_grids}). Illuminated by a BPASS population with $\Zstar=10^{-5}$ and an age of 12.6\,Myr, a mixture of a low-density radiation-bounded component and dense matter-bounded clouds, truncated at the edge of the He$^{++}$ zone, reproduces all the lines within their uncertainties, including \heii and \civ, except He\,\textsc{i}$\lambda4471$, and gives $12+\log(\mathrm{O/H})=6.37$, close to that of the SED fit. Once the nebula is allowed to leak, the line ratios alone therefore do not require metal-free stars. The continuum however is hardly matched by this scenario: this population has $\log\xi_{\rm ion}=25.1$ on its stellar continuum and gives EW(\hb)$\,\leq160$\,\AA\ in a radiation-bounded nebula, 0.4\,dex and a factor of three below the observed values (25.52 and $\simeq480$\,\AA), and any leakage widens both differences by $1/(1-f_{\rm esc})$. Hence, it cannot supply the ionizing photons that the Balmer lines require per unit UV luminosity, which the Pop~III.1 stars of the fiducial model do (Section~\ref{sec:xiion}).

\section{Conclusions}
\label{sec:conclusions}

We have presented deep JWST/\NIRSpec spectroscopy of AMORE6-B, a strongly lensed galaxy at $z=5.725$, and compared it with photoionization and population-synthesis models to shed light on the nature of such an extremely metal-poor, strong-\heii-emitting system. Our main results are as follows.

\begin{itemize}
  \item The spectrum shows \heii$\lambda1640$ at $35\sigma$, with EW$_0=15.6\pm1.8$\,\AA, and \heii$\lambda4686$ at $5.5\sigma$, with \heii$\lambda4686$/\hb$\,=0.080\pm0.015$ and EW$_0=35\pm8$\,\AA. The only metal lines detected are \oiii$\lambda5007$, at $10\sigma$ but weak ($R_3=0.249\pm0.025$), and \civ, at $8\sigma$; there are no low-ionization metal lines (Section~\ref{sec:measurements}, Fig.~\ref{fig:total_spectrum}, Table~\ref{tab:line_fluxes}).
  \item The hydrogen and helium lines indicate a negligible attenuation and a low density, and favour high temperatures, $T_{\rm e}\simeq2\times10^{4}$\,K or above, with $T_{\rm e}<1.9\times10^{4}$\,K excluded at 95\% confidence. The helium mass fraction, $Y=0.226^{+0.022}_{-0.024}$, agrees with the primordial value within $1\sigma$ (Section~\ref{sec:helium}, Fig.~\ref{fig:helium}).
  \item With $12+\log(\mathrm{O/H})\simeq6.1$ from the $R_3$ calibrations and $6.34\pm0.12$ from the SED fit ($Z\simeq2.5$ and $4.5\times10^{-3}\,Z_\odot$), AMORE6-B is the most metal-poor galaxy whose oxygen abundance has been ever derived from emission lines, 0.7\,dex below the extrapolated mass--metallicity relation at $z>3$ (Sections~\ref{sec:empirical_oh} and \ref{sec:sed}, Fig.~\ref{fig:mzr_emp}). The \civ emission allows $-0.55\lesssim\log(\mathrm{C/O})\lesssim+0.4$ (95\%), at or preferentially above the plateau of metal-poor galaxies (and consistent with yields of PopIII SNe), the main systematics associated with the temperature and on the ionization correction factors (Section~\ref{sec:co}, Fig.~\ref{fig:co}).
  \item We compared all the available emission lines with the predictions from six families of models: simulated Pop~III galaxies of NEFERTITI at successive stages of enrichment, metal-free stellar populations with three IMFs, extremely metal-poor binary populations, very massive stars, metal-poor populations with X-ray binaries or ultraluminous sources, and massive black holes (AGN or DCBH). The lines favour metal-free stars with a top-heavy IMF, in the self-polluted or Pop~III-rich phases, or a metal-poor population with a hard X-ray source, and exclude metal-poor populations without such a source, very massive stars, less top-heavy Pop~III IMFs, and massive black holes (Sections~\ref{sec:seds}, \ref{sec:ionizing_source}, and \ref{sec:inference}, Figs.~\ref{fig:model_comparison} and \ref{fig:line_evidence}).
  \item The full SED of AMORE6-B is best reproduced with a star-formation history of a NEFERTITI Pop~III galaxy, with $1.3\times10^{5}\,M_\odot$ of Pop~III.1 stars formed within the last 4\,Myr, ionizing gas at $Z\simeq4.5\times10^{-3}\,Z_\odot$ ($12+\log(\mathrm{O/H})=6.34$) and letting 86\% of their ionizing photons escape (Section~\ref{sec:sed}, Table~\ref{tab:sed_scenarios}, Fig.~\ref{fig:sed}). 
  Under such a scenario, the pair-instability supernovae exploded so far have released $\simeq3\times10^{3}\,M_\odot$ of metals, of which the ionized gas holds only 0.2--2\%, consistent with the observed abundance if the metals have been mostly expelled and/or have not yet fully mixed from the hot supernova ejecta. 
  A single Pop~III.1 burst, with or without a hard X-ray source, is as likely as the fiducial history, but would still require a previous generation of stars that enriched the metal, which mass the data cannot constrain.
  \item Under our analysis, alternative scenarios without metal-free stars are  less likely: X-ray binaries in an older Pop~III.2 population require $2\times10^{6}\,M_\odot$ of metal-free stars, several times the largest Pop~III starbursts in simulations, whose metals must have been almost entirely expelled; X-ray binaries in a metal-poor population require stars too young to host compact objects and an X-ray flux above the typical sensitivity of the \textit{Chandra} data of the field; a massive black hole fits the SED only in an obscured, type-2-like configuration, but is then disfavoured by the emission lines alone (Sections~\ref{sec:sed} and \ref{sec:alternatives}, Appendix~\ref{app:sed_scenarios}).
  \item The \heii-to-Balmer line ratio allows us to constrain the most likely IMF slope and characteristic mass of the Pop~III stars hosted by AMORE6. The inferred Pop~III IMFs are consistent with constraints from both near-field cosmology, which disfavor the flattest and most top-heavy IMFs, and far-field cosmology, through the Hebe galaxy, which disfavor the steepest and least top-heavy IMFs (Section~\ref{sec:imf}, Fig.~\ref{fig:imf}).
  \item The high escape fraction of ionizing photons required by the SED fit is supported by independent observables: the very blue UV slope ($\sim -3$), the narrow \lya line peaking only $\simeq80$\,km\,s$^{-1}$ from the systemic redshift, with $f_{\rm esc}({\rm Ly}\alpha)\gtrsim0.4$--0.7, and the continuum detected blueward of \lya. Such a high escape fraction at an age below 4\,Myr is not expected in simulations of present-day-like leakers (Section~\ref{sec:leakage}, Fig.~\ref{fig:lya}).
\end{itemize}

In conclusion, AMORE6-B shows the combination of properties expected for a `self-enriched' galaxy dominated by Pop~III stars: a hard ionizing radiation that only metal-free stars with a top-heavy IMF can supply, gas enriched to $\simeq0.5\%$ of the solar abundance by the first pair-instability supernovae, and a stellar population young and compact enough to let most of its ionizing photons escape, turning AMORE6-B into a benchmark for the search for the first stars. 
Excluding the remaining alternatives definitively will require a more complete modelling, which treats self-consistently the star-formation history, the chemical enrichment and mixing of the gas, its geometry and the escape of ionizing photons, and the contribution of accreting sources.

\section*{Data and software availability}

The \JWST/\NIRSpec data of programme GO~9645 and the \JWST/\NIRCam and \textit{HST} imaging of Abell~2744 used in this work are available from the Mikulski Archive for Space Telescopes (MAST) at the Space Telescope Science Institute. The \NIRCam mosaics and empirical PSFs are those of the UNCOVER and Medium Bands, Mega Science third data release \citep{Bezanson2024,Weaver2024,Suess2024}. The archival VLT/X-shooter (ESO programme 115.29G6.001) and \MUSE \citep{Richard2021} data are available from the ESO Science Archive Facility. The lens model of \citet{Bergamini2023b} is publicly available. The line fluxes, slit-loss corrections, photoionization grids, and SED-fit posteriors derived in this work will be shared on reasonable request to the corresponding author.

The \NIRSpec data were reduced with the \NIRSpec GTO pipeline \citep{CurtisLake2026,Scholtz2026DR4} (CRDS context \texttt{jwst\_1584.pmap}). This work also made use of \textsc{unite} \citep{Hviding+2025,Hviding2026}, \texttt{msafit} \citep{deGraaff2024}, \textsc{grizli} \citep{Brammer2023grizli}, \textsc{Lenstool} \citep{Jullo2007}, \textsc{PyNeb} \citep{Luridiana2015}, \Cloudy \citep{Gunasekera2025}, \textsc{bagpipes} \citep{Carnall2018}, \textsc{nautilus} \citep{Lange2023}, HOMERUN \citep{Marconi2024}, and of the stellar population models of \textsc{Yggdrasil} \citep{Zackrisson2011}, BPASS \citep{Eldridge2017,StanwayEldridge2018}, and Starburst99 \citep{Leitherer1995}.

\section*{Acknowledgements}

We are extremely grateful to the DAJE team for proposing and leading the GO~9645 programme.
Generative-AI tools (Anthropic Claude Opus 5.5) were used to assist with code development, figure preparation, and language editing; all the scientific analysis and interpretation, numerical results, references to literature work, and final text have been validated and remain under the full responsibility of the authors.

\appendix
\onecolumngrid

\section{Lensing, astrometry, and slit-loss systematics}
\label{app:lensing}

\paragraph{Lens-model posterior.}
We use the \textsc{Lenstool} model of \citet{Bergamini2023b}, described in
Section~\ref{sec:lensmodel}.  Each sample of the published chain specifies all
180 dPIE mass components through the individually optimised haloes and the
cluster-member scaling relations.  Our implementation reproduces the
\textsc{Lenstool} magnification map around AMORE6 to better than $10^{-5}$.
From 5000 posterior samples we recover
$\mu_\mathrm{A}=39.2^{+4.4}_{-3.7}$ and
$\mu_\mathrm{B}=77.8^{+8.2}_{-6.9}$ at the model-predicted image positions,
consistent with the published values.

\paragraph{Image-position residual and astrometry.}
The model-predicted images are displaced from the observed positions by
$0.15''$ for A and $0.23''$ for B.  Evaluating the same posterior at the
observed positions instead gives
$\mu_\mathrm{A}=33.5^{+4.5}_{-3.8}$ and
$\mu_\mathrm{B}=99.5^{+19.7}_{-14.3}$; the difference is caused by the steep
magnification gradient near the critical curve and the lens-model residual,
not by the imaging astrometry.  Within $20''$ of AMORE6, UNCOVER and
GLASS-JWST agree to $\simeq0.01''$ per coordinate; 72 F444W stellar centroids
and the three secure \textit{Gaia}~DR3 matches give median radial offsets of
$0.005''$ and $0.009''$, respectively.  The published coordinate of AMORE6-B
is $0.098''$ from the GO~9645 target coordinate, whereas the latter lies only
$0.02$--$0.04''$ from B in the foreground-subtracted NIRCam images.  This is a
local difference in source-position definition, not a global WCS error.

\paragraph{Realised position in the MSA.}
The delivered flight metafiles for the G140M, G235M, and G395M visits all
encode the same position of AMORE6-B in their three nods:
$(x,y)=(0.662,0.570)$, $(0.668,0.581)$, and $(0.656,0.558)$ in shutter-pitch
coordinates.  These correspond to across-shutter offsets
$u=-0.0436$, $-0.0452$, and $-0.0419''$, respectively, where the open
half-width is approximately $0.10''$.  An apparent additional displacement of
$\simeq56$\,mas arose when the fixed APT coordinate was projected through the
post-acquisition science WCS.  The first MSATA exposure and the acquisition
confirmation differ by $38.4$\,mas east, $6.7$\,mas south, and $52.2''$ in
roll, whereas the confirmation and first science exposure agree to $1.5$\,mas.
Removing this common acquisition attitude rotation reduces the apparent
APT--metafile displacement across the shutter to $<0.4$\,mas in every nod.
The 56-mas value was thus the successful MSATA attitude correction counted a
second time as source motion, and not a measured offset of AMORE6-B within its
shutter.  Independently, the H$\gamma$--\ha velocity difference gives a
Balmer-emission position $+0.002\pm0.033''$ relative to the metafile value.
We therefore use the flight-metafile coordinates for the wavelength zero
point, path loss, slit-loss correction, and line-spread function.

\paragraph{Differential magnification.}
We recomputed the magnification using the same Gaussian morphologies adopted
for the slit-loss calculation: the two-component F115W model for the UV
continuum and the continuum-subtracted \hb$+$\oiii and \ha maps for the line
emission.  For a surface-brightness distribution $I(\boldsymbol\theta)$ and a
weight $W$ representing either the whole image or the PSF-convolved open
shutter, we evaluate
\begin{equation}
  \mu_W =
  \frac{\int W I\,\mathrm{d}^2\theta}
       {\int W I/|\mu(\boldsymbol\theta)|\,\mathrm{d}^2\theta},
\end{equation}
averaging the shutter result over the three flight nods.  At the
model-predicted position the whole-image values are
$78.4^{+7.9}_{-6.8}$, $81.7^{+8.3}_{-7.1}$, and
$84.5^{+8.6}_{-7.4}$ for the UV, \hb$+$\oiii, and \ha morphologies,
respectively.  The corresponding shutter-weighted values are
$80.1^{+8.2}_{-6.9}$, $82.9^{+8.4}_{-7.3}$, and
$84.2^{+8.6}_{-7.4}$.  Thus the shutter samples a UV magnification only
$2.1\%$ above the whole-image value.  For total-flux line-to-continuum ratios,
the \hb$+$\oiii and \ha emission are magnified by $4.1\%$ and $7.7\%$ more
than the UV continuum; lines assigned the same morphology are unaffected.
These effects are small compared with the $\pm0.10$-dex
absolute-magnification systematic.  Moreover, the observed B/A ratios,
$1.44\pm0.42$ in \hb and $1.38\pm0.05$ in F115W \citep{Morishita2026}, are
below the model value $1.98^{+0.19}_{-0.17}$ and favour, if anything, a smaller
$\mu_\mathrm{B}$.  We retain $\mu_\mathrm{B}=77.8^{+8.2}_{-6.9}$ and the
symmetric $\pm0.10$-dex systematic for absolute quantities.

\begin{figure*}
  \centering
  \includegraphics[width=\textwidth]{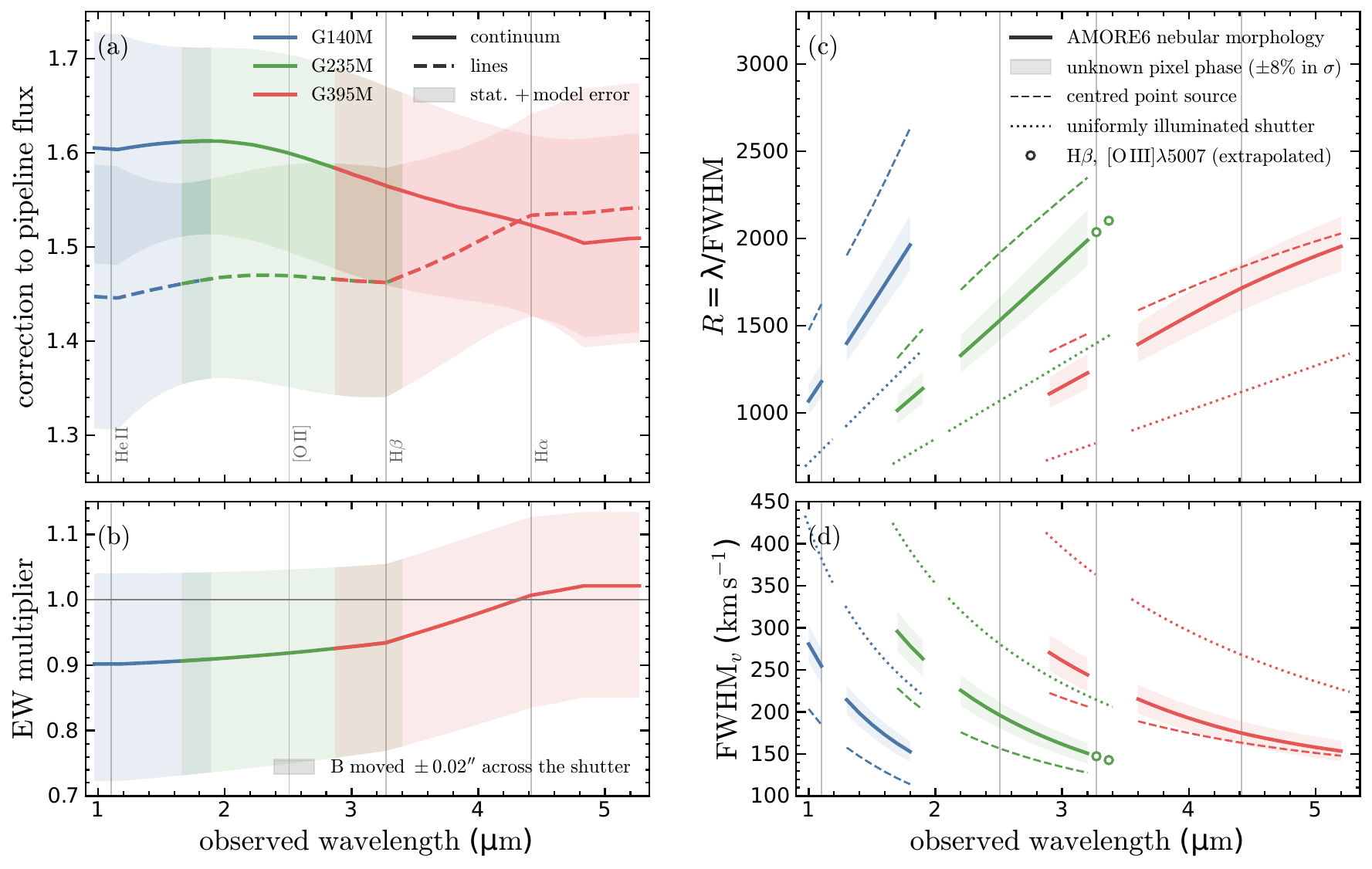}
  \caption{Slit-loss corrections and effective line-spread function of
  AMORE6-B at the flight-metafile positions for the 5-pixel extraction.
  \textit{(a)} Multiplicative correction to the pipeline-based fluxes --- which already account for the position-dependent point-source path loss --- for 
  the F115W continuum morphology (solid) and the nebular morphology (dashed).
  The latter uses the \hb$+$\oiii map up to \hb and interpolates to the \ha map
  at longer wavelengths.  The shaded regions include the statistical and
  morphology-model uncertainties.  \textit{(b)} Correction to an equivalent
  width measured directly from the pipeline spectrum; the band shows the
  effect of a $\pm0.02''$ displacement across the shutter.  \textit{(c,d)}
  Gaussian-equivalent resolving power and velocity FWHM from \texttt{msafit}
  for the nebular morphology (solid), a centred point source (dashed), and a
  uniformly illuminated shutter from the STScI dispersion files (dotted).
  The light bands show the $\pm8\%$ uncertainty in LSF width from the unknown
  detector pixel phase.  Open circles mark the short extrapolations at \hb and
  \oiii$\lambda5007$ in G235M beyond the 3.2-$\mu$m limit of the PSF library.
  All three gratings use their delivered flight metafiles.}
  \label{fig:slitloss_lsf}
\end{figure*}

\paragraph{Slit losses and effective LSF.}
Relative to the position-dependent point-source path loss already applied in
the reduction, the morphology-aware line corrections are 1.426, 1.424, and
1.482 at \heii$\lambda1640$, \hb, and \ha, respectively.  The corresponding
corrections to equivalent widths measured from the spectrum are 0.902, 0.934,
and 1.007.  Moving the morphology by $\pm0.02''$ changes the
UV EW correction by at most $-20\%$ to $+15\%$, but changes ratios between the
tabulated nebular-line corrections by no more than $4\%$.  The principal
remaining morphology assumption is that the rest-UV lines follow the
\hb$+$\oiii map; if \heii$\lambda1640$ instead followed the compact UV
continuum, its correction relative to \hb would increase by approximately
$11\%$.

An independent \texttt{msafit} calculation agrees with the flight-pathloss
throughputs to within $0.4$--$7.9\%$, for which we retain an 8\% model
uncertainty.  The same forward models give phase-averaged,
morphology-dependent LSFs.  The Gaussian-equivalent resolving powers are
$R=1181$ at \heii$\lambda1640$ in G140M, $R=2036$ at \hb and $R=2101$ at
\oiii$\lambda5007$ in G235M, and $R=1714$ at \ha in G395M, corresponding to
FWHM values of 254, 147, 143, and 175\,km\,s$^{-1}$.  The G235M values at \hb
and \oiii$\lambda5007$ are short extrapolations beyond the available PSF
library.  Because the extended-source profile is narrow and non-Gaussian in
detector pixels, its sampled width depends on the unknown pixel phase; we
therefore allow an 8\% LSF-width nuisance term in the kinematic fits.

\section{Robustness of the helium abundance derivation}
\label{app:helium}

We tested the sensitivity of the joint fit of the hydrogen and helium lines (Section~\ref{sec:helium}) to the assumptions that the data do not fully constrain. With the broader temperature prior $1\leq T_{\rm e}/10^{4}\,{\rm K}\leq4$, the posterior rises towards $4\times10^{4}$\,K ($T_{\rm e}>1.9\times10^{4}$\,K at 95\% confidence) with $\chi^{2}=22$, and the other parameters are unchanged within their uncertainties ($Y=0.224$, $A_V=0.03$\,mag, $y^{++}/y=0.14$); the tests below were run with that prior. Because $T_{\rm e}$ is not measured, we repeated the fit with $T_{\rm e}$ fixed at $2\times10^{4}$\,K: $Y$ is unchanged ($0.227^{+0.021}_{-0.023}$), as it is when the density is also restricted to $1.5\leq\log(n_{\rm e}/{\rm cm^{-3}})\leq2.5$ ($Y=0.226$). Fixing $C_{395}=1$ gives $Y=0.224$ and $A_V=0.02^{+0.02}_{-0.01}$\,mag, and removing the G395M H$\gamma$ measurement, which leaves $C_{395}$ to its prior, gives $Y=0.224$ and $A_V=0.04^{+0.04}_{-0.03}$\,mag: with \heii$\lambda1640$ in the fit, the attenuation is set mainly by the ratio of the two He\,\textsc{ii} lines and no longer depends on the relative flux scale of G235M and G395M. Leaving out \heii$\lambda1640$ gives $A_V=0.09^{+0.07}_{-0.06}$\,mag, $y^{++}=0.0081\pm0.0015$, and $Y=0.224$; in that case $A_V$ rests on \ha/\hb and on $C_{395}$, and fixing $C_{395}=1$ lowers it to $0.03^{+0.03}_{-0.02}$\,mag. Allowing for collisional excitation of H\,\textsc{i}, with the neutral fraction $\xi$ as a free parameter and the rates adopted by \citet{Aver2021jcap}, gives $\log\xi<-3.2$ (97.5th percentile) and $Y=0.239^{+0.026}_{-0.027}$, so that we neglect it. Treating as upper limits only the lines that the measurement fit classifies as non-detections, so that He\,\textsc{i}$\lambda4471$ (${\rm S/N}=2.3$) enters with a Gaussian likelihood, gives $Y=0.226$. The largest change comes from the complete-nod-pattern resampling control of Section~\ref{sec:measurements}, whose fluxes give $Y=0.245^{+0.017}_{-0.019}$ but a poorer fit, $\chi^{2}=36$ for the same 18 measurements. The difference arises mainly from \hb, which is 4\% fainter in that control and sets the normalization $F_{0}$, and from He\,\textsc{i}$\lambda5876$, which is 4\% brighter; we adopt the fiducial fluxes, whose fit is better. The statistical uncertainty of $Y$ is dominated by the weakness of the He\,\textsc{i} lines: besides the H8+He\,\textsc{i}$\lambda3889$ blend, only He\,\textsc{i}$\lambda5876$ is measured at more than $3\sigma$ (${\rm S/N}=3.8$), and six of the seven resolved He\,\textsc{i} lines are only upper limits.

\section{Photoionization grids and priors}
\label{app:model_grids}

\paragraph{Nebular setup.}
The models are computed with \Cloudy version 25.00 \citep{Gunasekera2025}. Each model is a plane-parallel, constant-density cloud illuminated by the ionizing spectrum, with the ionization parameter defined at the illuminated face. The cloud is radiation bounded: the calculation stops where the electron fraction falls below 1\%, with additional stops at a hydrogen column of $10^{24}$\,cm$^{-2}$ and at a temperature of 100\,K, following the single-cloud setup of HOMERUN \citep{Marconi2024}. Lines that form partly in the neutral zone beyond the ionization front ([O\,\textsc{i}]$\lambda6300$ and part of \sii and [N\,\textsc{ii}]) are therefore lower limits. The gas abundances are the solar set of \Cloudy (GASS10) scaled to the required oxygen abundance, without grains or depletion; carbon and nitrogen are then rescaled to the C/O and N/O of \citet{Nicholls2017} at that oxygen abundance, with nitrogen raised by a further 0.2\,dex as in HOMERUN, which at $12+\log(\mathrm{O/H})\leq7$ gives $\log(\mathrm{C/O})=-0.80$ and $\log(\mathrm{N/O})=-1.53$. Pristine gas is represented by metals at $10^{-6}$ of the solar abundance, because a pure hydrogen--helium composition does not converge thermally in \Cloudy; the metal lines of these models are $<10^{-3}$ of those at $12+\log(\mathrm{O/H})=5$. Helium is kept at the GASS10 abundance, He/H$\,=0.085$, about 5\% above the primordial value \citep{Aver2021jcap}, a difference negligible against the uncertainties of the helium lines. The lines are recorded as emergent intensities from a single iteration, and the He\,\textsc{i} multiplets are summed over their components. Equivalent widths are measured against the incident plus nebular continuum of the same model, with the lines removed, in a window of $\pm40$\,\AA\ around each line; for the massive-black-hole family the continuum is that of the accretion disc alone, without a host stellar population, which is why its equivalent widths are small despite its strong \heii. The grids used in the comparison of Section~\ref{sec:inference} and in the SED fits (Section~\ref{sec:sed}) have the same setup, except for N/O at the value of \citet{Nicholls2017} without offset, a spherical geometry with an inner radius of $10^{21}$\,cm (effectively plane parallel), and a stop at 3000\,K for the clouds illuminated by X-ray sources, whose X-ray-heated zone does not converge thermally; they span $-3.5\leq\log(\Zgas/Z_\odot)\leq-2$ and $-3.5\leq\log U\leq0$ in steps of 0.5, at $n_{\rm H}=10^{3}$\,cm$^{-3}$, for stellar ages up to 50\,Myr.

\begin{figure*}
  \centering
  \includegraphics[width=\textwidth,height=0.84\textheight,keepaspectratio]{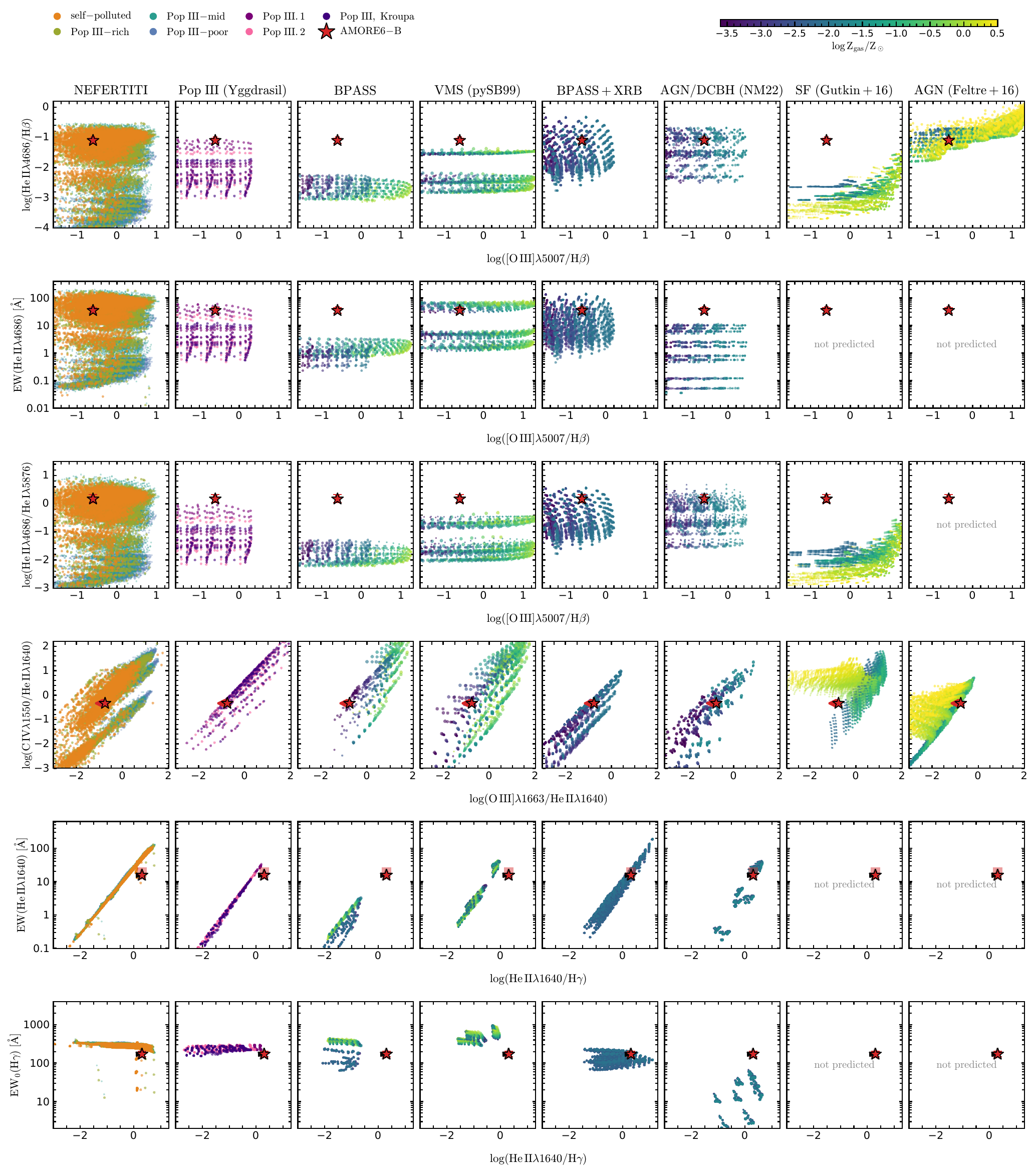}
  \caption{Every model of every family of Table~\ref{tab:sed_families} in the six planes of Fig.~\ref{fig:model_comparison}, one column per family and one row per plane. The NEFERTITI galaxies \citep{Rusta2025}, at all $\log U$, are coloured by evolutionary stage and the \textsc{Yggdrasil} populations by IMF; the other families share one colour scale of gas metallicity, and where a grid has several densities the point size grows with $n_{\rm H}$. The BPASS column includes the $\Zstar=10^{-5}$ population at 1\,Myr and the $\Zstar=10^{-4}$ populations at 1--10\,Myr; the very-massive-star column the $\Zstar=0$ populations at 1 and 3\,Myr and the $\Zstar=10^{-5}$ population at 1\,Myr; the BPASS+XRB column both disc temperatures with $0.01\leq f_Q\leq1$; the accretion column the nine \citet{NakajimaMaiolino2022} continua with both stopping criteria. The last two columns are the published star-forming and AGN grids of \citet{Gutkin2016} and \citet{Feltre2016}, which do not give equivalent widths. AMORE6-B as in Fig.~\ref{fig:model_comparison}.}
  \label{fig:model_grids_appendix_main}
\end{figure*}

\paragraph{Grid axes.}
The base grid has $12+\log(\mathrm{O/H})\in\{\mathrm{pristine},5.0,5.5,6.0,6.5,7.0\}$, $\log U\in\{-3,-2,-1,-0.5\}$, and $n_{\rm H}\in\{10^{2},10^{4},10^{6}\}$\,cm$^{-3}$ for each of the 13 ionizing spectra (two blackbodies at $8\times10^{4}$ and $10^{5}$\,K, which stand in for a Pop~III spectrum of a given hardness; six \textsc{Yggdrasil} Pop~III spectra; two BPASS populations; two very-massive-star populations; the AGN continuum). The accretion branch runs the nine \citet{NakajimaMaiolino2022} continua on the same axes. The BPASS $\Zstar=10^{-5}$ (1\,Myr) and the very-massive-star populations were in addition run on a finer grid ($\log U$ from $-3.5$ to $-0.5$ in steps of 0.5, $\log(\Zgas/Z_\odot)$ from $-2$ to 0 in steps of 0.25 and, for the very massive stars, down to $-6$, $n_{\rm H}$ from $10^{2}$ to $10^{7}$\,cm$^{-3}$ in decades), used for the tracks of Fig.~\ref{fig:model_comparison} and for the BPASS column of Fig.~\ref{fig:model_grids_appendix_main}.

\paragraph{Ionizing spectra.}
The ionizing spectra are described in Section~\ref{sec:seds}. In the diagnostic grid, the \textsc{Yggdrasil} spectra are taken at the tabulated ages closest to 1 and 3\,Myr, the BPASS populations at $\Zstar=10^{-5}$ and $10^{-3}$ and ages of 1 and 3\,Myr, and the very-massive-star populations at 1\,Myr (and 2 and 3\,Myr at $\Zstar=0$). The massive-black-hole continua use the default UV and X-ray slopes of \Cloudy ($\alpha_{\rm uv}=-0.5$, $\alpha_{\rm x}=-1$). All spectra enter \Cloudy as tabulated $\nu F_\nu$ between 10\,\AA\ and 10\,$\mu$m, normalised by the ionization parameter. \citet{NakajimaMaiolino2022} stop their clouds at a neutral column of $10^{21}$\,cm$^{-2}$ rather than at the ionization front; all nine continua were also run with that criterion (648 models). The X-ray-heated, partially ionized zone beyond that column contributes 2--30\% of \hb (medians over the grid), so that every line rises relative to \hb by 0.01--0.4\,dex, the ratios between high-ionization lines are unchanged, and [O\,\textsc{i}]$\lambda6300$, [N\,\textsc{ii}]$\lambda6584$, and \oii fall by up to 3, 2, and 1\,dex, respectively. The NEFERTITI grids of \citet{Rusta2025,Rusta2026}, the star-forming grid of \citet{Gutkin2016}, and the AGN grid of \citet{Feltre2016} are used as published, without recomputation; the first provides equivalent widths of \heii$\lambda1640$ only, which is why that line replaces \heii$\lambda4686$ in the equivalent-width panels.

\begin{figure*}
  \centering
  \includegraphics[width=\textwidth,height=0.80\textheight,keepaspectratio]{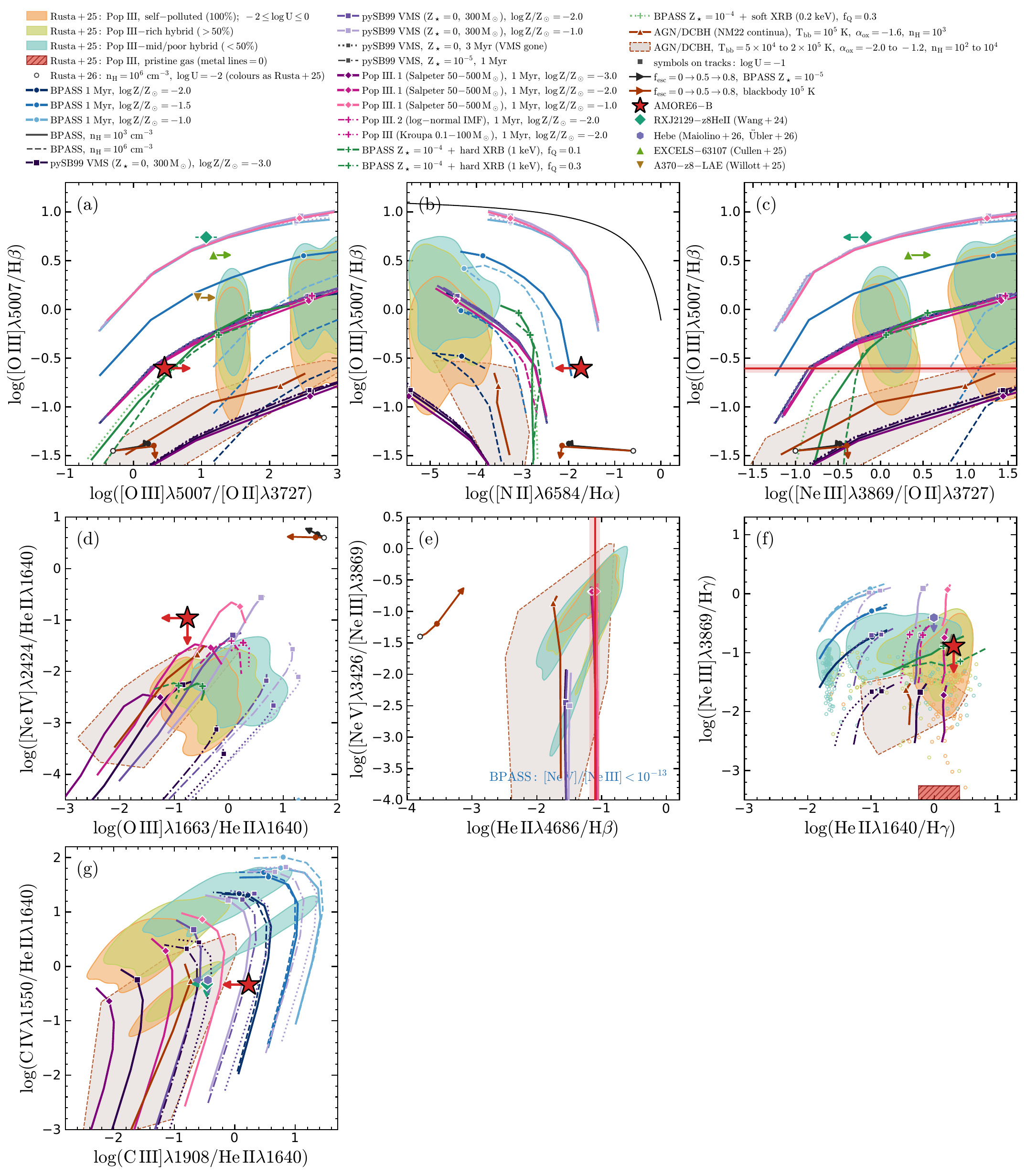}
  \caption{As Fig.~\ref{fig:model_comparison}, for the remaining diagnostic planes: $R_3$ against \textit{(a)} O32, \textit{(b)} [N\,\textsc{ii}]$\lambda6584$/\ha, with the maximum-starburst line of \citet{Kewley2001}, and \textit{(c)} \neiii$\lambda3869$/\oii; \textit{(d)} [Ne\,\textsc{iv}]$\lambda2424$/\heii$\lambda1640$ against O\,\textsc{iii}]$\lambda1663$/\heii$\lambda1640$; \textit{(e)} [Ne\,\textsc{v}]$\lambda3426$/\neiii$\lambda3869$ against \heii$\lambda4686$/\hb; \textit{(f)} \neiii$\lambda3869$/H$\gamma$ against \heii$\lambda1640$/H$\gamma$; \textit{(g)} \civ/\heii$\lambda1640$ against \ciii/\heii$\lambda1640$, with the $3\sigma$ X-shooter limit on \ciii (Section~\ref{sec:measurements}). Where only one axis is measured or constrained, AMORE6-B is drawn as a band.}
  \label{fig:model_comparison_extra}
\end{figure*}

\begin{figure*}
  \centering
  \includegraphics[width=\textwidth,height=0.90\textheight,keepaspectratio]{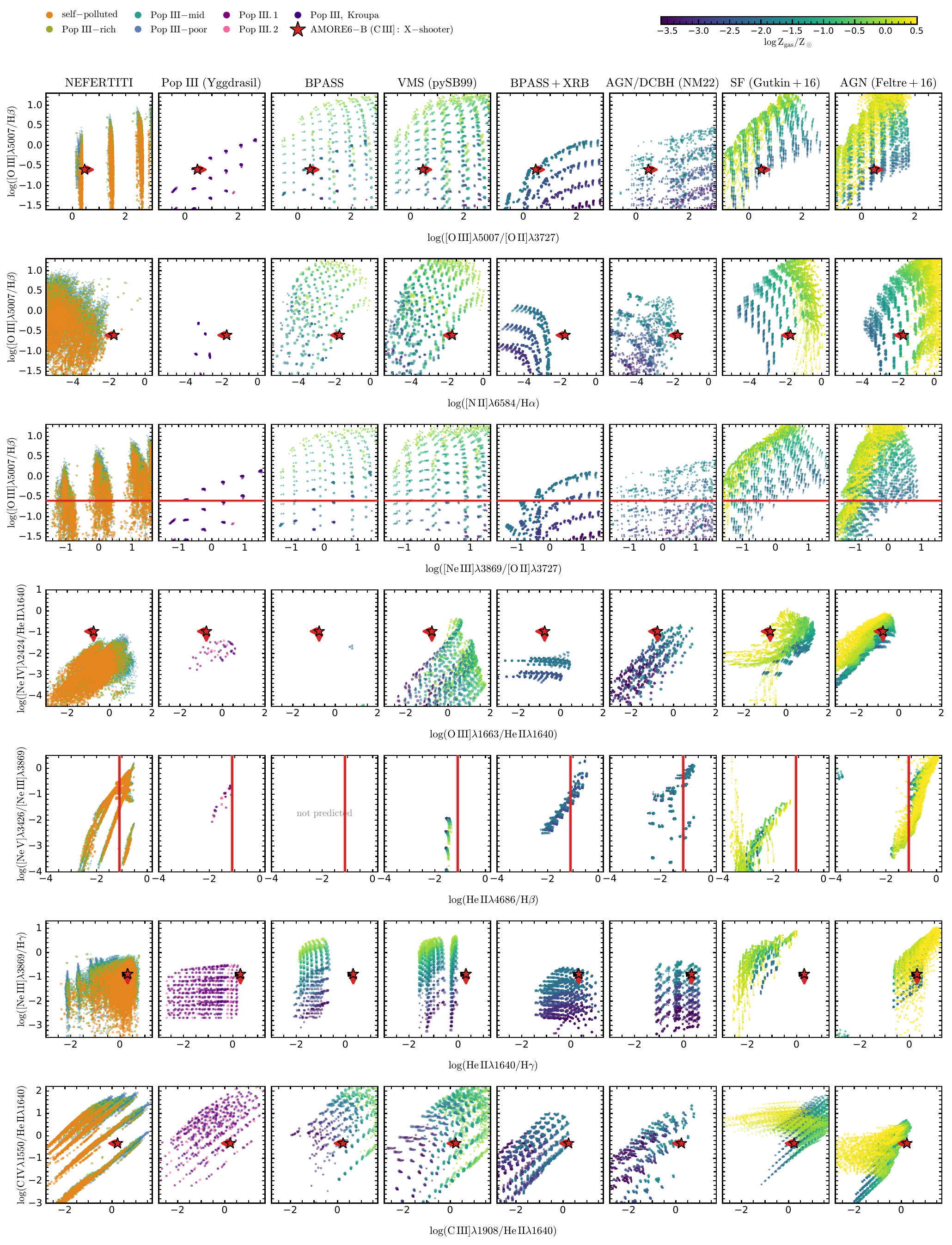}
  \caption{As Fig.~\ref{fig:model_grids_appendix_main}, for the seven planes of Fig.~\ref{fig:model_comparison_extra}.}
  \label{fig:model_grids_appendix_extra}
\end{figure*}

\paragraph{Density and collisional de-excitation.}
The sensitivity of each forbidden line to density in the grid follows its critical density: \oii$\lambda\lambda3727,3730$ and \sii$\lambda\lambda6716,6731$ are suppressed above $\simeq10^{3}$--$10^{4}$\,cm$^{-3}$, \oiii$\lambda5007$ and \neiii$\lambda3869$ above $7\times10^{5}$ and $\simeq10^{7}$\,cm$^{-3}$, \oiii$\lambda4363$ above $\simeq3\times10^{7}$\,cm$^{-3}$, and the UV semi-forbidden lines (O\,\textsc{iii}]$\lambda1663$, N\,\textsc{iv}]$\lambda1486$, \civ) remain unaffected up to $\gtrsim10^{10}$\,cm$^{-3}$. In the grid, raising $n_{\rm H}$ from $10^{4}$ to $10^{7}$\,cm$^{-3}$ lowers $R_3$ by 1.0--1.4\,dex at fixed abundance and $U$, raises \oiii$\lambda4363/\lambda5007$ from $\simeq0.02$ to $\simeq0.7$, and leaves \neiii$\lambda3869$/\hb within 0.2\,dex of its low-density value; these are the behaviours exploited in the model comparison of Section~\ref{sec:inference}. Figs.~\ref{fig:model_grids_appendix_main} and \ref{fig:model_grids_appendix_extra} show every model of every family of Table~\ref{tab:sed_families} in the thirteen planes of Figs.~\ref{fig:model_comparison} and \ref{fig:model_comparison_extra}, without envelopes or parameter cuts, so that those figures can be read as a synthesis of these.

\paragraph{Leakage of ionizing photons.}
The leakage vectors of Fig.~\ref{fig:model_comparison} are computed from radiation-bounded models at $\log(\Zgas/Z_\odot)=-2$, $\log U=-1$, and $n_{\rm H}=10^{3}$\,cm$^{-3}$ by truncating the cloud at the depths where 50\% and 80\% of the hydrogen-ionizing photons escape, for the BPASS $\Zstar=10^{-5}$ population and for the $10^{5}$\,K blackbody. Mixtures of radiation-bounded clouds with clouds truncated at these depths lie along the same vectors; clouds truncated closer to the He$^{++}$ zone, which leak almost all the hydrogen-ionizing photons, move the ratios much further (see below).

\begin{figure}
  \centering
  \includegraphics[width=0.5\textwidth]{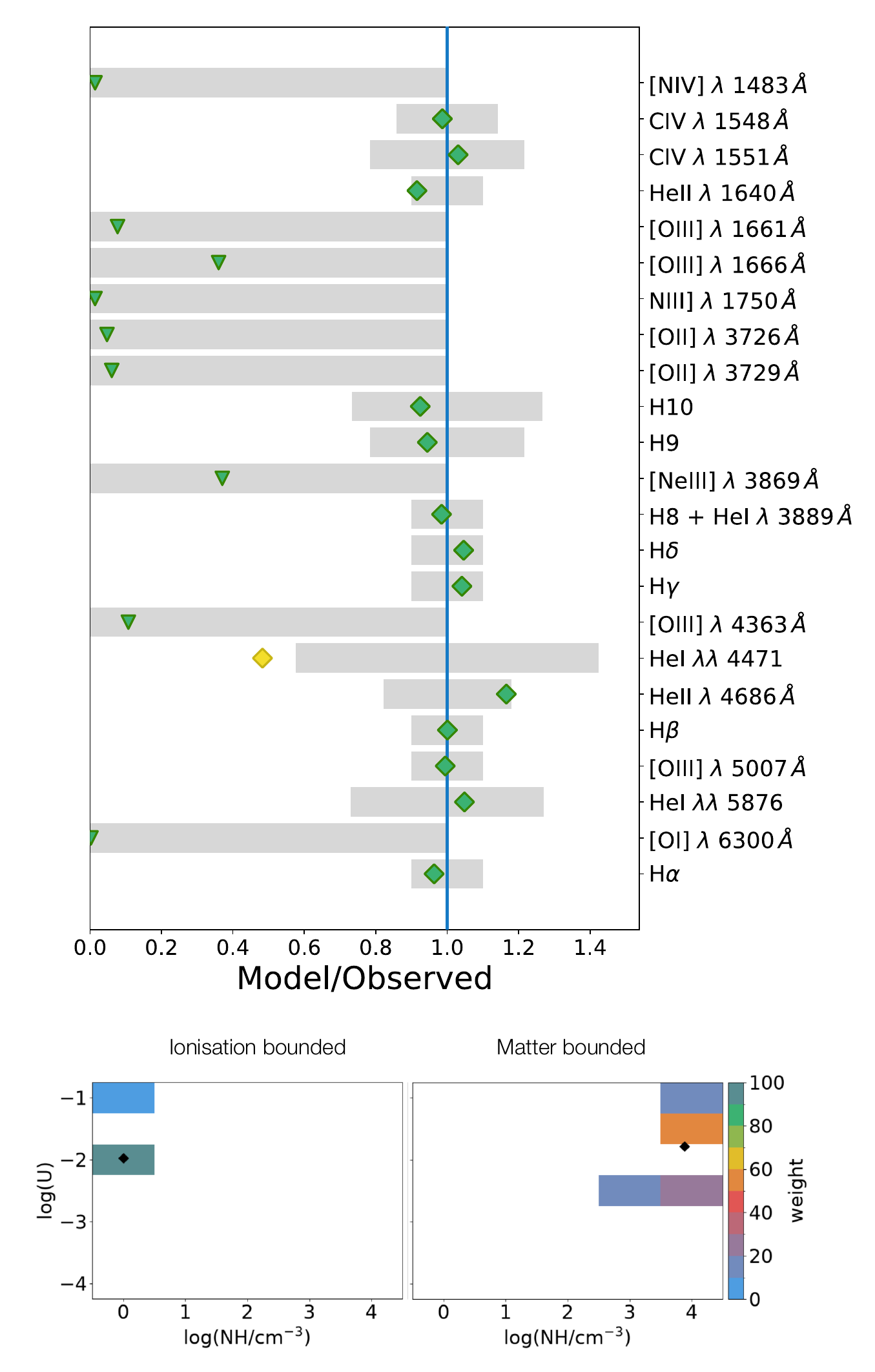}
  \caption{Multi-cloud HOMERUN fit of the emission lines of AMORE6-B with radiation-bounded and matter-bounded clouds, illuminated by a BPASS population with $\Zstar=10^{-5}$ and an age of 12.6\,Myr. \textit{Top:} ratio of the model to the observed flux for each line (diamonds: detections; triangles: upper limits). The grey bars show the observational uncertainty, or the range below the upper limit; the yellow diamond marks the line outside its range. \textit{Bottom:} weights of the single-cloud models in the mixture, in the plane of density and ionization parameter, for the radiation-bounded (left) and matter-bounded (right) clouds; the black diamonds mark the weighted means. The fit uses the line fluxes alone: the continuum of this population is incompatible with the observed $\xi_{\rm ion}$ and EW(\hb) (Section~\ref{sec:leakage}).}
  \label{fig:homerun}
\end{figure}

\paragraph{Multi-cloud fit with matter-bounded clouds.}
HOMERUN \citep{Marconi2024} models the line fluxes as a linear combination of single-cloud \Cloudy models, each normalised to the same \hb luminosity, with non-negative weights determined by least squares; the weight of a cloud is therefore its contribution to the total \hb. Following \citet{Moreschini2026}, the library includes radiation-bounded clouds and matter-bounded clouds stopped where the optical depth at 4\,Ryd reaches 0.5, which maximizes the \heii emission and lets almost all the hydrogen-ionizing photons escape. The ionizing source is a BPASS v2.3 binary population with $\Zstar=10^{-5}$ \citep{Byrne2022}. The best fit (Fig.~\ref{fig:homerun}) is obtained at an age of 12.6\,Myr, with radiation-bounded clouds at $n_{\rm H}\simeq1$\,cm$^{-3}$ and $\log U\simeq-2$ and matter-bounded clouds at $n_{\rm H}=10^{3}$--$10^{4}$\,cm$^{-3}$ and $-2.5\lesssim\log U\lesssim-1$. All the lines except He\,\textsc{i}$\lambda4471$ are reproduced within their uncertainties, and the oxygen abundance is $12+\log(\mathrm{O/H})=6.37$. The fit uses only the line fluxes. The continuum properties of the population quoted in Section~\ref{sec:leakage} ($\log\xi_{\rm ion}=25.1$, $\beta_{\rm UV}=-2.9$, and EW(\hb)$\,\leq160$\,\AA\ for a radiation-bounded nebula) are computed from the BPASS v2.2.1 spectra with the same IMF and metallicity.

\paragraph{Evidence of the model families.}
For a family with models $i=1,\dots,N$, the evidence of Section~\ref{sec:inference} is $\mathcal{Z}=N^{-1}\sum_i\mathcal{L}_i$, i.e.\ each model of the grid (or each NEFERTITI galaxy) has the same prior weight, and no interpolation is applied. The likelihood of a model is Gaussian in the line fluxes, $\mathcal{L}_i\propto\exp(-\chi^{2}/2)$, with $\chi^{2}$ summed over the 16 lines of Table~\ref{tab:line_fluxes} other than \hb (detected or not, with their measured fluxes and uncertainties) and \hb itself; the model fluxes are the predicted ratios to \hb times a normalisation, which is fitted. Each flux has an additional 10\% calibration uncertainty, and each model ratio a fractional uncertainty $\sigma_{\rm mod}$ (0.1\,dex in the fiducial case). The attenuation (SMC curve) and the scale of G140M relative to the other gratings are marginalised over, the former with the split-normal posterior of Section~\ref{sec:attenuation} and the latter with a Gaussian prior of mean 1 and width 0.15. The equivalent widths of \hb, \heii$\lambda4686$, and \heii$\lambda1640$ enter as Gaussian terms in their logarithm. For the BPASS and very-massive-star families computed on the finer grid, which give only EW(\hb), EW(\heii$\lambda4686$) is derived as in Fig.~\ref{fig:model_comparison}. In the combinations of BPASS with an X-ray or accretion source, the two nebulae are computed separately at the same $\Zgas$ and $U$ and added, after scaling each to its share of the hydrogen-ionizing photons; $f_Q$ takes nine values evenly spaced in $\log f_Q$ between $-2$ and 0, and the continuum of the non-stellar source is included in the equivalent widths.

\section{SED fits of the alternative scenarios}
\label{app:sed_scenarios}

\begin{figure}[h]
  \centering
  \includegraphics[width=0.48\textwidth]{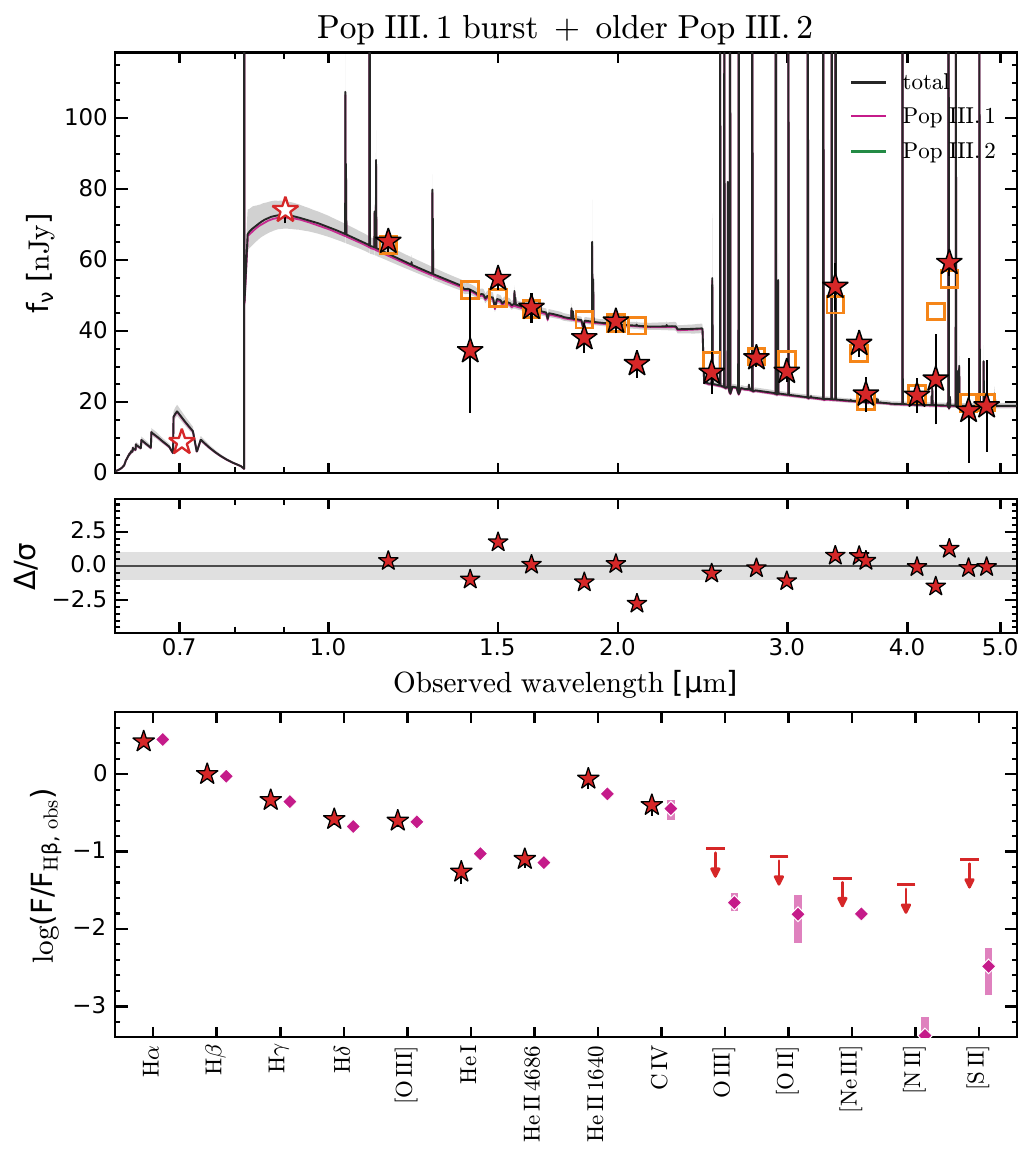}\hfill
  \includegraphics[width=0.48\textwidth]{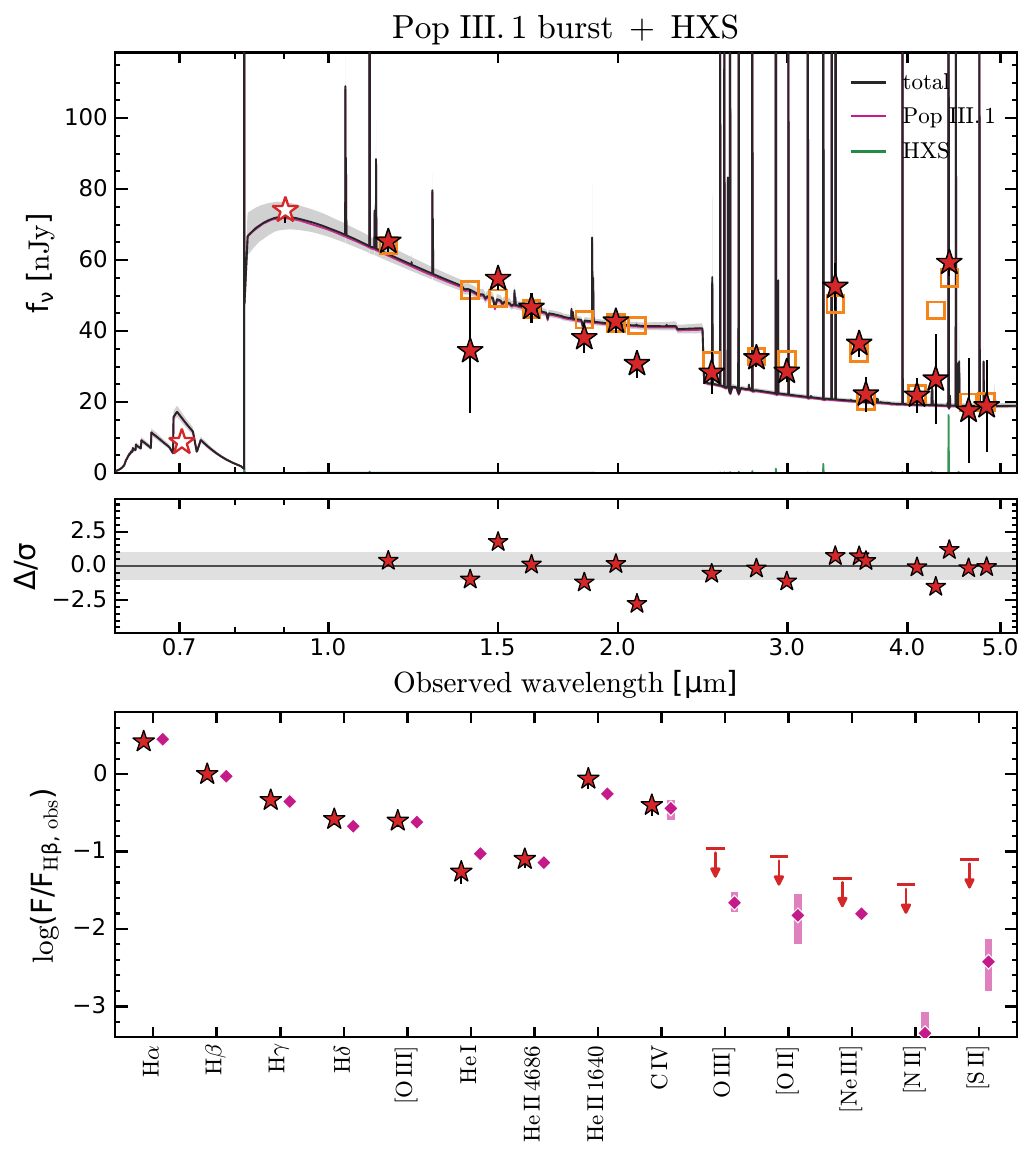}\\[4pt]
  \includegraphics[width=0.48\textwidth]{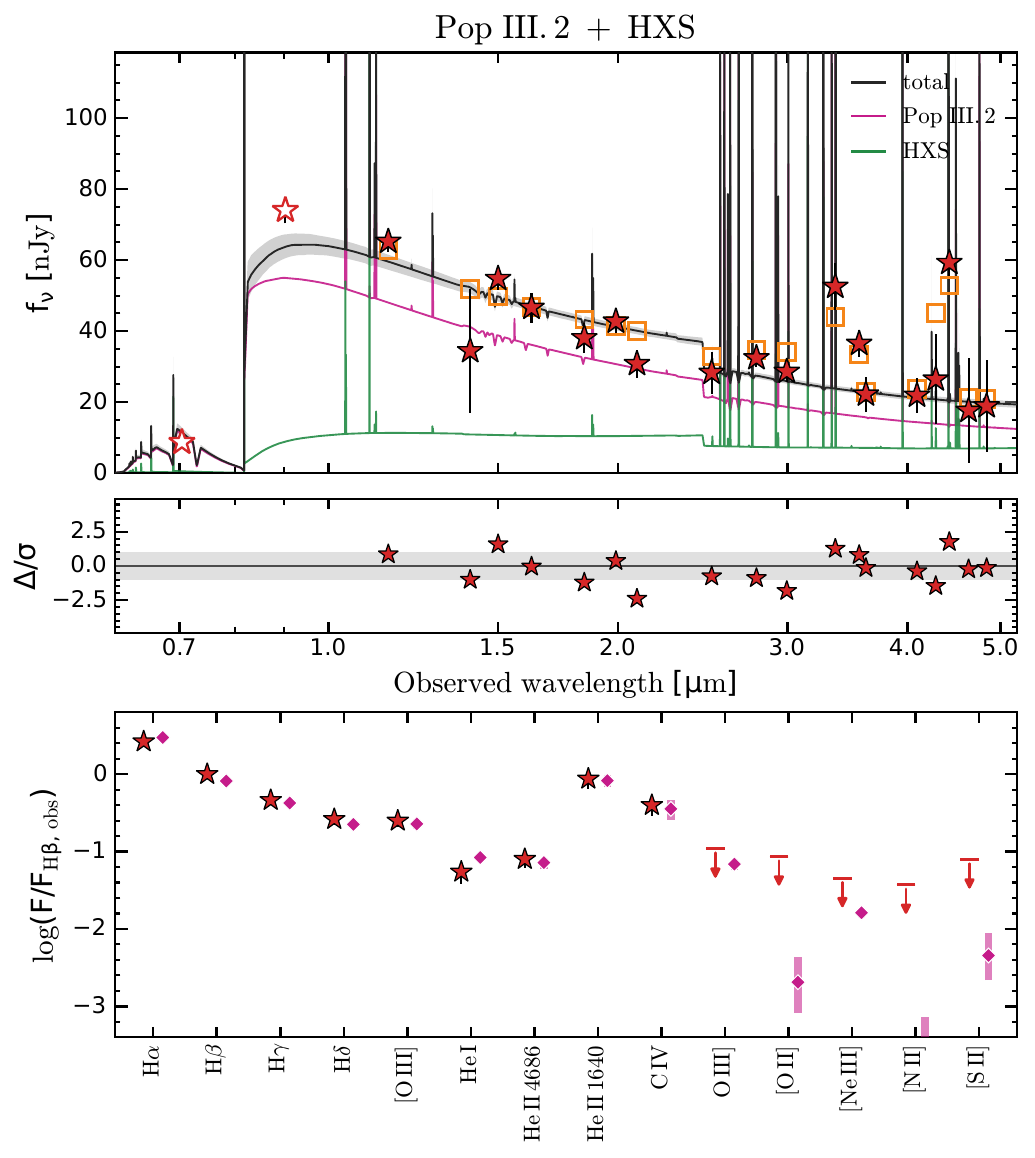}\hfill
  \includegraphics[width=0.48\textwidth]{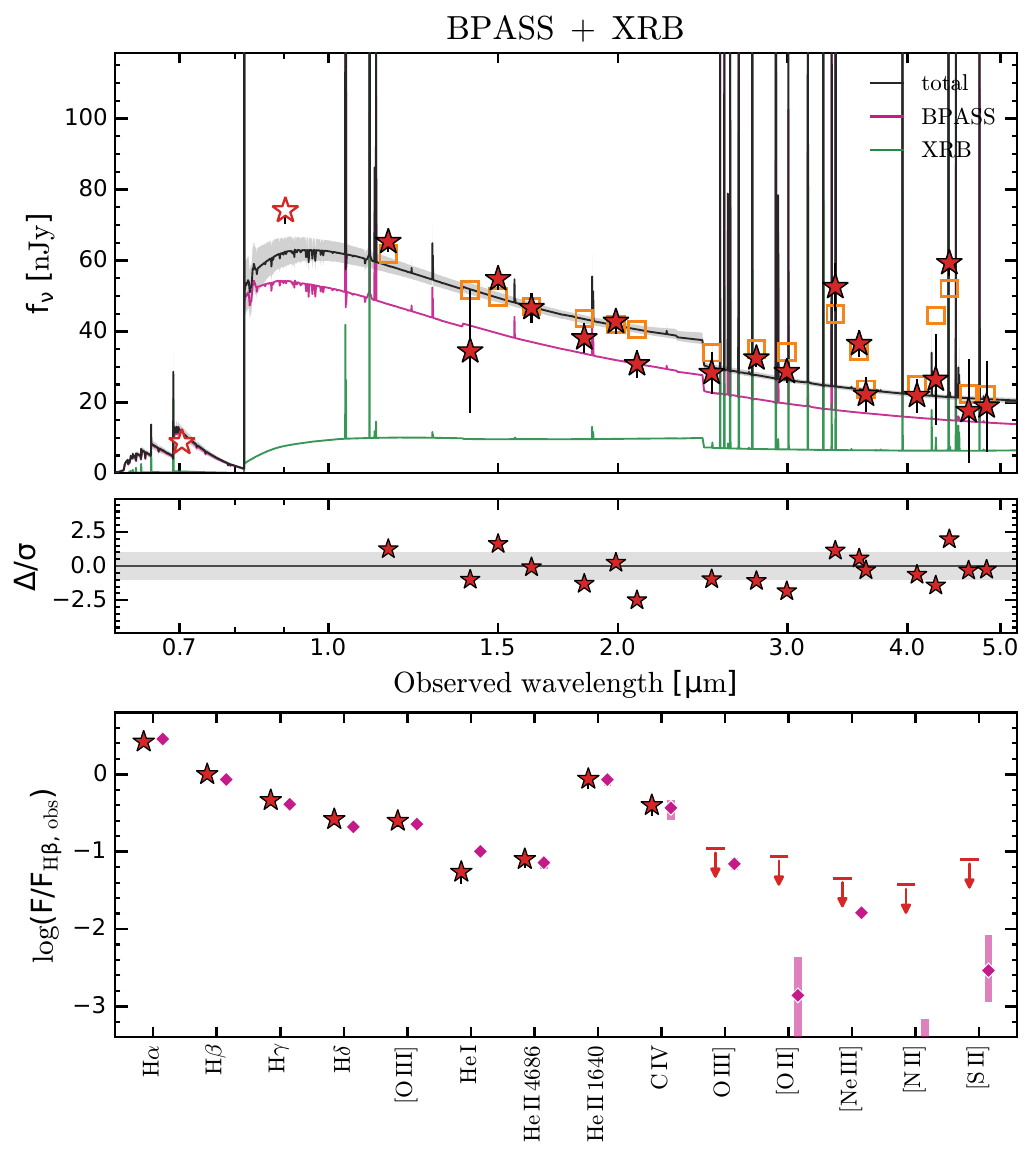}
  \caption{SED fits of the alternative scenarios with a stellar or X-ray-binary source of \heii (Table~\ref{tab:sed_scenarios}). In each panel, top: observed photometry (red stars; open stars are not fitted), median model (black) with its 16th--84th percentile band (grey), its two components (magenta: the young stellar population; green: the older population or the X-ray source, each with its nebula), and model photometry (open squares), with the residuals below; bottom: line fluxes relative to the observed \hb, observed (red stars, with arrows for $3\sigma$ upper limits) and predicted (magenta diamonds, with the 16th--84th percentile range). HXS: hard X-ray source; XRB: X-ray binaries.}
  \label{fig:sed_alt_stars}
\end{figure}

\begin{figure}[h]
  \centering
  \includegraphics[width=0.48\textwidth]{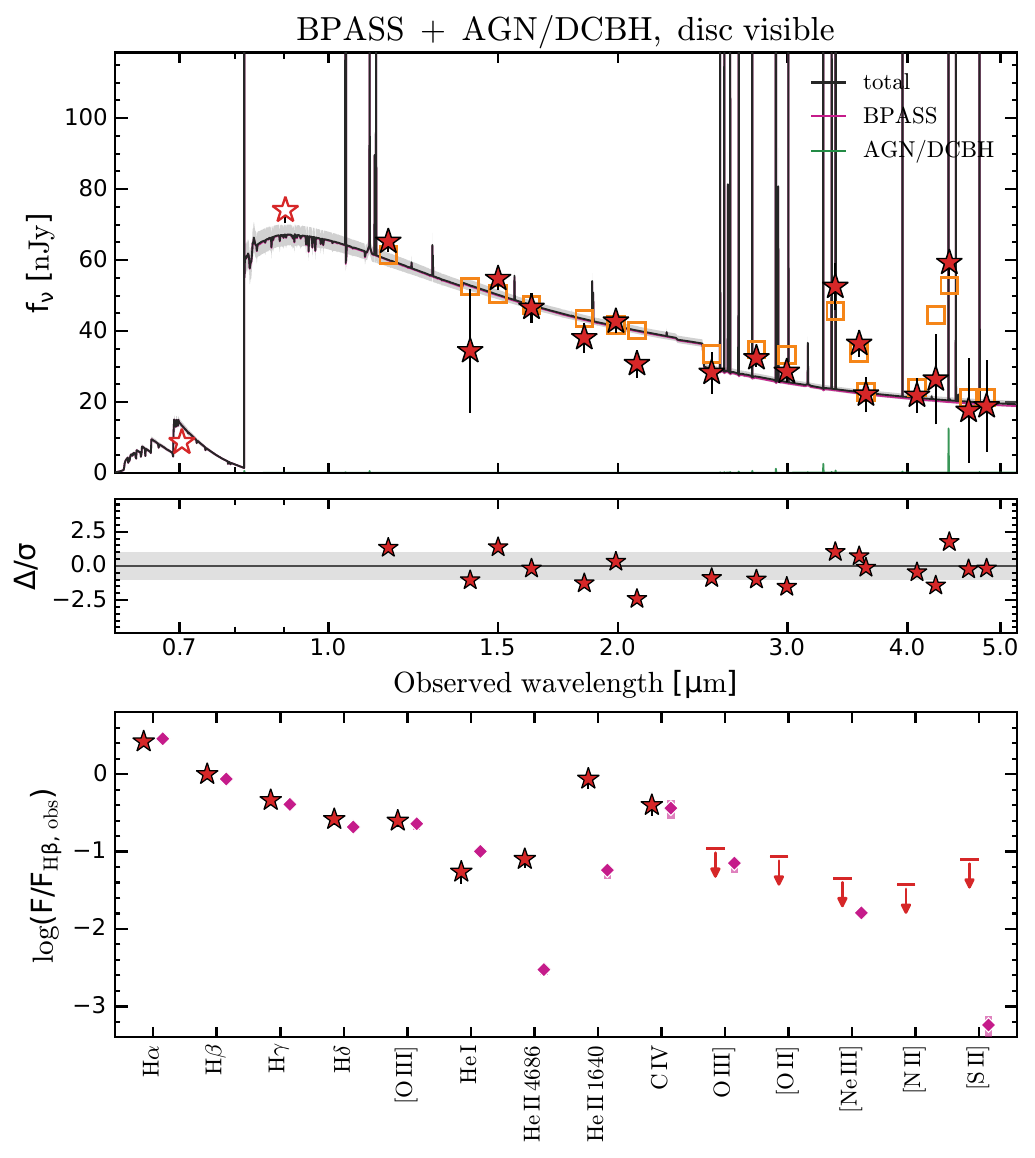}\hfill
  \includegraphics[width=0.48\textwidth]{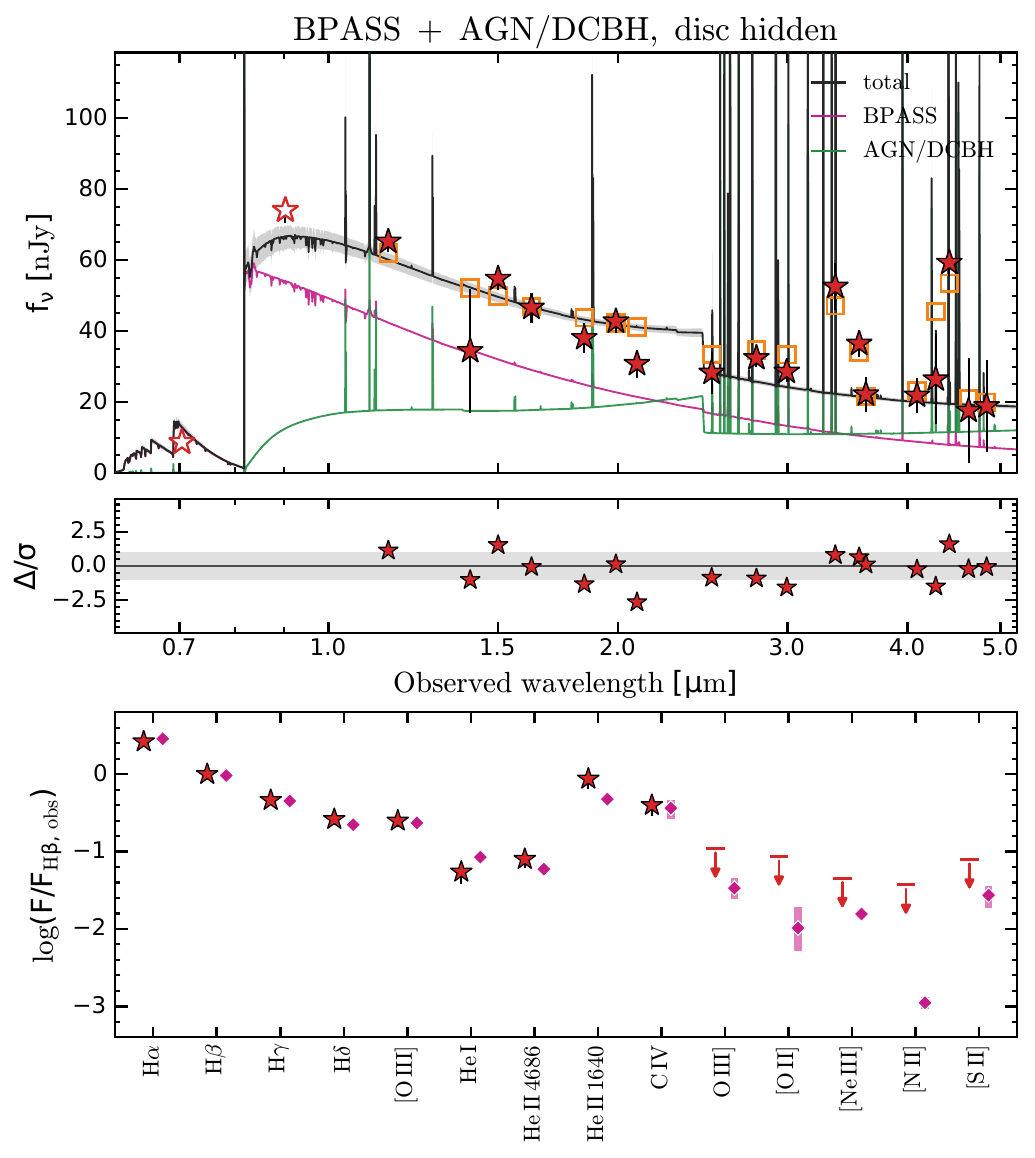}
  \caption{As Fig.~\ref{fig:sed_alt_stars}, for the BPASS population with the massive black hole (AGN or DCBH), with the accretion disc visible (left) or hidden (right). With the disc visible the fit switches the black hole off.}
  \label{fig:sed_alt_bh}
\end{figure}

Figs.~\ref{fig:sed_alt_stars} and \ref{fig:sed_alt_bh} show the SED fits of the six scenarios of Table~\ref{tab:sed_scenarios} other than the fiducial one (Fig.~\ref{fig:sed}), with the same data and setup (Section~\ref{sec:sed}).

\clearpage
\twocolumngrid
\bibliographystyle{apsrev4-1}
\bibliography{references}

\end{document}